\documentclass[aps,prb,floatfix,amsmath,amssymb,preprint,eqsecnum,nofootinbib]{revtex4}
\usepackage{graphicx}
\usepackage{dcolumn}
\usepackage{bm}
\usepackage{epstopdf}

\def\ben{\begin{equation}}
\def\een{\end{equation}}

\let\a=\alpha \let\bb=\beta \let\g=\gamma \let\dd=\delta 
   \let\kk=\kappa
    
\let\s=\sigma \let\tt=\tau

\let\w=\omega \let\G=\Gamma

\def\nn{\nonumber}

\let\pa=\partial
\def\be{\begin{equation}}
\def\ee{\end{equation}}
\def\beq{\begin{equation}}
\def\eeq{\end{equation}}
\def\ba{\begin{array}}
\def\ea{\end{array}}

\def\dalemb#1#2{{\vbox{\hrule height .#2pt
       \hbox{\vrule width.#2pt height#1pt \kern#1pt
               \vrule width.#2pt}
       \hrule height.#2pt}}}

\newcommand{\bea}{\begin{eqnarray}}
\newcommand{\eea}{\end{eqnarray}}

\def\Lag{{\mathcal{L}}}

\def\ocal{{\mathcal{O}}}

\usepackage{setspace} 
\begin{document}

\title{Quantum critical response at the onset of\\ spin density wave order in two-dimensional metals}

\author{Sean A. Hartnoll}
\affiliation{Department of Physics, Harvard University, Cambridge MA
02138}
\author{Diego M. Hofman}
\affiliation{Department of Physics, Harvard University, Cambridge MA
02138}
\author{Max A. Metlitski} 
\affiliation{Department of Physics, Harvard University, Cambridge MA
02138}
\author{Subir Sachdev}
\affiliation{Department of Physics, Harvard University, Cambridge MA
02138}
\date{May 31, 2011 \\
\vspace{1.6in}}

\begin{abstract}
We study the frequency dependence of the electron self energy and the optical conductivity 
in a recently developed field theory
of the spin density wave quantum phase transition in two-dimensional metals. 
We focus on the interplay between the Fermi surface `hot spots' and the 
remainder of the `cold' Fermi surface. Scattering of electrons off the fluctuations of the spin density wave order
parameter, $\phi$, is strongest at the hot spots; we compute 
the conductivity due to this scattering in a rainbow approximation. We point out the importance of composite operators,
built out of products of the primary electron or $\phi$ fields: these have important effects also
away from the hot spots. The simplest composite operator, $\phi^2$, leads to non-Fermi liquid behavior on the entire
Fermi surface. We also find an intermediate frequency window in which the cold electrons loose their quasiparticle form due to 
effectively one-dimensional scattering processes. The latter processes are part of umklapp scattering which leads to
singular contributions to the optical conductivity at the lowest frequencies at zero temperature.
\end{abstract}
\maketitle

\section{Introduction}
\label{sec:intro} 

Many experimental developments in the past few years point to the quantum criticality associated with the onset
of spin density wave order in metals as a crucial phenomenon in the physics of modern correlated electron materials \cite{norman,mv,pt}; 
some recent experimental reports may be found in Refs.~\onlinecite{keimer,dai,louis,kartsovnik}.
It is also interesting that such a quantum critical point leads naturally to a mechanism for higher temperature
superconductivity with $d$-wave-like spin-singlet pairing \cite{scalapino,fink,mazin,kuroki,jiangping,Metlitski:2010zh}.

Theoretically, the problem of the quantum criticality of metals  
has been the focus of much study for several decades \cite{hertz,moriya,millis,advances,vrmp}. In recent work, it is has become
clear that the quantum critical theories are strongly coupled in two spatial dimensions \cite{sungsik,metnem,Metlitski:2010vm}. A $1/N$ expansion, 
where $N$ is the
number of fermion flavors, can be used to organize perturbation theory and derive renormalization group equations
at low orders in $1/N$. However, there are new classes of infrared divergences that appear at higher orders
from fermion excitations on the Fermi surface, and these are not suppressed by the powers of $1/N$ expected
from counting fermion flavors. Nevertheless, much can be deduced about the nature of the strongly-coupled theory,
assuming that the general framework of the field theory survives.

For the onset of spin density wave order in two-dimensional metals, the critical theory \cite{ChubukovShort,advances,Metlitski:2010vm}
is expressed in terms
of fermions at special `hot spots' on the Fermi surface, which are coupled to the fluctuations of a bosonic field, $\phi$,
representing the spin density wave order: an explicit action for this theory appears 
in Section~\ref{sec:lowtheory} for the case of commensurate antiferromagnetic ordering at the wavevector $(\pi, \pi)$ on the square
lattice. We will restrict our attention to this commensurate case in the present paper. 
The hot spots are identified as points on the Fermi surface which are separated by the 
spin density wave ordering wavevector (see Fig.~\ref{fig:hotspots} below). The fermionic excitations at the hot spots lose their quasiparticle character from strong coupling to $\phi$, justifying their prominent role
in the critical theory. 

The transport properties of metals near a spin density wave transition are also clearly of interest. The early
theoretical work \cite{rice,rosch} concluded that these were unlikely to be dominated by the physics of the fermions near the
hot spots. Instead, the cold fermions on the remainder of the Fermi surface would `short-circuit' the electrical current,
and so dominate the electrical conductivity. More recently, the quantum critical conductivity of the hot spot
fermions has been considered \cite{advances} at frequencies or temperatures large enough
so that the momentum dependence of the fermion self energy could be neglected. We will re-examine
the quantum critical conductivity from $\phi$ scattering in the present paper, and find significant differences from the earlier results.
We also note a recent study of the conductivity at non-zero temperatures on the ordered side of the critical point \cite{millispg}: we will
not address this regime here, and will limit our study to the quantum critical point.

It is important to note that the dichotomy between `cold' and `hot' regions of the Fermi surface is intrinsically a weak-coupling
concept, and assumes that the fermion damping can be organized in terms of scattering from quanta of $\phi$ fluctuations. More generally, we should consider the full set of composite operators of the field theory, made up
of multiple primary fermion or $\phi$ operators, and determine how they relax the current of fermions at all points on the Fermi surface. An example of the importance of such composite operators appeared in the work of Pelissetto {\em et al.\/} 
\cite{vicari}, who examined the simpler problem of the onset of spin density wave order in a $d$-wave superconductor.
Here the low energy fermionic excitations reside only at special `nodal points' in the Brillouin zone,
and these are not generically connected by the ordering wavevector. Thus, the analog of the `hot spots'
is absent in this problem, and one would initially conclude that the nodal fermionic excitations are
not strongly scattered by the $\phi$ fluctuations. However, it was shown that a composite operator linked to the square of $\phi$,
which measured Ising-nematic ordering, did 
broaden the nodal fermions, so that the nodal quasiparticles were ultimately only marginally defined near
the spin density wave quantum critical point. For this problem, the critical theory was ultimately under good analytic
control, allowing accurate determination of the influence of composite operators.

This paper will describe various spectral functions of the spin density wave quantum critical point at zero temperature ($T=0$).
We will address the fermion self energy both on and off the hot spots, and the frequency dependent conductivity $\sigma (\Omega)$
{\em i.e.\/} the optical conductivity. However, we will not address the difficult issues associated with $T>0$, and in particular
frequencies with $\Omega < T$. The latter regime is clearly of great experimental importance, but requires an analysis of relaxational
processes which we will not undertake. Such finite $T$ transport has been addressed away from the critical 
point, \cite{mtu,maebashi1,maebashi2,yanase,kontani}
and we hope our critical-point results below at $T=0$ will serve as a prelude to the corresponding analysis at $T>0$.

We will begin in Section~\ref{sec:lowtheory} by recalling the low energy theory of the spin density wave quantum critical point in two spatial
dimensions. In the leading gradient expansion, the fermion energy disperses linearly as a function of wavevector, and so is particle-hole symmetric
about the Fermi surface. As a consequence of this particle-hole symmetry and the ordering wavevector being $(\pi,\pi)$ the effective action acquires an emergent (SU(2))$^{4}$ pseudospin symmetry.\cite{Metlitski:2010vm}
An important point, reviewed in Section~\ref{sec:lowtheory}, is that the electrical current transforms as a vector under
this pseudospin symmetry, while the total momentum is a pseudospin scalar. Consequently, when we perturb the system electrically and create
an electrical current, the resulting state has vanishing momentum, allowing the electrical current to relax to zero even in the absence of any
impurities \cite{nernst}. Thus umklapp scattering processes are implicitly included within our continuum theory.
Provided we regulate the low energy theory in a manner which protects the pseudospin symmetry, the d.c. conductivity will be finite at $T>0$.
This is an attractive feature of the theory allowing, in principle, computation of a $T$-dependent resistivity which depends upon
interactions alone in a non-Fermi liquid.

Section~\ref{sec:hotspot} will begin our computation of the conductivity of the fermions from interactions alone.
We will consider electron scattering 
from $\phi$
fluctuations. We will do this within the framework of a conventional rainbow approximation, computing self energies and corresponding vertex
corrections, while retaining full momentum and frequency dependence. 
We find that at $T=0$ the low frequency optical conductivity takes the form,
\beq \sigma(\Omega) = C_0 \frac{i}{\Omega + i \epsilon} + C_1 + C_2 (-i \Omega)^{r_0} \label{condintro}\eeq
where $r_0 > 0$ is a computable exponent whose value depends only upon the Fermi surface geometry, and is specified
in Fig.~\ref{fig:r0}.
The $C_0$ term in Eq.~(\ref{condintro}) is the non-dissipative Drude contribution, while the constant $C_1$ term is the first dissipative correction. These two terms have the same form as in a Fermi liquid with umklapps, and are dominated by the contribution of cold fermions away from the hot spot. Unlike in certain critical theories which do not posses a Fermi surface, the constant $C_1$ is non-universal. We will explicitly demonstrate that $C_1$ remains finite in the `weak coupling' limit of the theory in Section~\ref{sec:lowtheory}.

The $C_2$ term in Eq.~(\ref{condintro}) is the quantum-critical hot spot contribution to the optical conductivity, which includes the 
contributions of some umklapp processes. Vertex corrections lead to a positive exponent $r_0 > 0$, thus suppressing this term relative to the cold contribution. The result (\ref{condintro}) is too small to 
explain the optical conductivity in the hole-doped cuprates \cite{marel1,marel2}, which has a negative exponent. Our theoretical results here
are at variance with earlier treatments \cite{advances}.

We note a recent numerical study of the optical conductivity near a spin density wave transition \cite{tremblay}, which included only the
first term in the set of rainbow vertex corrections which we have summed. In this approximation, our analysis shows that the singular
term in Eq.~(\ref{condintro}) reduces to $\sigma (\Omega) \sim \log (1/\Omega)$, but the numerical study does not appear to have
the dynamic range to observe this.

Section~\ref{sec:gen} will turn our focus to scattering off composite operators, built out of products of the primary fields 
of the low energy theory of Section~\ref{sec:lowtheory}.
We will describe the general structure of the fermion self energy corrections due to such operators, and the corresponding contributions to the
conductivity. We will also introduce a number of specific composite operators, whose 
fluctuations will be explored in the subsequent sections.
The simplest is the square of the order parameter, $\phi^{2}$, analogs of which were considered in Ref.~\onlinecite{vicari}. 
A second class of composite operators is associated with pairs of fermions: important among them is the Cooper pair operator
and a $2 k_{F}$ charge density wave (CDW) operator (which has an Ising-nematic component), 
fluctuations of which are enhanced near the quantum critical point \cite{Metlitski:2010zh,Metlitski:2010vm}. We emphasize that all these
operators are generated directly from our continuum theory in Eq.~(\ref{eq:theaction}), and do not require reference to the underlying lattice model.

Section~\ref{sec:selfen} will present our results on self energy of the fermions from composite operator scattering, 
focusing on the previously `cold' regions of the Fermi surface away from the hot spots. We will find that scattering off the $\phi^{2}$ operator leads
to fermion self energy which behaves as $\Omega^{3/2}$, assuming leading order scaling dimensions from the low energy field theory. Thus this simplest composite operator is already sufficient to give non-Fermi liquid behavior over the entire Fermi surface.

Also notable will be the contribution from scattering in the $2 k_F$ channel. Here, for fermions in the vicinity of the hot spots, we find an intermediate energy window in which the Fermi surfaces may be regarded as flat, giving rise to effectively one-dimensional Luttinger liquid-like divergences, which destroy the cold fermionic quasiparticles. Whether such divergences sum up to give a power-law behavior of the electron Green's function, as in a true Luttinger liquid, or lead to an instability is subject to further investigation. At lowest energy, the one-dimensional divergences will be cut off by the Fermi surface curvature, which therefore controls the width of the intermediate frequency regime and the possible crossover to coherent Fermi liquid behavior. We further describe how processes saturating the Fermi surface curvature scale change the low energy scaling dimension of $2k_F$ operators.

Section~\ref{sec:cond} will extend the computations of Section~\ref{sec:selfen} to the electrical conductivity. For low-momentum
scattering processes, the contributions to the optical conductivity are less singular than those to the self energy, because they do
not lead to an appreciable degradation of the electrical current. However, this does not apply to the scattering off $2k_{F}$ fluctuations,
which lead to umklapp processes. We estimate that such processes give rise to a power law optical conductivity
as $\Omega \rightarrow 0$ (at $T=0$), 
\beq \delta \sigma(\Omega) \sim (-i \Omega)^{{-b_\kappa}/({2 + b_\kappa})}, \label{sigmaneg}
\eeq
with a negative exponent controlled by the interplay of one-dimensional divergences (noted in Section~\ref{sec:selfen}) and Fermi surface curvature. 
At leading order, we find $b_\kappa = 1$, but higher order renormalization group flows\cite{Metlitski:2010vm} are expected to renormalize
this to smaller values, as we will discuss in Section~\ref{sec:2kfself}. Such $2 k_F$ umklapp contributions are a promising avenue for understanding
the optical conductivity data,\cite{marel1,marel2} given the negative exponent in Eq.~(\ref{sigmaneg}).
 
\section{Low energy theory}
\label{sec:lowtheory}

\begin{figure}[h]
\begin{center}
\includegraphics[height=280pt]{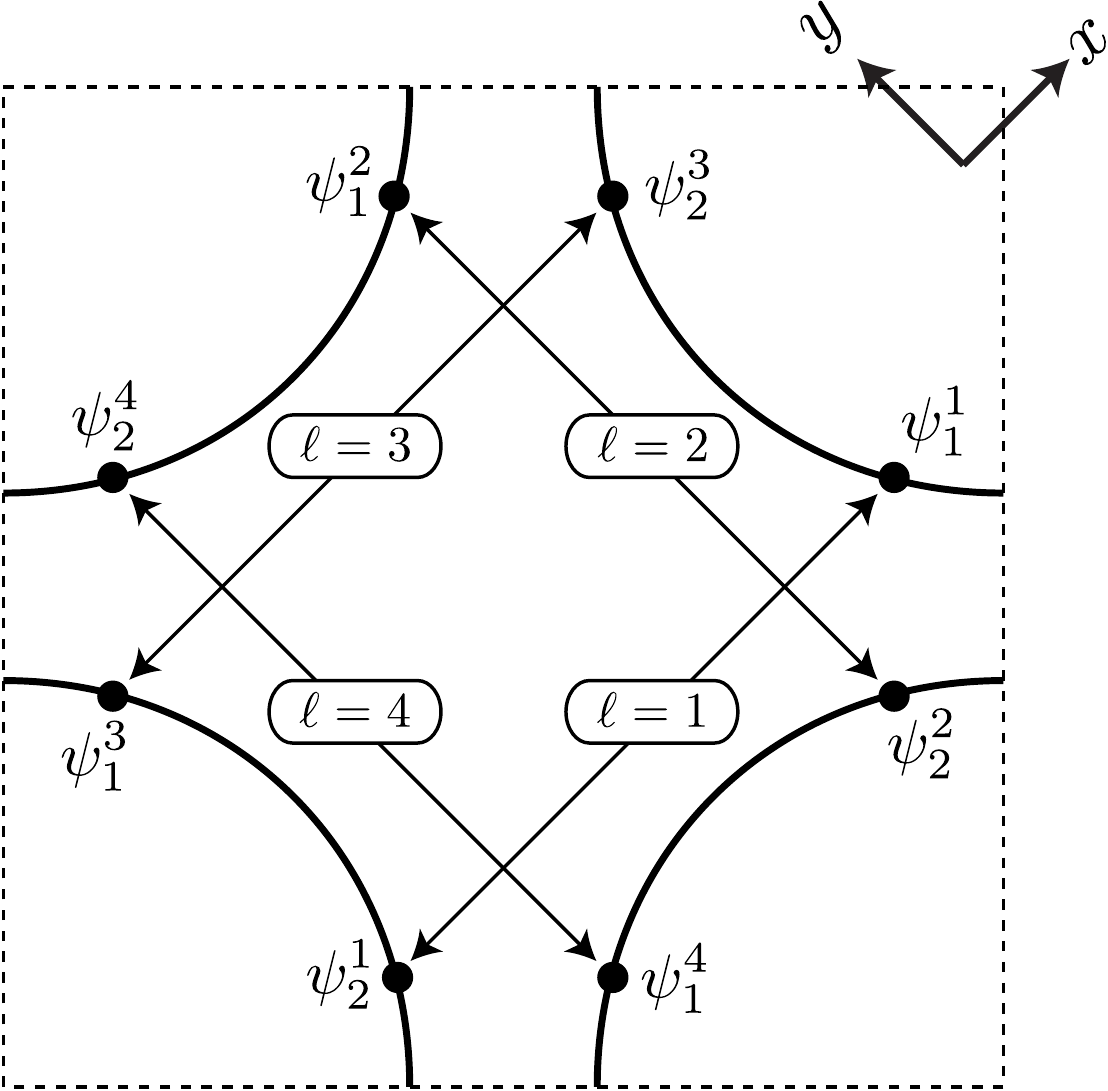}
\caption{The four pairs of hot spots and locations of the fermion fields $\Psi^\ell_a$, related to the fields in the 
figure by Eq.~(\ref{Psipsi}). \label{fig:hotspots}}
\end{center}
\end{figure}

\begin{figure}[h]
\begin{center}
\includegraphics[width=4in]{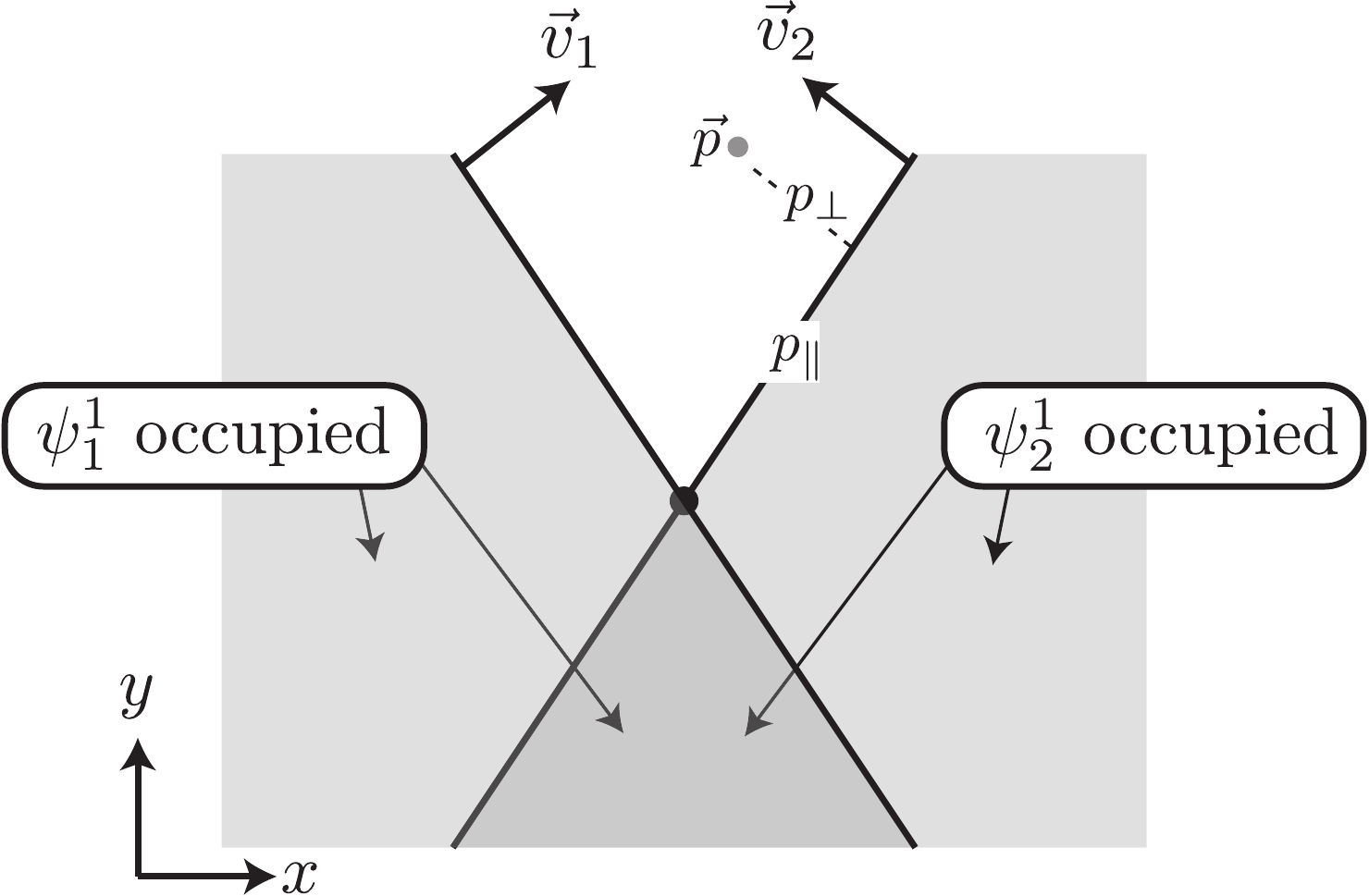}
\caption{(From Ref.~\onlinecite{Metlitski:2010zh}) Configuration of the $\ell=1$ pair of hot spots, with the momenta
of the fermion fields measured from the common hot spot at $\vec{k}=0$, indicated by the dark filled circle. The Fermi
velocities $\vec{v}_{1,2}$ of the $\psi_{1,2}$ fermions are indicated. The momentum components of the $\psi_2^1 (\vec{p})$ fermion parallel
($p_\parallel$) and orthogonal ($p_\perp$) to the Fermi surface are indicated.}
\label{fig:fermions}
\end{center}
\end{figure}

Our starting point will be the following (imaginary time) low energy effective field theory of fermionic excitations living at pairs of hot spots and interacting via bosonic fluctuations of a $(\pi,\pi)$ spin density wave \cite{Metlitski:2010vm}
\bea
\Lag & = & \frac{N}{2 c^2} (\pa_\tau \phi)^2 + \frac{N}{2} \left(\vec \pa \phi \right)^2 + \frac{N u}{4} (\phi^2)^2\nonumber \\
& & + \sum_{a,\ell} \frac{1}{2} \Psi^{\dagger \ell}_a \left(\pa_\tau - i \vec v_a^{\,\ell} \cdot \vec \pa \right) \Psi^\ell_a + \sum_\ell \frac{\lambda}{2} \phi \cdot \left(\Psi^{\dagger \ell}_1 \tau \Psi^{\ell}_2 + \Psi^{\dagger \ell}_2 \tau \Psi^{\ell}_1 \right) \,. \label{eq:theaction}
\eea
In this Lagrangian, $\phi$ is a three component boson and each $\Psi^\ell_a$ is a four component spinor. The mass of the boson has be tuned to zero. The label $\ell$ runs over the four pairs of hot spots while $a$ runs over the two patches of each pair. The geometry of the Fermi surface and hot spots is shown in Fig.~\ref{fig:hotspots}. The four components of the spinor may be labeled by the pair of two component indices $\{\sigma, \alpha\}$, where $\sigma$ is a spin index and $\alpha$ is a particle-hole index. We have suppressed these indices in the above Lagrangian. The Pauli matrices $\tau$ act on the spin indices. Also suppressed is a fermion flavor index running from $1$ to $N$. All interactions are diagonal in flavor. The spatial derivatives $\vec \pa$ act in two-dimensional space with coordinates $\{x,y\}$ and $\vec v_a^{\,\ell}$ are the Fermi velocities of each hot spot.
In terms of the fermions illustrated in Figs.~\ref{fig:hotspots} and~\ref{fig:fermions}, the fermions $\Psi^{\ell}_{a}$ are given by
\beq \Psi^{\ell}_a = \left(\begin{array}{c} \psi^{\ell}_a\\ i \tau^2 \psi^{\ell \dagger}_a\end{array}\right) \label{Psipsi} \eeq
so that they satisfy the hermiticity  condition
\beq i \tau^2 \left(\begin{array}{cc} 0 & -1\\ 1 & 0\end{array}\right) \Psi^{\ell}_a = \Psi^{\ell \ast}_a . \label{Herm}\eeq

The advantage of writing the action in terms of four component fermions is that it makes manifest an emergent $(SU(2))^4$ `pseudospin' symmetry \cite{Metlitski:2010vm}. For each $\ell$ we have an $SU(2)$ symmetry with the generators $T^m_\ell$ acting as
\be
SU(2)_\ell : \quad T^m_\ell \Psi^\ell_a = i \sigma^m \Psi^\ell_a \,, 
\ee
and leaving all other fields invariant.
Here $\sigma^m$ are Pauli matrices acting on the particle-hole indices of the spinor. The symmetry therefore has the following charge densities and currents
\be\label{eq:currents}
J^m_{\ell\tau} = \frac{i}{2} \sum_a \Psi^{\dagger \ell}_a \sigma^m \Psi^{\ell}_a \,, \qquad \vec{J}^m_{\ell} = \frac{1}{2} \sum_a \vec v^{\,\ell}_a \Psi^{\dagger \ell}_a \sigma^m \Psi^{\ell}_a \,.
\ee
The diagonal subgroup of each $SU(2)$ describes the conservation of fermion number at each pair of hot spots. Higher order kinetic terms in the action describing e.g. the curvature of the Fermi surface will break the $SU(2)$ symmetries down to this subgroup. These and other symmetry breaking terms are irrelevant in the low energy scaling limit.

The Fermi velocities at one pair of hot spots can be parameterised by
\be
\vec v^{\,\ell=1}_1 = (v_x, v_y) \,, \qquad \vec v^{\,\ell=1}_2 = (- v_x, v_y) \,.
\ee
The remaining velocities are given by 90 degree rotations
\be
\vec v^{\,\ell}_a = R_{\pi/2}^{\ell-1} \vec v^{\,\ell=1}_a \,.
\ee
Later it will also be convenient to introduce the ratio and the modulus
\be
\a \equiv \tan \varphi \equiv \frac{v_y}{v_x} \,, \qquad v = |\vec v| \,.
\ee
Here $0 < 2 \varphi < \pi$ is the angle between the Fermi surfaces at the hot spot. 

The fermion and boson propagators are constrained by both the pseudospin symmetry and by the form of the scaling limit. The symmetry requires the fermion propagators to take the form
\be
\langle \Psi^\ell_{a\s\a}(x) \Psi^{\ell' \dagger}_{b\s'\bb}(x') \rangle = - \dd^{\ell \ell'} \dd_{ab} \dd_{\s\s'} \dd_{\a\bb} G^\ell_a(x-x') \,.
\label{eq:Gpseud}\ee
In the low energy scaling limit, the fermion and boson two point functions are characterised by the dynamical critical exponent $z$ and anomalous dimensions $\eta_\psi$ and $\eta_\phi$. Thus
\bea\label{eq:scalingGD}
G^{-1}(\w,\vec p \,) & = & p^{z/2 - \eta_\psi} \widetilde G^{-1}\left(\frac{\w}{p^z}, \hat p \right) \,, \\
D^{-1}(\w,\vec p \,) & = & p^{2 - \eta_\phi} \widetilde D^{-1}\left(\frac{\w}{p^z}, \hat p \right) \,. 
\eea
Here $p = |\vec p|$. Much of the interesting physics we will describe below is due to fermions that are within the scaling regime, but far from the hot spots. If $p_\perp$ is the distance to the Fermi surface and $p_\parallel$ the distance along the Fermi surface to the hot spot, then this condition requires $p_\perp, \w^{1/z} \ll p_\parallel$. In this `lukewarm' region one expects the quasiparticle form\footnote{A possible violation of the quasiparticle form in Eqs.~(\ref{eq:lukewarm}), (\ref{eq:vZ}) will be discussed in section \ref{sec:2kfself}.}
\be\label{eq:lukewarm}
G(\w,\vec p \,) = \frac{Z(p_\parallel)}{i \w - v_F(p_\parallel) p_\perp} \qquad , \qquad (p_\perp, \w^{1/z} \ll p_\parallel) \,,
\ee
with
\be\label{eq:vZ}
v_F(p_\parallel) \propto p_\parallel^{z-1} \,, \qquad Z(p_\parallel) \propto p_\parallel^{z/2 + \eta_\psi} \,.
\ee

Working in the rainbow approximation and taking the limit $N \to \infty$ one finds that $z=2$ and $\eta_\psi = \eta_\phi = 0$.
The explicit forms of the correlators were obtained in Refs.~\onlinecite{advances, Metlitski:2010vm}. The boson has
\be\label{eq:bosonD}
D^{-1}(\w,\vec p \,) = N \left( \g |\w| + \vec p^{\,2} \right) \,,
\ee
with
\be\label{eq:gam}
\gamma = \frac{4 \lambda^2}{2\pi v_x v_y} \,,
\ee
while the fermions have
\beq G^{-1}_a(\omega, \vec{p}) = i \omega -\vec{v}_a \cdot \vec{p} + \frac{1}{N} \frac{3 v \sin 2 \varphi}{8} i \, \text{sgn}(\w) \left(\sqrt{\gamma |\omega| + (\hat{v}_{\bar{a}} \cdot \vec{p})^2} - |\hat{v}_{\bar{a}} \cdot \vec{p}|\right) \,. \label{eq:fermionG0}\eeq
Here, $\bar{1} = 2$ and $\bar{2} = 1$ and we have suppressed the hot spot index $\ell$. In the expression (\ref{eq:fermionG0}) we have not explicitly written the real part of the self energy which renormalises the velocities $v_x$ and $v_y$. Note that the tree level analytic term $i \omega$ in the propagator is suppressed at low energy compared to the dynamically generated self energy. We have kept the tree level term here as a UV regulator as will be discussed in more detail below. For future reference, it will be convenient to introduce the momentum scale,
\beq \Lambda = \frac{3 v \gamma \sin 2 \varphi}{16 N} = \frac{3 \lambda^2}{4 \pi N v} \label{Lambdadef}\eeq
and the associated energy scale $\Lambda_\omega = \Lambda^2/\gamma$. $\Lambda$ and $\Lambda_\omega$ are the momentum and energy scales at which the dynamically generated self energy becomes comparable to the tree level analytic term $i \omega$. The numerical factors in Eq.~(\ref{Lambdadef}) are inserted for future convenience.

We may also represent Eq.~(\ref{eq:fermionG0}) as
\be\label{eq:fermionG}
G^{-1}(\w,\vec p \,) =  i \omega - v p_\perp + \frac{1}{N}\frac{3 v \sin 2 \varphi}{8}  i \, \text{sgn}(\w) \left(\sqrt{\g |\w| + \left(\hat v_1 \times \hat v_2 \, p_\parallel\right)^2} - \left| \hat v_1 \times \hat v_2 \, p_\parallel\right| \right)  \,.
\ee
Here again $p_\perp = \hat v \cdot \vec p$ is the momentum in the direction orthogonal to the Fermi surface while $p_\parallel$ is the distance to the hot spot and is the momentum in the direction orthogonal to $\hat v$.  Furthermore, assuming for concreteness that the propagator in question is for the $\Psi_1$ fermions, in the self energy we have approximated
\be
\hat v_2 \cdot \vec p \quad \to \quad |\hat v_1 \times \hat v_2| \, p_\parallel =  \sin 2 \varphi \,p_\parallel \,.
\ee
In general $\hat v_2$ is not orthogonal to $\vec v_1$ and so this replacement is not exact. However, the dependence on the $\vec v_1$ component of the momentum in the self energy is subleading in powers of $N$ in the full propagator compared to the tree-level $v p_\perp$ dependence in (\ref{eq:fermionG}). Thus we can and have projected out the component of $\hat v_2$ parallel to $\hat v_1$ for the purposes of this paper, see e.g. the computations in Ref.~\onlinecite{Metlitski:2010vm}.

The correlators (\ref{eq:bosonD}) and (\ref{eq:fermionG}) will be used at various points in the remainder of this paper. From them one can read off the constants of proportionality (\ref{eq:vZ}) in the lukewarm region. Note that in the present approximation, the lukewarm fermions have a Fermi liquid like $\omega^2$ damping rate,
\be
G(\w,\vec p \,) = \frac{Z(p_\parallel)}{i \w - v_F(p_\parallel) p_\perp - i a(p_\parallel) \omega^2 \mathrm{sgn}(\omega)} \qquad , \qquad (p_\perp, \sqrt{\gamma\omega} \ll p_\parallel) \,,
\label{eq:lukewarmdamp}\ee
with $a(p_\parallel) \sim p^{-2}_\parallel$. We will demonstrate in Sec.~\ref{sec:selfen} that scattering off composite operators qualitatively modifies the damping rate of cold (and hence also lukewarm) fermions. Such corrections, however, appear only at higher order in $1/N$.

We should note that the large $N$ expansion used to obtain the propagators (\ref{eq:bosonD}) and (\ref{eq:fermionG}) breaks down both at sufficiently low energy scales and at higher loop order: The RG flow equations at leading nontrivial order in $N$ flow to an $\a=0$ IR fixed point, at which Fermi surfaces at each hot spot pair are parallel \cite{Metlitski:2010vm}. At sufficiently low energies the anomalous dimensions become order one and the large $N$ expansion breaks down. The RG flow can be controlled only in the regime $1/\sqrt{N} \ll \a \ll 1$. In this regime all anomalous dimensions acquire small nonzero corrections. In particular $z \neq 2$. At higher loop order the large $N$ expansion breaks down for the reasons first identified in Ref.~\onlinecite{sungsik} for a 2+1 dimensional Fermi surface coupled to a gauge field. Loops involving fermions on the Fermi surface give extra powers of $N$ that disrupt the
na\"ive large $N$ scaling of diagrams. We will essentially ignore this last complication; perhaps the expansion can be controlled using a generalization of the methods developed in Refs.~\onlinecite{nayak1, nayak2, mross}. Alternatively, a strong coupling approach to metallic criticality may be possible using the holographic correspondence, along the lines of e.g. Refs.~\onlinecite{Faulkner:2011tm, Goldstein:2009cv, Hartnoll:2010gu}.

\section{Hot spot conductivity: rainbow approximation}
\label{sec:hotspot}

As discussed in Section~\ref{sec:intro}, we first consider the conductivity due to scattering off the primary field, $\phi$.
This scattering is strongest at the hot spots. As shown in Ref.~\onlinecite{Metlitski:2010vm}, the hot spot theory remains strongly coupled even when one takes the 
number of fermion flavors $N \to \infty$. In this section, we treat the hot spot theory in the `rainbow' approximation. Furthermore, we take the limit $N \to \infty$ in order to simplify the calculations. A similar approach was considered previously in Ref.~\onlinecite{advances}. A key difference with Ref.~\onlinecite{advances} is that we take into account the rainbow corrections to the current vertex, which are necessary to satisfy the Ward identity associated with charge conservation. 

The four $SU(2)$ currents of the theory (\ref{eq:currents}) necessarily have orthogonal correlators
\be\label{eq:su2currents}
- \int d^3x \langle J_\mu^{\, \ell\,m}(z) J_\nu^{\, \ell'\,n}(0)\rangle e^{-i \vec{q} \vec{x}} e^{i \omega \tau} \equiv \dd^{mn} \dd^{\ell \ell'} \Pi^{\ell}_{\mu \nu}(\w,\vec q) \,.
\ee
Here $\mu,\nu$ are space-time indices. We focus on the electrical conductivity which is related to the $m=3$ component of the $SU(2)$ current correlator (\ref{eq:su2currents}) summed over all of the hot spots. (From here on, all the pseudospin indices $m, n$ are set to 3, unless otherwise noted.) Specifically, the electrical conductivity at real frequency $\Omega$ and vanishing momentum is
\be\label{eq:hotc}
\sigma_{ij}(\Omega) = \left. \frac{\sum_\ell \Pi^\ell_{ij}(\w,0)}{\w} \right|_{i \w = \Omega + i \epsilon} \,.
\ee
In relating the conductivity and current correlator in this way, we are neglecting the tadpole or `diamagnetic' term, which gives rise to a $\delta$-function at zero frequency in the real part of the conductivity and a $1/\Omega$ behavior in the imaginary part. Such a contribution is always present at zero temperature. Below, we will be interested in corrections to this behavior.   
 
It will also be convenient for our purposes to define the current vertex $\Gamma^{\ell}_{\mu a}$
\beq \int d^3z d^3x \langle J^{\ell}_\mu(z) \psi^{\ell'}_{\sigma a}(x) \psi^{\dagger \ell'}_{\sigma' b}(0)\rangle
e^{-i q z} e^{-i p x} = \delta_{\sigma \sigma'} \delta_{a b} \delta^{\ell \ell'} \Gamma^{\ell}_{\mu a}(q, p) G^{\ell}_a (p) G^{\ell}_a (p+q) \label{vertdef}\eeq
Note that the correlator vanishes for $\ell \neq \ell'$ by pseudospin symmetry. The diagonal subgroup of the $SU(2)$ symmetry at each hot spot pair implies the following Ward identities,
\bea -q_\tau \Pi^{\ell}_{\tau \nu}(q) + q_i \Pi^{\ell}_{i \nu} (q) &=& 0 \\
-q_\tau \Gamma^{\ell}_{\tau  a}(q, p) + q_i \Gamma^{\ell}_{i a}(q, p) &=& G^{\ell}_a(p)^{-1} - G^{\ell}_a(p+q)^{-1} 
\label{WardGamma}\eea
In particular, the density-density correlation function at zero wavevector vanishes,
\beq \Pi^{\ell}_{\tau \tau}(\omega, 0) = 0\eeq 
We concentrate below on the hot spot pair with $\ell = 1$. Observe that,
\bea J^{1}_\tau &=& i (\psi^{1 \dagger}_1 \psi^1_1 + \psi^{1 \dagger}_2 \psi^1_2) \nn\\ 
J^1_x &=& v_x (\psi^{1\dagger}_1 \psi^1_1 - \psi^{1\dagger}_2 \psi^1_2)\nn\\
J^1_y &=& v_y (\psi^{1 \dagger}_1 \psi^1_1 + \psi^{1 \dagger}_2 \psi^1_2)\eea
We see that $J^1_y$ is proportional to $J^1_\tau$. Therefore, $\Pi^1_{yy}(\omega, 0) = 0$. Moreover, $\Pi^1_{xy}(\omega, 0) = 0$ by reflection symmetry, $x \to -x$. Thus, the only non-trivial element of the conductivity tensor at zero wavevector is $\Pi^1_{xx}$. 

To compute $\Pi^1_{xx}$, we first evaluate the corresponding current vertex $\Gamma^1_{x a}$. For briefness, we drop the hot spot index below; $\ell = 1$ is assumed unless otherwise noted.  It is convenient to define the symmetric and antisymmetric vertices $\Gamma^{\pm}_{a}$ as
\beq \int d^3z d^3x \langle (\psi^{\dagger}_1 \psi_1 \pm \psi^{\dagger}_2 \psi_2) (z) \psi_{\sigma a}(x) \psi^{\dagger}_{\sigma' b}(0)\rangle
e^{-i q z} e^{-i p x} = \delta_{\sigma \sigma'} \delta_{a b} \Gamma^{\pm}_{a}(q, p) G_a (p) G_a (p+q) \eeq
Then, $\Gamma_{\tau a} = i \Gamma^{+}_a$, $\Gamma_{x a} = v_x \Gamma^-_a$ and $\Gamma_{y a} = v_y \Gamma^+_a$ and the Ward identity (\ref{WardGamma}) reads,
\beq (- i q_\tau + v_y q_y) \Gamma^+_{a}(q,p) + v_x q_x \Gamma^-_a(q,p)= G^{-1}_a(p) - G^{-1}_a(p+q) \label{Wardpm}\eeq
Also, we may express $\Pi_{xx}$ as,
\beq \Pi_{xx}(q) = 2 N v^2_x \sum_a (-1)^{a+1} \int \frac{d^3p}{(2\pi)^3} \Gamma^-_a(q,p) G_a(p) G_a(p+q) \label{Pixx}\eeq

\begin{figure}[h]
\begin{center}
\includegraphics[height = 2in]{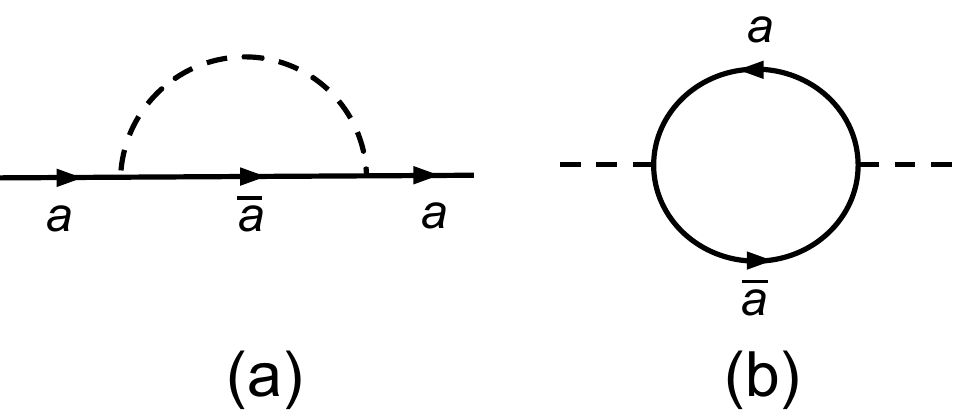}
\caption{(a) Fermion and (b) boson self-energies in the rainbow approximation. Solid lines are fermion Green's functions,
while dashed lines are $\phi$ Green's functions. All internal lines are full self-consistent propagators. 
In this section we are dropping the hot-spot index $\ell$, and so $\psi_1 \equiv \psi_{1}^1$ and $\psi_{\bar{1}}
\equiv \psi_2^1$.}\label{fig:propsrainbow}
\end{center}
\end{figure}

We will treat the fermion propagator and the current vertex in the rainbow approximation. The rainbow approximation for the boson and fermion propagators amounts to performing a self-consistent Hartree-Fock + RPA calculation, see Fig.~\ref{fig:propsrainbow}, and gives the following integral equations,
\beq (G^{\ell}_a(p))^{-1} = i \omega - \vec{v}^{\ell}_a \cdot \vec{p} - \Sigma^{\ell}_a(p) \label{Sigmadef}\eeq
\beq D^{-1}(q) = N \vec{q}^2 +  {\it \Pi}(q) \label{Pidef}\eeq
\beq \Sigma^{\ell}_a(p) = 3 \lambda^2 \int \frac{d^3 l}{(2\pi)^3} G^{\ell}_{\bar{a}}(p+l) D(l) \label{SigmaRainbow}\eeq
\beq {\it \Pi}(q) = 2 N \lambda^2 \sum_{\ell,a} \int \frac{d^3 l}{(2\pi)^3} G^{\ell}_a(l+q) G^{\ell}_{\bar{a}}(l)\label{PiRainbow}\eeq
The solution to these integral equations in the limit $N \to \infty$ is given in Eqs.~\ref{eq:bosonD}, \ref{eq:fermionG0}.\cite{advances}

\begin{figure}[h]
\begin{center}
\includegraphics[width = 7in]{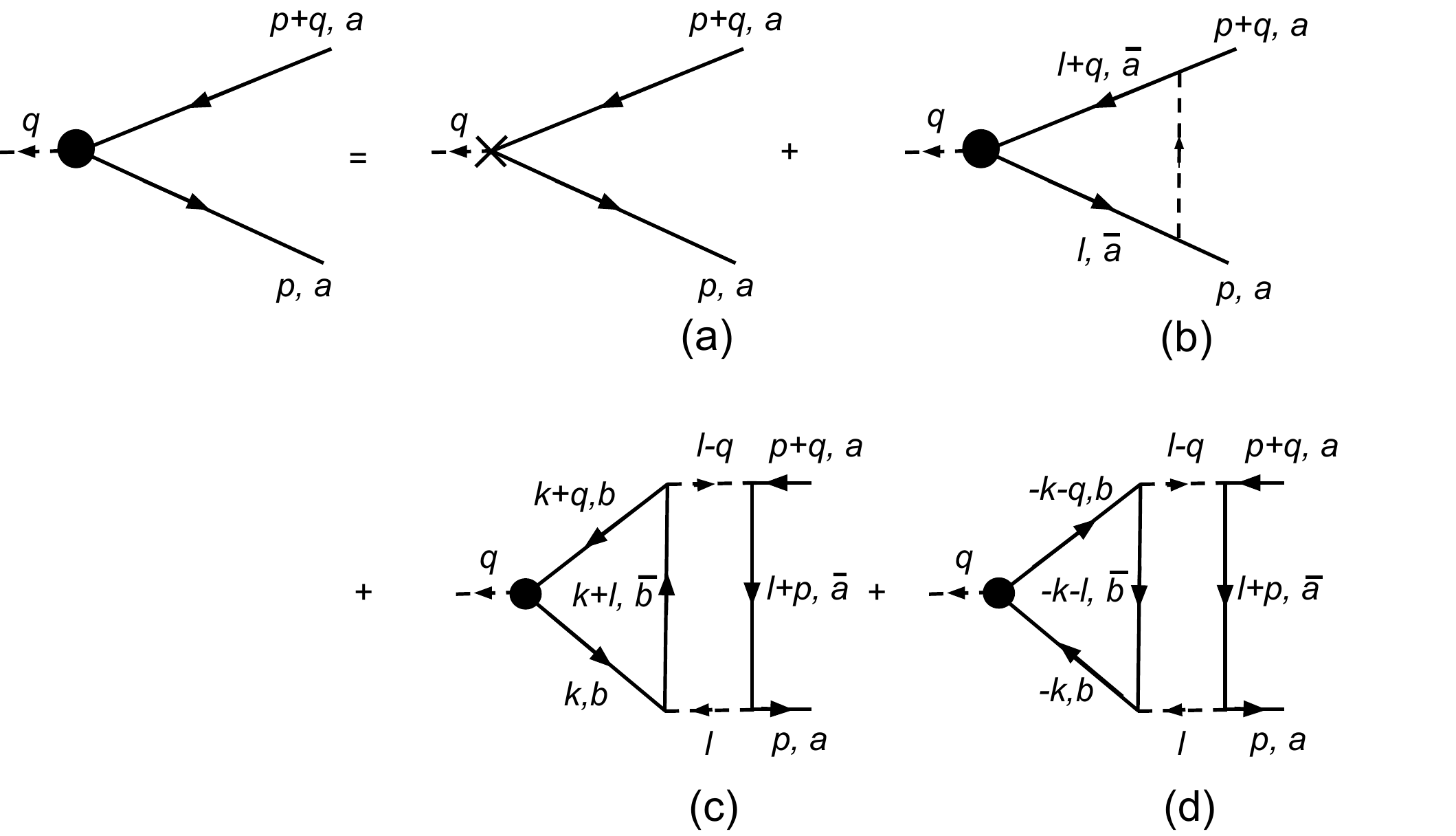}
\caption{Diagrams for the current vertex $\Gamma^{\pm}_a$ in the rainbow approximation. The cross denotes the bare vertex and the solid circle, the full vertex. The `Aslamazov-Larkin' type diagrams in (c) and (d) cancel against each other due to pseudospin symmetry.
As in Fig.~\ref{fig:propsrainbow}, $\psi_1 \equiv \psi_{1}^1$ and $\psi_{\bar{1}}
\equiv \psi_2^1$.
\label{fig:rainbow}}
\end{center}
\end{figure}

We would like to treat the current vertex at the same level of approximation as the boson and fermion propagators. This can be achieved by inserting the bare current vertex into every bare fermion propagator which appears when Eqs.~(\ref{Sigmadef})-(\ref{PiRainbow}) are iterated. The resulting  dressed current vertex satisfies an integral equation which is diagramatically shown in Fig.~\ref{fig:rainbow}. We note that the two `Aslamazov-Larkin'-type diagrams in Figs.~\ref{fig:rainbow} c) and d) cancel with each other due to the pseudospin symmetry. Indeed, the triangle portion of the two diagrams involves the correlator $\langle J^{\ell}_{\mu} \phi \phi\rangle$. The current $J^{\ell}_\mu$ transforms as a pseudospin vector, while the field $\phi$ transforms as a pseudospin scalar, therefore, $\langle J^{\ell}_{\mu} \phi \phi\rangle = 0$. Thus, the `Aslamazov-Larkin' diagrams in the particle-hole Fig.~\ref{fig:rainbow} c) and particle-particle Fig.~\ref{fig:rainbow} d) channels cancel with each other and we are left only with the rainbow corrections in Fig.~\ref{fig:rainbow} b). 

\begin{figure}[h]
\begin{center}
\includegraphics[width = 5in]{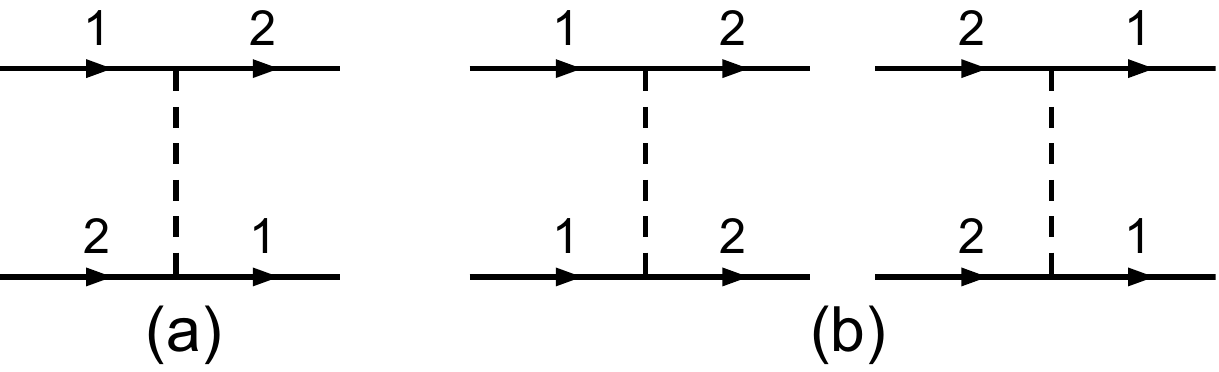}
\caption{Scattering processes involving fermions from the hot spot pair $\ell = 1$. The diagram in (a) corresponds to `regular' scattering, which conserves the current. The two diagrams in (b) correspond to `umklapp' scattering, which conserves the lattice momentum but not the current. 
Note that $\psi_2 \equiv \psi_{\bar{1}}$.
\label{fig:scatt}}
\end{center}
\end{figure}

\begin{figure}[h]
\begin{center}
\includegraphics[height = 220pt]{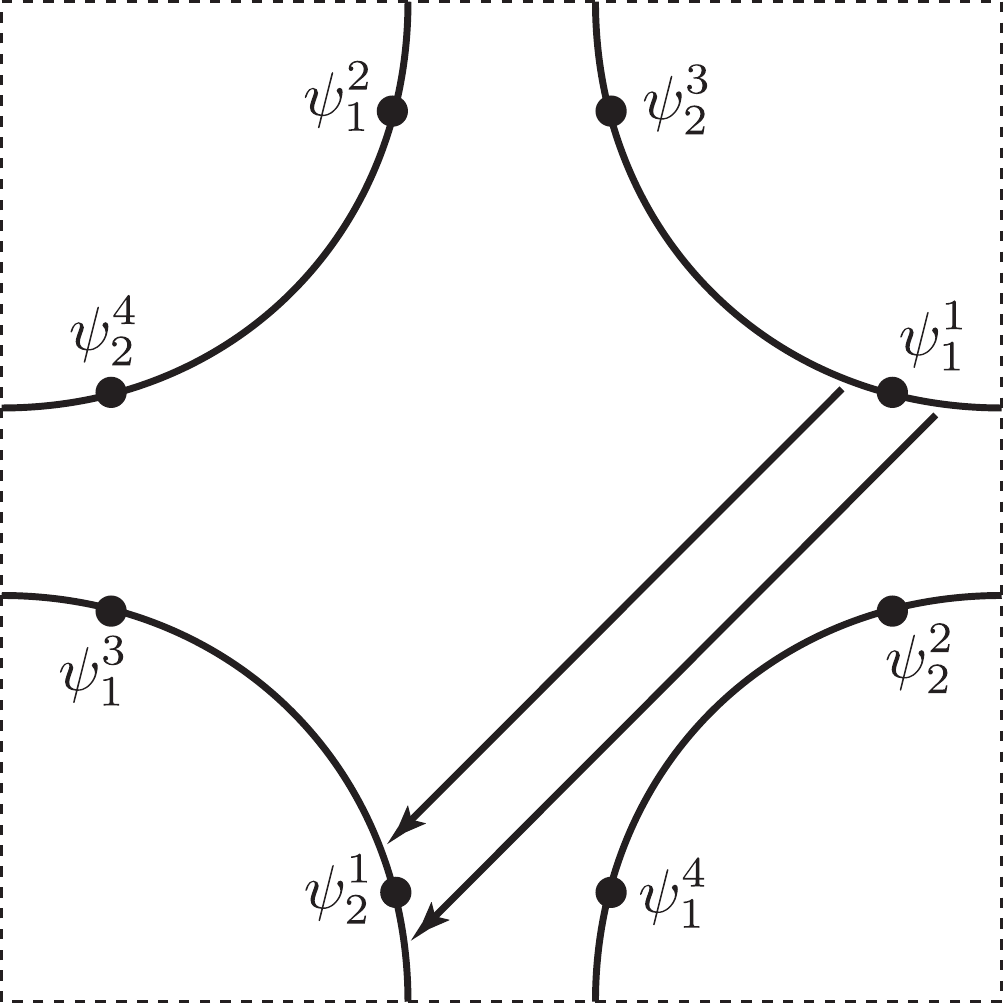}
\caption{Umklapp scattering processes in Fig.~\ref{fig:scatt}b. Such processes degrade the electric current and lead to a finite optical conductivity\label{fig:umklappd}}
\end{center}
\end{figure}

We note that the presence of umklapp processes in our theory is secretely hidden in the fact that the Aslamazov-Larkin diagrams cancel with each other and so can be ignored. Indeed, our theory allows for both types of scattering shown in Fig.~\ref{fig:scatt}. (For simplicity, we temporarily consider processes involving the hot spot pair $\ell = 1$ only.) The process in Fig.~\ref{fig:scatt}a) is a `regular' scattering process, which conserves not only the electron momentum, but also the electron current. If only such processes were allowed, the optical conductivity at finite frequency would vanish.\footnote{In principle, the electron current is not strictly conserved by `regular' scattering processes once a finite Fermi surface curvature is accounted for in the definition of the electrical current. However, the resulting corrections to the optical conductivity are expected to be of higher order than those due to umklapp processes and won't be considered in this section.} On the other hand, the processes in Fig.~\ref{fig:scatt}b) (see also  Fig.~\ref{fig:umklappd}) are  `umklapp' scattering processes, which flip the $x$ component of the electron current - hence, a finite optical conductivity is expected.  At the level of the rainbow approximation, the main difference between the two cases is the following. If only the regular processes in Fig.~\ref{fig:scatt}a) are allowed, then the index $b$ of the Aslamazov-Larkin diagram in the particle-hole channel, Fig.~\ref{fig:rainbow} c), is constrained to be $b = a$, while in the particle-particle channel, Fig.~\ref{fig:rainbow} d), $b = \bar{a}$. Hence, there is no longer a cancellation between the two Aslamazov-Larkin diagrams. Instead, in the umklapp-free case, the Aslamazov-Larkin diagrams, Fig.~\ref{fig:rainbow} c),d), and the rainbow diagram, Fig.~\ref{fig:rainbow} b) would conspire to give a vanishing optical conductivity. On the other hand, in the case when umklapps are allowed, this conspiracy between the Aslamazov-Larkin and rainbow graphs is broken. The pseudospin symmetry technically simplifies the calculation, allowing one to ignore the Aslamazov-Larkin diagrams alltogether, and the remaining rainbow graphs give a finite optical conductivity as we will explicitly verify below.

Note, finally, that although for illustrative purposes we have limited the above discussion to hot spot pair $\ell = 1$ only, in fact, umklapp processes involving different pairs of hot spots also contribute to the optical conductivity. At the level of the rainbow approximation, all hot spot pairs contribute to the boson polarization in Fig.~\ref{fig:propsrainbow} b), and hence scattering between hot spots with different $\ell$ is accounted for in the fermion self energy Fig.~\ref{fig:propsrainbow} a). The umklapp nature of this scattering is again hidden in the fact that it contributes to the fermion self energy, but not directly to the current vertex corrections. Indeed, the rainbow diagrams in Fig.~\ref{fig:rainbow} b) involve only one pair of hot spots, while the Aslamazov-Larkin diagrams Figs.~\ref{fig:rainbow} c), d) cancel for fermions from any hot spot $\ell'$ running in the internal loop.

With the above remarks in mind, to obtain the current vertex we have to sum the series of rainbow diagrams in Fig.~\ref{fig:rainbow} b). This sum obeys the following integral equation,
\bea \Gamma^{+}_a(q, p) &=& 1 + 3 \lambda^2 \int \frac{d^3 l}{(2\pi)^3} G_{\bar{a}}(l) G_{\bar{a}}(l + q) D(l-p) \Gamma^+_{\bar{a}} (\omega,l)\nn\\
\Gamma^{-}_a(q, p) &=& (-1)^{a+1} + 3 \lambda^2 \int \frac{d^3 l}{(2\pi)^3} G_{\bar{a}}(l) G_{\bar{a}}(l + q) D(l-p) \Gamma^-_{\bar{a}} (q,l) \label{Gminit}\eea
A detailed analysis of Eqs.~(\ref{Gminit}) is performed in Appendix \ref{sec:appvert}. We summarise the results of this analysis below.

Let us set $q = (\omega, 0)$ and without loss of generality assume $\omega > 0$. It is convenient to change variables to $l_a = \hat{v}_a \cdot \vec{l}$. As shown in Appendix \ref{sec:appvert}, in the limit $N = \infty$, to determine $\Gamma^{\pm}_a(\omega,p)$ for all $p$ it is sufficient to know its behavior for $p_a = 0$, i.e. we need to find the current vertex with external momentum on the Fermi surface. Moreover, introducing the variable $\nu = p_\tau + \omega$ we can restrict our attention to $0 < \nu < \omega$. Thus, definining,
\bea \Gamma^+_a(\omega, \nu, p_a = 0, p_{\bar{a}} = p) &\equiv& \Gamma^+(\omega, \nu, p) \label{Gammapdef}\\
\Gamma^-_a(\omega, \nu, p_a = 0, p_{\bar{a}} = p) &\equiv& (-1)^{a+1} \Gamma^-(\omega, \nu, p)\eea
we obtain, see Appendix~\ref{sec:appvert} for details,
\bea \Gamma^{\pm}(\omega, \nu, p) &=& 1 \pm 2 \pi \gamma \sin 2 \varphi \int \frac{d l}{2\pi} \int_0^{\w} \frac{d \nu'}{2 \pi} 
\frac{1}{\sqrt{\gamma \nu' + l^2} + \sqrt{\gamma (\omega - \nu') + l^2} - 2 |l| + \frac{\gamma \omega}{2 \Lambda}}\nn \\
&& \frac{1}{(l - p \cos 2 \varphi)^2 + \sin^2 2 \varphi(\gamma |\nu' - \nu| + p^2)} \Gamma^{\pm}(\omega, \nu', l) \label{GammaF}\eea
where the UV momentum scale $\Lambda$ is given by Eq.~(\ref{Lambdadef}). 

Let us check the consistency of Eq.~(\ref{GammaF}) with the Ward identity. From Eq.~(\ref{Wardpm}) at $\vec{q} = 0$,
\beq \Gamma^+_a(\omega, p) = 1 + \frac{3 v \sin 2 \varphi}{8 N \omega}\left(\mathrm{sgn}(p_\tau + \omega)
(\sqrt{\gamma |p_\tau + \omega| + p^2_{\bar{a}}} - |p_{\bar{a}}|) - \mathrm{sgn}(p_\tau)
(\sqrt{\gamma |p_\tau | + p^2_{\bar{a}}} - |p_{\bar{a}}|)\right) \label{Gammapsol}\eeq
which satisfies Eq.~(\ref{GammaF}). Note that at low momentum $p \ll \Lambda$ and frequency $\omega \ll \Lambda_\omega$, the second term in Eq.~(\ref{Gammapsol}) dominates over the bare vertex $\Gamma^{+,0} = 1$, so that $\Gamma^+(\omega, \nu, p)$ can be written in the scaling form,
\beq \Gamma^+(\omega, \nu, p) \propto \left(\frac{\omega}{\Lambda_\omega}\right)^{-1/2} \gamma^+\left(\frac{\nu}{\omega}, \frac{p}{\sqrt{\gamma \omega}}\right)\eeq
with
\beq \gamma^+(x,y) = \sqrt{x+y^2} + \sqrt{1-x+y^2} - 2|y|\eeq
Observe that by summing the rainbow diagrams we have generated a large anomalous dimension for the vertex $\Gamma^+$. 

\begin{figure}[h]
\begin{center}
\includegraphics[height=180pt]{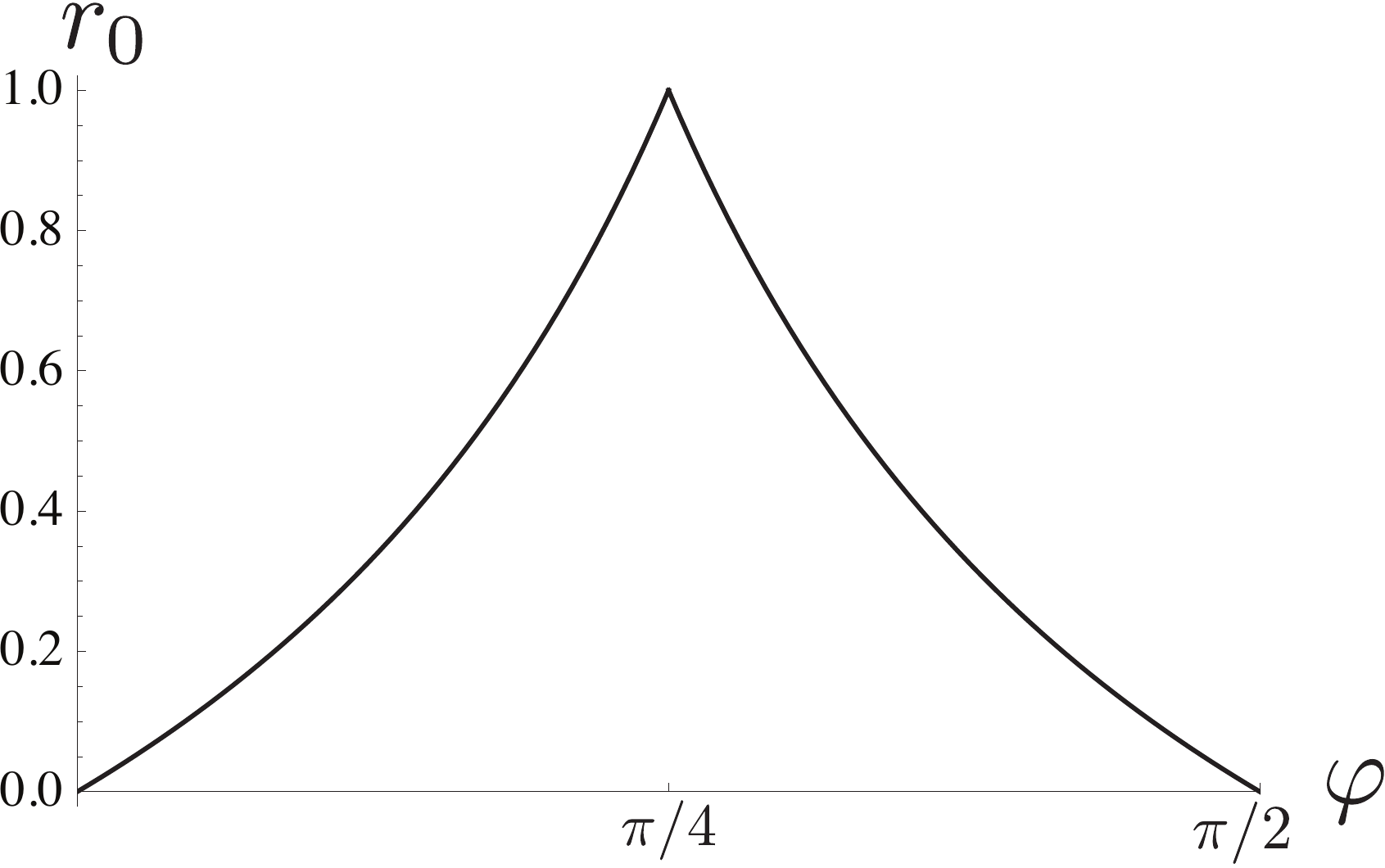}
\caption{The exponent $r_0$, Eq.~(\ref{r0}), associated with the current vertex $\Gamma^-$, as a function of the angle $2 \varphi$  between the Fermi surfaces.}
\label{fig:r0}
\end{center}
\end{figure}

We next proceed to discuss the vertex $\Gamma^-$. Unlike with $\Gamma^+$, one cannot extract $\Gamma^-$ at $\vec{q} = 0$ from the Ward identity Eq.~(\ref{Wardpm}) without also knowing the low $\vec{q}$ behaviour of $\Gamma^+$. Hence, we must actually solve Eq.~(\ref{GammaF}) for $\Gamma^-$. As shown in Appendix \ref{sec:appvert}, at low frequency and momentum $\Gamma^-$ assumes the following scaling form,
\beq \Gamma^-(\omega, \nu, p) \sim \left(\frac{\omega}{\Lambda_\omega}\right)^{r_0/2} \gamma^-\left(\frac{\nu}{\omega}, \frac{p}{\sqrt{\gamma \omega}}\right)\label{Gmscal2}\eeq
with the exponent $r_0$ given by
\beq r_0 = \left\{\begin{array}{lr} {\displaystyle \frac{2 \varphi}{\pi - 2 \varphi}}, & \quad 0 < \varphi < \pi/4\\
\\
{\displaystyle \frac{\pi - 2\varphi}{2 \varphi}}, & \quad \pi/4 < \varphi < \pi/2 \end{array}\right.\label{r0}\eeq
Fig.~\ref{fig:r0} shows the behavior of the exponent $r_0$ as a function of $\varphi$. Note that $0 < r_0 \leq 1$.
Hence, the vertex $\Gamma^-$ acquires an anomalous dimension. Unlike $\Gamma^+$, which is enhanced at low energy, $\Gamma^-$ is suppressed. We briefly note that when $\varphi = \pi/4$, $r_0 = 1$, the scaling form (\ref{Gmscal2}) is modified by an additional factor $\log \Lambda_\omega/\omega$. 

We have not been able to analytically determine the full form of the scaling function $\gamma^-(x,y)$ entering Eq.~(\ref{Gmscal2}). However, we have extracted its asymptotic behavior for $y \to \infty$ (i.e. $p \gg \sqrt{\gamma \omega}$) We will shortly demonstrate that this asymptotic behavior controls the frequency dependence of the optical conductivity. As shown in Appendix \ref{sec:appvert}, 
\beq \gamma^-(x,y) = g_0 y^{r_0} + g_1(x) y^{r_0 -2} + g_2 y^{r_2} + \cdots, \quad y \to \infty \label{gammamaspt}\eeq
Here,
\beq r_2 =  -2 - r_0 \label{r2} \eeq
The leading term in Eq.~(\ref{gammamaspt}) ensures that $\Gamma^-$ has a finite static limit, $\Gamma^-(p) = \lim_{\omega, \nu \to 0} \Gamma^-(\omega, \nu, p)$ with
\beq \Gamma^-(p) \sim \left(\frac{p}{\Lambda}\right)^{r_0}\eeq
We have also kept two subleading terms in Eq.~(\ref{gammamaspt}), which turn out to control the optical conductivity beyond the simple Drude response. Of particular interest is the last term in Eq.~(\ref{gammamaspt}) with the exponent $r_2$, which as we will see, gives rise to the `quantum critical' contribution to the optical conductivity. As will be discussed below, the relationship (\ref{r2}) between the exponents $r_0$ and $r_2$, which appears to come out accidentally in our calculations, is actually crucial for a consistent renormalization group (RG) interpretation of our results.

We now proceed from the vertex correction to conductivity itself. Starting with Eq.~(\ref{Pixx}) and retracing the steps that led from (\ref{Gminit}) to (\ref{GammaF}), we obtain
\beq \Pi_{xx}(\omega) = \frac{8 N^2}{3  \sin^2 \varphi} \int \frac{dl}{2\pi} \int_{0}^\omega \frac{d \nu}{2\pi} \frac{1}{\sqrt{\gamma  \nu + l^2}+\sqrt{\gamma (\omega - \nu) + l^2} -  2 |l| + \frac{\gamma \omega}{2 \Lambda}} \Gamma^-(\omega, \nu, l)\label{PixxGm}
\eeq
Substituting the scaling form (\ref{Gmscal2}) into (\ref{PixxGm}) and cutting off the integral over $l$ at $l \sim \Lambda$ we obtain,
\beq \Pi_{xx}(\omega) \sim  N^2 \omega \left(\frac{\omega}{\Lambda_\omega}\right)^{r_0/2} \int_0^{\sqrt{\Lambda_\omega/\omega}}dy \int_0^1 dx \frac{1}{\sqrt{x + y^2} + \sqrt{1 -x +y^2} - 2 y} \gamma^-(x,y) \label{Pixxgm}\eeq
Now, as shown in Appendix~\ref{sec:appvert}, for $\mu \gg 1$,
\beq \int_0^{\mu} dy \int_0^1 dx \frac{1}{\sqrt{x + y^2} + \sqrt{1 -x +y^2} - 2 y} \gamma^-(x,y) = a_0 \mu^{r_0+2} + a_1 \mu^{r_0} + a_2 \mu^{r_2 + 2} + \cdots \label{sumrule}\eeq
Hence, 
\beq \Pi_{xx} \sim N^2 \Lambda_\omega \left(a_0 + a_1 \frac{\omega}{\Lambda_\omega} + a_2 \left(\frac{\omega}{\Lambda_\omega}\right)^{(r_0-r_2)/2}\right)\eeq
and the optical conductivity at real frequency $\Omega$ is given by
\bea
\lefteqn{\sigma(\Omega) \sim N^2 \Lambda_\omega \left(a_0 \frac{i}{\Omega + i \epsilon} + a_1 \frac{1}{\Lambda_\omega} + a_2 \left(\frac{-i \Omega}{\Lambda_\omega}\right)^{(r_0-r_2-2)/2}\right)} \nn \\
& & = N^2 \Lambda_\omega \left(a_0 \frac{i}{\Omega + i \epsilon} + a_1 \frac{1}{\Lambda_\omega} + a_2 \left(\frac{-i \Omega}{\Lambda_\omega}\right)^{r_0}\right) \label{sigmaf}\eea
where we've used Eq.~(\ref{r2}) in the last step. 

Let us discuss the three terms in Eq.~(\ref{sigmaf}). The term with the coefficient $a_0$ when added with the diamagnetic tadpole term renormalizes the total Drude weight. This term is clearly cut-off dependent and non-universal. 

Next, consider the term with the coefficient $a_1$, which gives a constant contribution to the optical conductivity. It naively appears that the cut-off dependence cancels in this term. However, in reality, this term is non-universal. Indeed, we know that in a Fermi liquid with umklapps the real part of the optical conductivity tends to a constant as $\Omega \to 0$. Hence, we expect that umklapp scattering of Fermi liquid like quasiparticles away from the hot spots will renormalize the $a_1$ term rendering it non-universal. At a technical level this non-universality appears in our calculation in the following way. The cut-off dependence in Eq.~(\ref{Pixxgm}) comes from two sources: i) the normalization of the current vertex, Eq.~(\ref{Gmscal2}), ii) the cut-off on the integral over the momentum $l$ along the Fermi surface. The argument given above indicates that we should not generally expect a universal cancellation between these two cut-offs in the $a_1$ term.

The final term with the coefficient $a_2$ in Eq.~(\ref{sigmaf}) is the critical hot spot contribution in the rainbow approximation. Unlike the first two terms in Eq.~(\ref{sigmaf}), which are present in a Fermi liquid, this term appears only at the critical point. Since $r_0 > 0$, this term is always suppressed compaired to the cold Fermi liquid contribution. In contrast, Chubukov {\em et al.} \cite{advances} found $r_{0}=-1/2$
in the corresponding regime. There are two effects taken into account in our calculations that lead to the difference:
i) the dependence of the fermion self energy on the momentum along the Fermi surface; ii) rainbow corrections to the current vertex.

It is interesting to discuss the $a_2$ term from the point of view of RG. Let $[J_x]$ be the scaling dimension of the current operator $J_x$. Then the current vertex $\Gamma(q,p)$ should have the dimension $[J_x] - 2 [\psi]$. The dimension of the fermion operator $[\psi] = 3/2$ in the rainbow approximation (all dimensions are with respect to momentum). Hence, from Eq.~(\ref{Gmscal2}), in the rainbow approximation,
\beq [J_x] = 3 + r_0 \eeq
where we've used $z = 2$. Similarly, the quantum critical contribution to the optical conductivity $\sigma(\Omega)$ should have the dimension $2[J_x] - d - 2 z = 2 r_0$, which is precisely the scaling of the $a_2$ term in Eq.~(\ref{sigmaf}). Hence, our scaling forms for the conductivity and the current vertex have a consistent RG interpretation. Technically, this is based on the relation (\ref{r2}), which seems to appear quite accidentally in our calculations.


Coming back to the constant $a_1$ `Fermi liquid' contribution to the conductivity, although this term is non-universal in the broad sense, it can still be calculated within our spin-fermion model if one treats the action (\ref{eq:theaction}) as a `semi-microscopic' theory rather than an effective theory valid only in the infra-red scaling limit. This is only meaningful when the `UV' scale of the theory $\Lambda$, Eq.~(\ref{Lambdadef}), is much smaller then the Brillouin zone size. For this to be true, we must either send the coupling constant $\lambda \to 0$ or $N \to \infty$. In such a weak coupling limit, only the fermions within a distance $\Lambda$ to the hot spot are strongly affected by the interactions with the spin density wave fluctuations and contribute to the constant term in the conductivity.

\begin{figure}[h]
\begin{center}
\includegraphics[width = 4in]{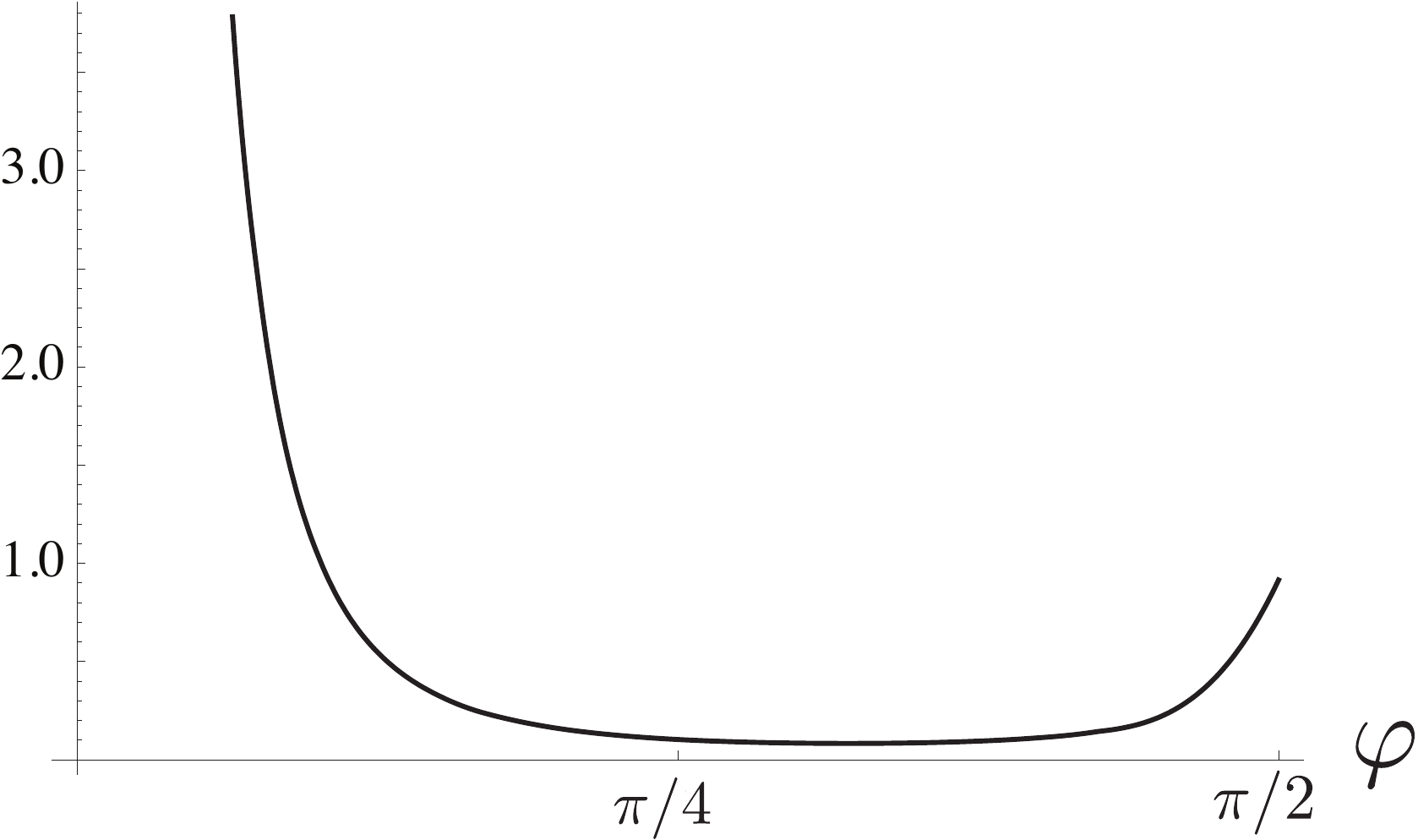}
\caption{The constant contribution to the conductivity as a function of the angle $2 \varphi$  between the Fermi surfaces.}
\label{fig:C1}
\end{center}
\end{figure}

To extract the constant contribution in the rainbow approximation and in the limit $N \to \infty$, we have to solve the integral equation (\ref{GammaF}) for the current vertex $\Gamma^-$ in the regime $p \sim \Lambda$ and then use Eq.~(\ref{PixxGm}) to determine the conductivity. The details of the calculation are presented in the Appendix \ref{app:const}. Here we only quote the result
\beq \mathrm{Re} \sigma_{ij}(\Omega) \to N^2 C_1(\varphi) \delta_{ij} \label{C1}\eeq
with the function $C_1(\varphi)$ shown in Fig.~\ref{fig:C1}. Note that even though we have taken the `weak-coupling' limit, the coupling constant $\lambda$ does not enter the expression for $C_1$. Moreover, Eq.~(\ref{C1}) is enhanced in $N$ compared with the naively expected scaling $\sigma \sim N$. Note that the conductivity diverges in the nested limits $\varphi \to 0$,\, $\varphi \to \pi/2$. 

We conclude this section by reminding the reader that the `Drude' pole (and associated delta function) in Eq.~(\ref{sigmaf}) should not be confused with the low frequency conductivity at $T>0$. By setting $T=0$ exactly in this paper we are effectively working in the limit $T \ll \Omega$, while the phenomenologically interesting d.c. conductivity is often $\Omega \ll T$. In the latter limit, a delta function in the electrical conductivity follows from translational invariance if the electric current and momentum operators have nonvanishing overlap. In our quantum critical regime the electrical current lies in an $SU(2)$ pseudospin triplet, while the momentum is a pseudospin singlet. Therefore there can be no overlap and a finite quantum critical d.c. conductivity is expected at $T>0$. A delta function will re-emerge due to the effect of irrelevant pseudospin symmetry breaking operators such as Fermi surface curvature terms. This delta function must then be resolved by additional physics such as impurity scattering or umklapp scattering beyond that already included in our theory\cite{roschandrei,maslov}.

\section{Self energy, vertex corrections and `cold' conductivity}
\label{sec:gen}
The anisotropic structure of criticality in our model -- the existence of hot spots -- means that as well as the quantum critical contribution to the conductivity, there is also the contribution of the `cold' fermions. We have already encountered such a contribution in Sec.~\ref{sec:hotspot}. At the level of the approximation in Sec.~\ref{sec:hotspot}, the cold fermions are well-defined quasiparticles with a Fermi liquid like $\omega^2$ damping rate, which give rise to a Drude peak in the optical conductivity, as well as a constant dissipative part at finite frequency. 

In the remainder of this paper we will investigate the effects of scattering of the cold fermions by quantum critical modes living at the hot spots. Since the cold contribution in Sec.~\ref{sec:hotspot} was found to dominate over the critical hot spot contribution at low frequency, it is crucial to understand the extent to which hot modes can disrupt the cold Fermi liquid behavior. This issue is similarly important at finite temperature, where a cold Fermi liquid would contribute a $T^{-2}$ temperature dependent DC conductivity, which is likely to short-circuit any phenomenologically interesting quantum critical temperature scaling of the conductivity \cite{rice, rosch}. 

In section \ref{sec:selfen} we will study the damping rate of cold fermions due to scattering by critical modes. We would like to see if it is possible to obtain a scattering rate that is stronger than the $\w^2$ result of Fermi liquid theory. The next question is the extent to which the scattering rate feeds through to the electrical conductivity. One must be wary of cancellations at low frequencies between the effects of self energy and vertex corrections to the conductivity \cite{MIT}. This section studies the connection between self energy and conductivity in a general setting that is independent of the explicit form, e.g. (\ref{eq:theaction}), of the hot spot theory.

\subsection{Self energy and vertex correction cancellation}
\label{sec:cancel}

Before considering specific scattering processes, it is helpful to set up a general formalism describing the interaction of cold fermions with a  (generically composite) operator $\ocal$ in the quantum critical theory. Since the momentum of critical fluctuations is small compared to the Fermi momentum of the cold fermions, it is sufficient at low energy to zoom into a patch of the cold Fermi surface. In the simplest case of a zero momentum, neutral scalar operator $\ocal$ the interaction is effectively described by the coupling
\be\label{eq:coupling1}
S_\text{1} = \lambda_1 \int d^3x \psi^\dagger(x) \psi(x) \ocal(x) \,.
\ee
We will see more concretely below how a local effective coupling emerges. In the patch, the cold fermions $\psi$ have the usual inverse propagator
\be\label{eq:cold}
G_\text{cold}^{-1}(\w,\vec p) = i \w - v^\star p_\perp - \frac{p_\parallel^2}{2 m^\star} \,.
\ee
Similarly to before, $p_\perp$ is the momentum perpendicular to the Fermi surface at the patch while $p_\parallel$ is the parallel momentum. In the patch the fermion has local Fermi velocity $v^\star$ and local Fermi surface curvature radius $m^\star v^*$. We are suppressing spin indices, their only effect is to add various factors of 2 below. The propagator for $\ocal$ is computed in the hot spot theory. In frequency space we will write
\be\label{eq:cwp}
C(\w,\vec p) = \int d\tt d^2x e^{i \w \tt - i \vec p \cdot \vec x} \langle \ocal(\tt, \vec x) \ocal(0,0) \rangle \,.
\ee
For the moment we will not need to make any assumptions about the form of this propagator.

\begin{figure}[h]
\begin{center}
\includegraphics[height=80pt]{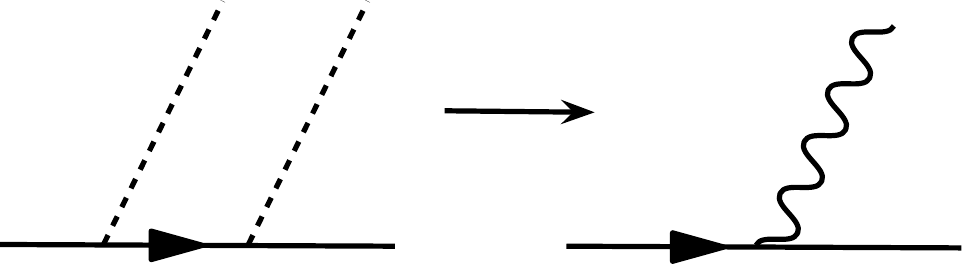}
\caption{Generating a local interaction between cold fermions and $\ocal = \phi^2$. As in Fig.~\ref{fig:rainbow}, a full line
is a fundamental fermion $\psi$, and a dashed line is a  
`fundamental boson' $\phi$. A wavy line represents a propagator for the operator $\ocal$.
The intermediate fermion on the left hand graph is necessarily very off shell.  \label{fig:phi2}}
\end{center}
\end{figure}

As an example of how the local interaction (\ref{eq:coupling1}) arises, consider the operator $\ocal = \phi^2$ of the critical theory (\ref{eq:theaction}). This is the relevant operator that drives the system away from criticality and  will be an important example later in our paper. The coupling of this operator to two cold fermions is shown in figure \ref{fig:phi2}. 
In figure \ref{fig:phi2} a cold fermion is scattered by a pair of hot bosons. Because the bosons are assumed to be hot, each bosonic scattering shifts the momentum of the fermion by the spin density wavevector. In the cold regions we are studying, away from the hot spots, this process takes the fermion far from the Fermi surface. Until being scattered back to the Fermi surface by the second boson, the fermion is hopelessly off shell. In the evaluation of figure \ref{fig:phi2} we can therefore approximate the propagators of these off shell fermions by a constant. Thereby collapsing the two bosonic operators to a point, we obtain the local effective interaction (\ref{eq:coupling1}) with the composite operator $\ocal = \phi^2$.

\begin{figure}[h]
\begin{center}
\includegraphics[height=90pt]{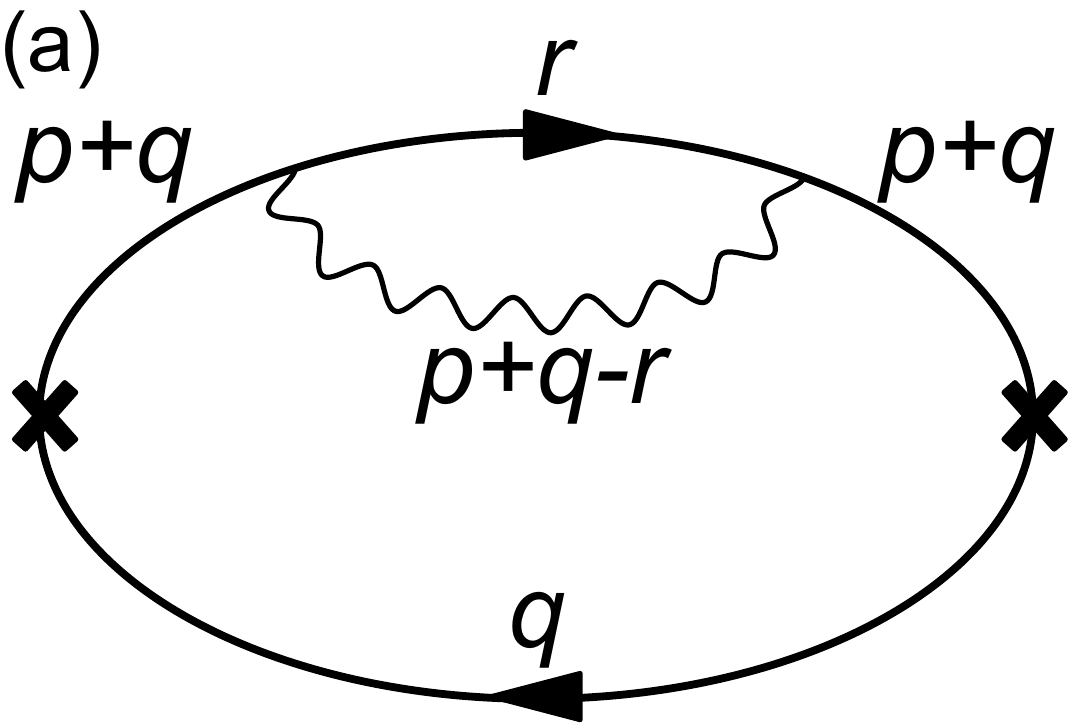}~~\includegraphics[height=90pt]{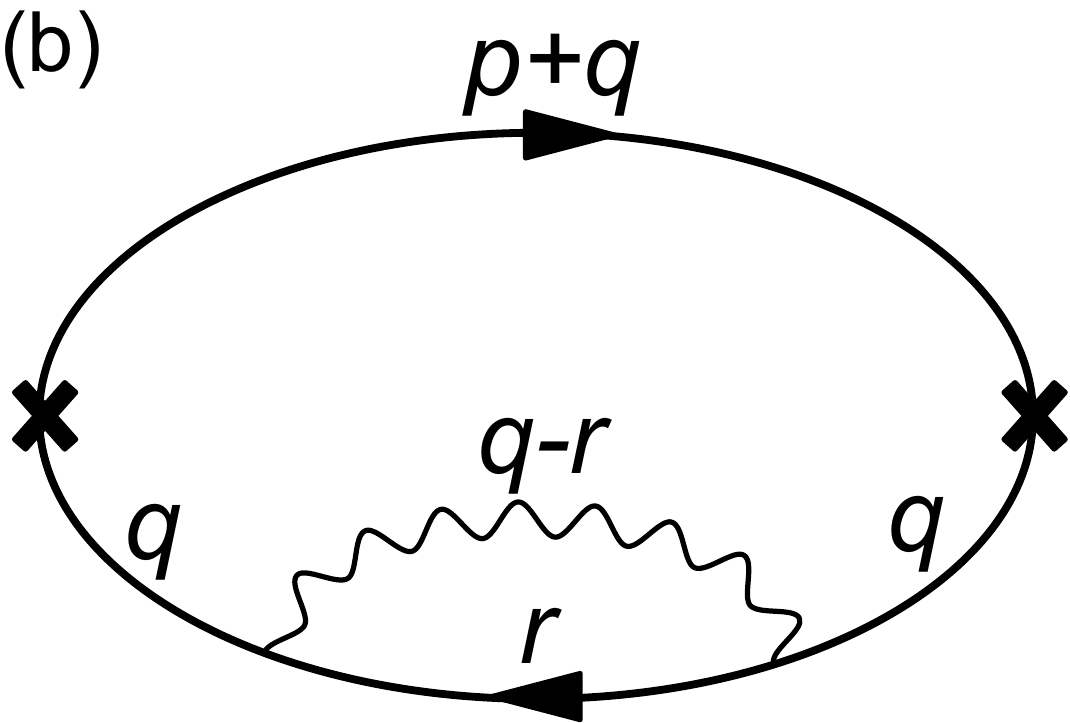}~~\includegraphics[height=90pt]{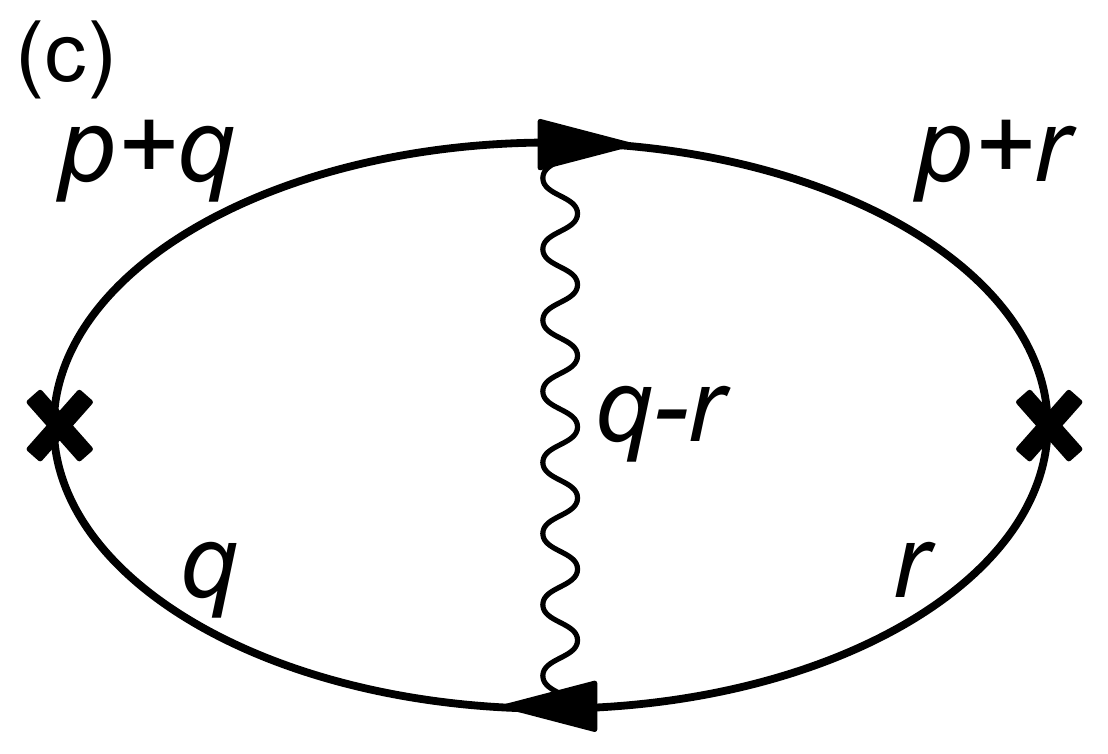}
\caption{The three leading order corrections in $\lambda_1$ to the conductivity due to cold fermions. The wavy lines denote propagators of the neutral bosonic quantum critical operator $\ocal$. \label{fig:twoloop}}
\end{center}
\end{figure}

The three Feynman diagrams shown in figure \ref{fig:twoloop} give the leading contribution to the conductivity due to the coupling $\lambda_1$ of (\ref{eq:coupling1}). The current operator corresponding to the cold fermions (\ref{eq:cold}) has components
\be\label{eq:coldcurrents}
J^\text{cold}_ \perp = v^\star \psi^\dagger \psi \,, \qquad J^\text{cold}_ \parallel = \frac{i}{m^\star} \psi^\dagger \pa_\parallel \psi \,.
\ee
Note that what is meant by the perpendicular and parallel directions, as well as the Fermi velocity, the curvature radius and indeed the coupling $\lambda_1$, is a function of the location of the patch on the Fermi surface. 

We remark that in the present section we are neglecting the four-fermi interactions between the cold fermions. One result of such Fermi liquid interactions is to make the finite frequency current vertex $\vec{\Gamma}(\omega\to 0, \vec{p} = 0)$, relevant for the calculation of the optical conductivity, different from the Fermi velocity $\vec{v}^*$. As already shown in Sec.~\ref{sec:hotspot} this effect is parametrically strong for lukewarm fermions in the neighbourhood of the hot spot. However, the replacement $\vec{v}^* \to \vec{\Gamma}$ in the current vertex does not qualitatively modify our discussion below and we will not further consider the interplay between Fermi liquid and hot scattering. 

In the cold fermion current (\ref{eq:coldcurrents}) the curvature term is down by a power of momentum compared to the Fermi velocity term. We can therefore drop the effects of curvature in a first computation. The first two graphs in figure \ref{fig:twoloop}, the self energy corrections, give the following contribution to the current correlator
\be\label{eq:ab}
\dd^{(a+b)} \Pi_{ij}^\text{cold}(\w) = 2 \lambda_1^2 \int \frac{d^3q}{(2\pi)^3}  \frac{d^3r}{(2\pi)^3}  v_i^\star v_j^\star G(q) G(r) G(p+q)
 \Big(G(p+q) C(p+q-r) + G(q) C(q-r) \Big) \,.
\ee
In this expression and in the following, the external 3-momentum is $p=(\w,0,0)$.
All the fermion propagators in this expression and the remainder of this section are those of the cold fermions (\ref{eq:cold}). The rightmost graph in figure \ref{fig:twoloop}, the vertex correction, gives
\be\label{eq:c}
\dd^{(c)} \Pi_{ij}^\text{cold}(\w) = 2 \lambda_1^2 \int \frac{d^3q}{(2\pi)^3}  \frac{d^3r}{(2\pi)^3}  v_i^\star v_j^\star G(q) G(r)
 G(p+q) G(p+r)  C(q-r)  \,.
\ee

Using the following identity twice, where $p_\tau=\w$ and the spatial components $\vec p = 0$,
\be\label{eq:identity}
G(q) G(p+q) = \frac{1}{i \w} \Big(G(q)-G(p+q) \Big) \,,
\ee
it is easy to show that the self energy and vertex corrections are equal and opposite
\bea\label{eq:equalopposite}
\lefteqn{\dd^{(a+b)} \Pi_{ij}^\text{cold}(\w) = - \dd^{(c)} \Pi_{ij}^\text{cold}(\w) =
- \frac{2 \lambda_1^2}{\w^2} \int \frac{d^3q}{(2\pi)^3}  \frac{d^3r}{(2\pi)^3}  v_i^\star v_j^\star 
G(q) G(r) } \nonumber \\
& &\qquad \qquad \qquad \qquad \qquad \times \Big(C(p+q-r) + C(-p+q-r) - 2 C(q-r) \Big) \,.
\eea
Therefore the total correction vanishes
\be
\dd \Pi_{ij}^\text{cold}(\w) = \dd^{(a+b)} \Pi_{ij}^\text{cold}(\w) + \dd^{(c)} \Pi_{ij}^\text{cold}(\w) = 0 \,.
\ee
This cancellation is essentially the same as that observed in Ref.~\onlinecite{MIT}. The identity (\ref{eq:identity}) integrates to a Ward identity relating vertex correction and self energy
\be
\dd \Gamma(p,r) = \frac{i}{\w} \Big(\Sigma(p+r) - \Sigma(r) \Big) \,.
\ee
Thus it is not surprising that the cancellation occurs in a range of different theories.
The physics behind the cancellation is that, having neglected the effect of Fermi surface curvature in (\ref{eq:coldcurrents}), the electrical current and charge density are proportional in each patch. Because the patches decouple at low energies, the current correlator is given by the density correlator summed over all patches.  Conservation of particle number requires that the density correlator vanish at zero momentum. The vanishing of the current correlator $\Pi_{ij}^\text{cold}(\w)$ then follows. 

The fact that the leading order self energy correction does not contribute to the conductivity implies that non-Fermi liquid self energies will typically not translate into non-Fermi liquid response. At the end of this section we will obtain the leading nonzero answer for the conductivity, due to curvature terms in the current, and in the following sections we will explicitly evaluate the self energy and hence the conductivity due to scattering off specific low scaling dimension quantum critical modes. Before these discussions, however, we will describe in the following subsection one important circumstance in which vertex and self energy corrections reinforce each other rather than cancel. The non-cancellation occurs for similar reasons to those discussed in section \ref{sec:hotspot} above, and will lead  to interesting `enhanced critical umklapp' modes that we discuss in detail below.

\begin{figure}[h]
\begin{center}
\includegraphics[height=90pt]{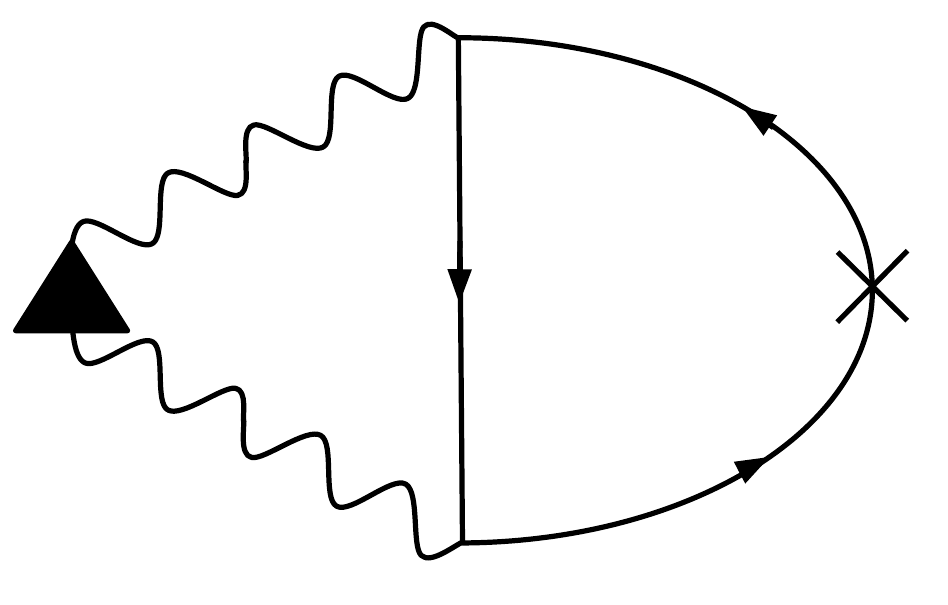}
\caption{The leading contribution to the cross-correlator of hot and cold fermion currents $\langle J^\text{hot} J^\text{cold}\rangle$. The triangle denotes the three point vertex $\langle J^\text{hot} \ocal \ocal \rangle$ in the critical theory. \label{fig:Jcross}}
\end{center}
\end{figure}

We conclude this section by noting that if we write the total current schematically as
\beq J = J^\text{cold} + J^\text{hot} \,,\eeq
then in addition to the purely cold contribution to the conductivity $\langle J^\text{cold} J^\text{cold} \rangle$ considered above and the purely hot contribution $\langle J^\text{hot} J^\text{hot} \rangle$, there generally exists a cross-term $\langle J^\text{cold} J^\text{hot}\rangle$. The leading contribution in $\lambda_1$ to this cross-term is schematically shown in Fig.~\ref{fig:Jcross} and requires as an input from the critical theory the three point function $\langle J^\text{hot} \ocal \ocal \rangle$. For the simplest case when $\ocal = \phi^2$, this three point function in the critical theory vanishes by pseudospin symmetry and so the cross-term disappears. However, for the Cooper pair and charge density wave operators considered in the following section the three point function is generally finite and, in a full calculation, the cross-term cannot be neglected. Nevertheless, as a first pass, we will ignore the cross-term in the present paper.

\subsection{Non-cancellation for scattering off neutral `$2 k_F$' CDW modes}
\label{sec:noncancellation}
In this section we extend the considerations in section \ref{sec:cancel} to the interaction of cold fermions with other types of composite operators. Namely, we consider charge $2$, zero momentum, Cooper pair operators and neutral, finite momentum, charge density wave operators. The fluctuations of both operators were found to be enhanced at the SDW critical point in Ref.~\onlinecite{Metlitski:2010vm}.

Let us begin with the Cooper pair operator $\ocal$, which couples fermions with opposite Fermi momentum,
\be\label{eq:lambda3}
S_\text{2} = \lambda_2 \int d^3x \Big( \psi^\dagger(x) \widetilde \psi^\dagger(x) \ocal(x) + \widetilde \psi (x) \psi(x) \ocal^\dagger(x) \Big) \,.
\ee
Here we use $\psi$ and $\widetilde \psi$ to denote cold fermion fields living in two antipodal patches of the Fermi surface. Note that the location of the antipodal patches on the Fermi surface can be arbitrary. We will shortly discuss how to generate the coupling (\ref{eq:lambda3}) in the theory (\ref{eq:theaction}).

Next, consider the coupling of cold fermions to a generic CDW operator $\ocal$,
\be\label{eq:lambda2}
S_\text{3} = \lambda_3 \int d^3x \Big( \psi^\dagger(x) \widetilde \psi(x) \ocal(x) + \widetilde \psi^\dagger(x) \psi(x) \ocal^\dagger(x) \Big) \,.
\ee
Here $\psi$ and $\widetilde \psi$ denote cold fermion fields in two patches separated by the wavevector $\vec{K}$ of $\ocal$. Generally, $\vec{K}$ only connects a finite number of points on the Fermi surface and so $\psi$ and $\widetilde \psi$ must reside in the vicinity of these points. However, if the Fermi surface was nested with the wavevector $\vec{K}$, i.e. $\epsilon(\vec{k} + \vec{K}) = - \epsilon(\vec{k})$, $\psi$ and $\widetilde \psi$ could reside anywhere on the Fermi surface. 

The example of the CDW that will be of interest to us below is the operator $\ocal = \psi^{1 \dagger}_2 \psi^3_2$ of the critical theory (\ref{eq:theaction}). We call this a `$2k_F$' operator since it connects opposite hot spots $\ell = 1$ and $\ell = 3$, so that its wavevector $\vec{K} = - 2 \vec{K}^{1}_2 = - 2 \vec{K}^1_1$, with $\vec{K}^{\ell}_a$ - the momentum of the hot spot associated with fermion $\psi^{\ell}_a$.\footnote{There is also an analogous operator connecting hot spots $\ell = 2$ and $\ell = 4$.} This operator was shown in Refs.~\onlinecite{Metlitski:2010vm, Metlitski:2010zh} to condense at zero temperature, contributing to a modulated Ising-nematic order. Our full Fermi surface is generally not nested with respect to the wavevector $\vec{K}$, hence the operator $\cal O$ will only couple efficiently to the fermions in the vicinity of the hot spots. However, within the critical theory (\ref{eq:theaction}) the Fermi surface {\it is} nested with respect to $\vec{K}$ by virtue of the pseudospin symmetry. Indeed, Eq.~(\ref{eq:Gpseud}) implies that $G^{\ell}_a(\omega, \vec{k}) = - G^{\ell}_a(-\omega,-\vec{k}) = -G^{-\ell}_a(-\omega, \vec{k})$, or measuring momenta in the full Brillouin zone 
\beq G(\omega, \vec{K}^{\ell}_a +\vec{k}) = -G(-\omega, -\vec{K}^{\ell}_a + \vec{k}) \,. \label{eq:Gnest}\eeq
Hence, if $\vec{k}$ is on the Fermi surface, so is $\vec{k} + \vec{K}$ and the Fermi surface is nested. Note that the nesting should not be confused with the Fermi surface being straight. Even though the bare theory (\ref{eq:theaction}) has a straight Fermi surface, the dressed Fermi surface possesses a finite curvature as discussed in Refs.~\onlinecite{advances,Metlitski:2010vm}. On the other hand, the nesting of the bare theory is protected by the pseudospin symmetry even when the effects of the interactions in the action (\ref{eq:theaction}) are taken into account. The nesting is destroyed only when we break the pseudospin symmetry by adding an explicit curvature term to the theory, which we schematically write as,
\beq \delta L = \kappa_0 |\nabla \psi|^2 \,. \label{eq:curvexp}\eeq
The coupling constant $\kappa_0$ will generally have an RG flow,
\beq \frac{d \kappa_0}{d \ell} = - b \kappa_0 \,. \label{eq:kappaflow}\eeq
At tree level, $b = 1$, and so we expect the curvature to be an irrelevant perturbation to the critical theory. Nevertheless, for lukewarm fermions away from hot spots this irrelevancy is dangerous as it eventually destroys the nesting. Let us estimate the energy scale at which the effects of the perturbation (\ref{eq:curvexp}) on lukewarm fermions a distance $k_\parallel$ from the hot spot kick in. From Eq.~(\ref{eq:scalingGD}), in the absence of the curvature perturbation, the natural scale of the fermion propagator is $G^{-1} \sim k^{z/2-\eta_\psi}_\parallel$. Hence, from the flow (\ref{eq:kappaflow}), the curvature correction to $G^{-1}$ scales as $\delta G^{-1} \sim \kappa_0 k^{z/2-\eta_\psi + b}_\parallel$. Assuming that lukewarm fermions remain well-defined quasiparticles, Eq.~(\ref{eq:lukewarm}), on the Fermi surface $G^{-1} \sim ({i \omega}/{k^z_\parallel}) k^{z/2-\eta_\psi}_\parallel$. Hence, the curvature correction $\delta G^{-1}$ becomes significant once $\omega \lesssim k^{z +b}_\parallel$. 
Therefore, there exists an energy window $k^{z+b}_\parallel \ll \omega \ll k^z_\parallel$ in which the lukewarm fermions can efficiently couple to the CDW operator $\ocal$.

\begin{figure}[h]
\begin{center}
\includegraphics[height=220pt]{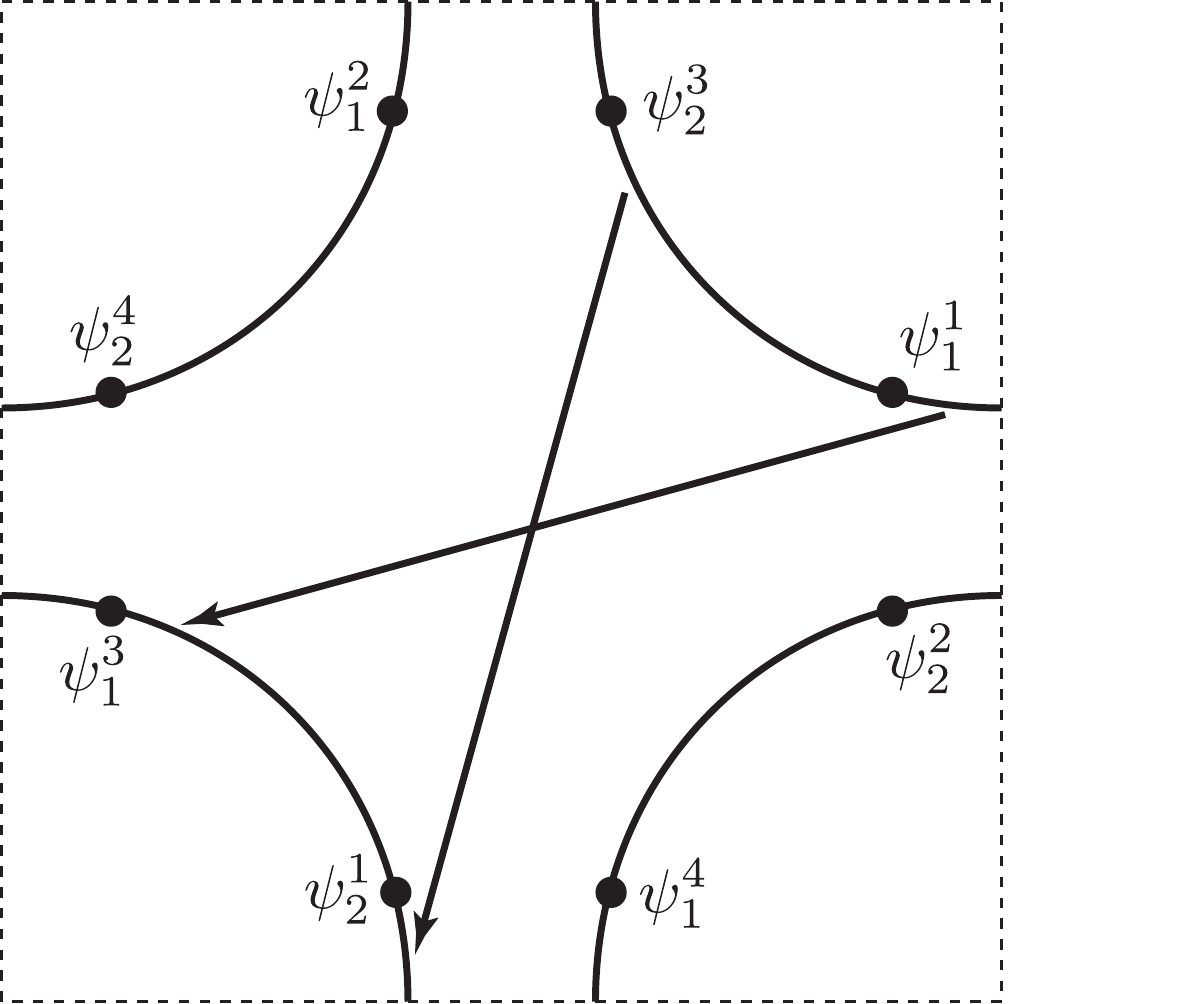}
\caption{The $2k_{F}$ operator generates the umklapp process shown here. The scattering is from the lukewarm regions near the hot spots,
and conserves total momentum upto a reciprocal lattice vector.
Note that this process arises after integrating
out high energy degrees of freedom from the effective theory in Eq.~(\ref{eq:theaction}). \label{fig:umklapp}}
\end{center}
\end{figure}

With the above remarks in mind, let us see how the effective coupling (\ref{eq:lambda2}) of the CDW operator $\ocal = \psi^{1 \dagger}_2 \psi^3_2$ to the lukewarm fermions can be generated in the theory (\ref{eq:theaction}).
Via an intermediate fundamental boson $\phi$ the fermions in $\ocal$ can be coupled to fermions $\psi = \psi^{3}_1$ and $\widetilde \psi = \psi^{1}_1$. This coupling is illustrated in Fig.~\ref{fig:psi2} below.
Because we wish to consider scattering by hot operators in the critical theory, we demand that the pair of fermions constituting $\ocal$ be hot.
The boson that connects the external cold (or lukewarm) fermions to the hot fermions is necessarily extremely off shell. Similarly to the off shell fermions in the previous subsection, we may replace its propagator by a constant. This then generates the local interaction (\ref{eq:lambda2}), as is illustrated in figure \ref{fig:psi2}.
\begin{figure}[h]
\begin{center}
\includegraphics[height=80pt]{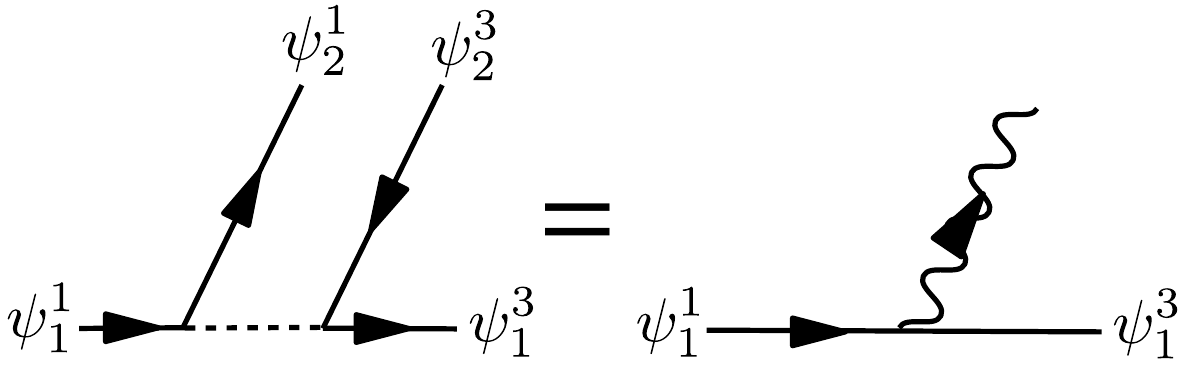}
\caption{Generating a local interaction of lukewarm fermions $\psi = \psi_1^3$ and $\widetilde \psi = \psi^{1}_1$ with the operator $\ocal = \psi^{1\dagger}_2 \psi^3_2$. The intermediate `fundamental' boson on the left hand side is necessarily very off shell. Wavy lines are propagators for the operator $\ocal$, as in Fig.~\ref{fig:phi2}. \label{fig:psi2}}
\end{center}
\end{figure}
To be completely explicit, we could write this coupling as the four fermion interaction
\be\label{eq:uneff}
S_\text{umklapp} \sim \int d^3x \psi^{3\dagger \mathrm{cold}}_1 \psi_1^{1 \mathrm{cold}} \psi^{1\dagger \mathrm{hot}}_2 \psi^{3 \mathrm{hot}}_2 + \mbox{c.c.} \,.
\ee
We show this scattering process in Fig.~\ref{fig:umklapp}. 
We see that in terms of the underlying microscopic fermions, this coupling does not conserve momentum and therefore describes umklapp scattering. We comment that numerical studies of the Hubbard model have also noted the importance
of umklapp processes in the breakdown of Fermi liquid theory \cite{ossadnik}.

The same argument holds for the scattering off the Cooper pair operator $\ocal = \psi^1_2 \psi^3_2$, leading to the interaction (\ref{eq:lambda3}). All that is necessary is to reverse e.g. the arrows of $\psi^3_1$ and $\psi^3_2$ propagators in figure \ref{fig:psi2}.

\begin{figure}[h]
\begin{center}
\includegraphics[height=90pt]{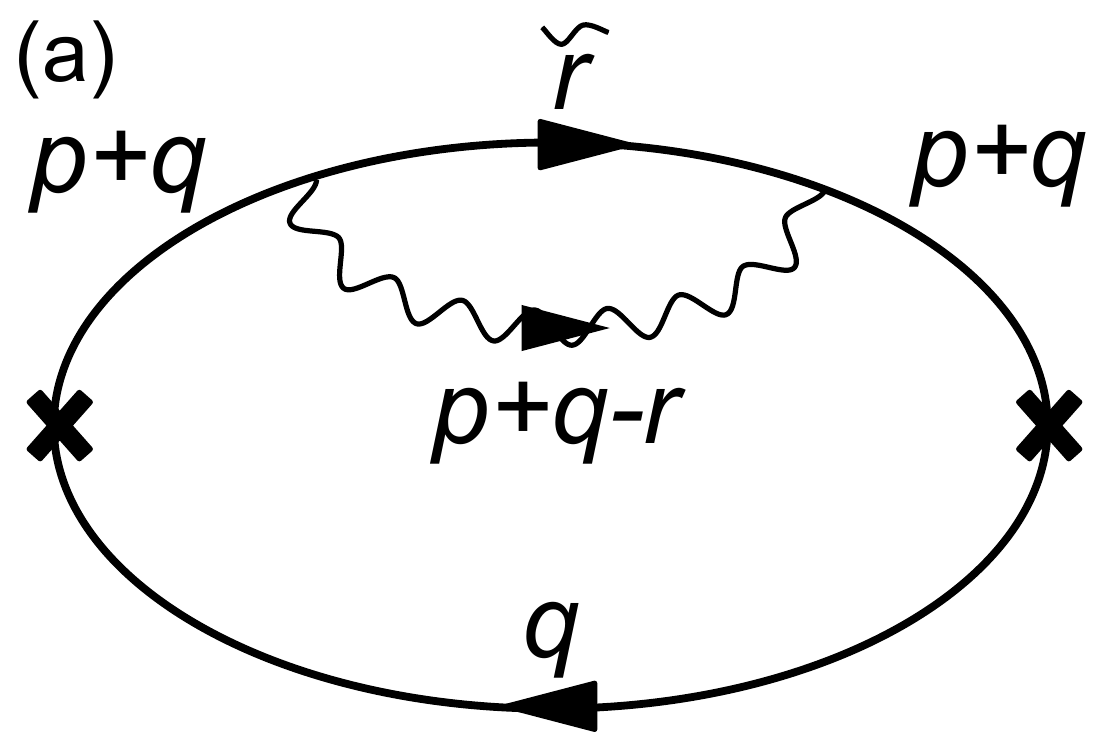}~\includegraphics[height=90pt]{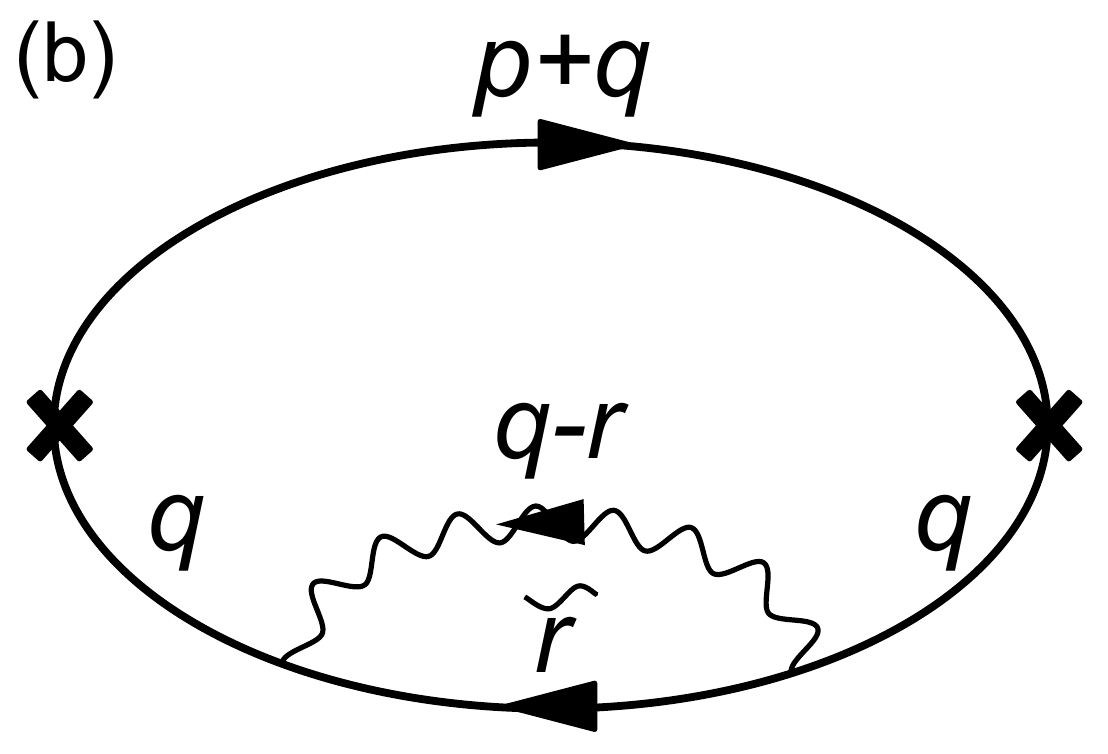}~\includegraphics[height=90pt]{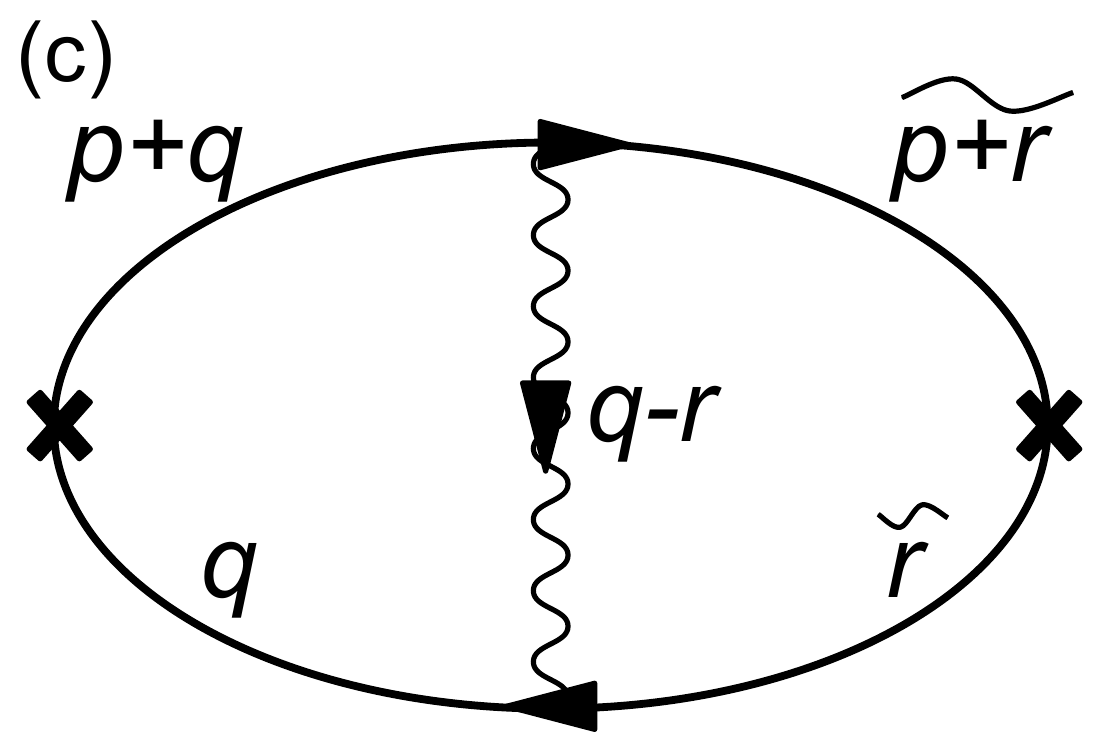}
\caption{The three leading order corrections in $\lambda_3$ to the conductivity due to cold fermions. Tildes over momenta indicate that the fermion propagator in question corresponds to the opposite patch. The hot operator $\ocal$ is now complex and so its propagators carry an arrow.\label{fig:twoloopnematic}}
\end{center}
\end{figure}

We now proceed to the discussion of optical conductivity. Consider scattering off CDW fluctuations first.
The diagrams describing the leading order corrections to the conductivity due to the interaction (\ref{eq:lambda2}) are shown in figure \ref{fig:twoloopnematic}. The wavy line now represents the two-point function of the CDW operator $C = \langle O O^{\dagger} \rangle$. 
The crucial fact to take into account in evaluating these diagrams is that the Fermi velocity is pointing in opposite directions in the two patches separated by the ordering wavevector $\vec{K}$. This is a consequence of the relation (\ref{eq:Gnest}). Therefore the current operator (neglecting as before the curvature term at low energies) should be written as
\be\label{eq:twopatchcurrent}
J^\text{cold}_ \perp = v^\star \Big( \psi^\dagger \psi - \widetilde \psi^\dagger \widetilde \psi \Big) \,.
\ee
Similarly to equations (\ref{eq:ab}) and (\ref{eq:c}) previously, the graphs of figure \ref{fig:twoloopnematic} give
\be
\dd^{(a+b)} \Pi_{ij}^\text{cold}(\w) = 2 \lambda_3^2 \int \frac{d^3q}{(2\pi)^3}  \frac{d^3r}{(2\pi)^3}  v_i^\star v_j^\star G(q) \widetilde G(r) G(p+q)
 \Big(G(p+q) C(p+q-r) + G(q) C(q-r) \Big) \,.
\ee
and
\be\label{eq:cnematic}
\dd^{(c)} \Pi_{ij}^\text{cold}(\w) = - 2 \lambda_3^2 \int \frac{d^3q}{(2\pi)^3}  \frac{d^3r}{(2\pi)^3}  v_i^\star v_j^\star G(q) \widetilde G(r) G(p+q) \widetilde G(p+r)  C(q-r)  \,.
\ee
Recall that the external 3-momentum is $p=(\w,0,0)$.
The crucial minus sign difference with (\ref{eq:c}) is due to the minus sign in the current (\ref{eq:twopatchcurrent}). Tilde indicates a Green's function for fermions on the opposite patch. The factors of 2 work out a little differently compared to section \ref{sec:cancel}: the Feynman diagrams have a reduced symmetry and we must be careful not to double count the patches.
The extra minus sign means that upon using the identity (\ref{eq:identity}), the self energy and vertex contributions add rather than cancel. The total result is
\bea\label{eq:nonzerok}
\lefteqn{\dd \Pi_{ij}^\text{cold}(\w) = \dd^{(a+b)} G_{ij}^\text{cold}(\w) + \dd^{(c)} G_{ij}^\text{cold}(\w)}   \\
&  & =  - \frac{4 \lambda_3^2}{\w^2} \int \frac{d^3q}{(2\pi)^3}  \frac{d^3r}{(2\pi)^3}  v_i^\star v_j^\star 
G(q) \widetilde G(r) \Big(C(p+q-r)+C(-p+q-r) - 2 C(q-r) \Big) \,. \nonumber
\eea

The absence of a cancellation means that the scattering off the CDW operator will directly influence the conductivity. We must, however, remember that due to the dangerously irrelevant curvature terms, such scattering is only efficient for lukewarm fermions in the neigbourhood of the hot spot. This issue will be discussed in more detail in sections \ref{sec:2kfself} and \ref{sec:cond2kf}, which will present a calculation of the fermion self energy and conductivity due to $2k_F$ scattering.  

\begin{figure}[h]
\begin{center}
\includegraphics[height=90pt]{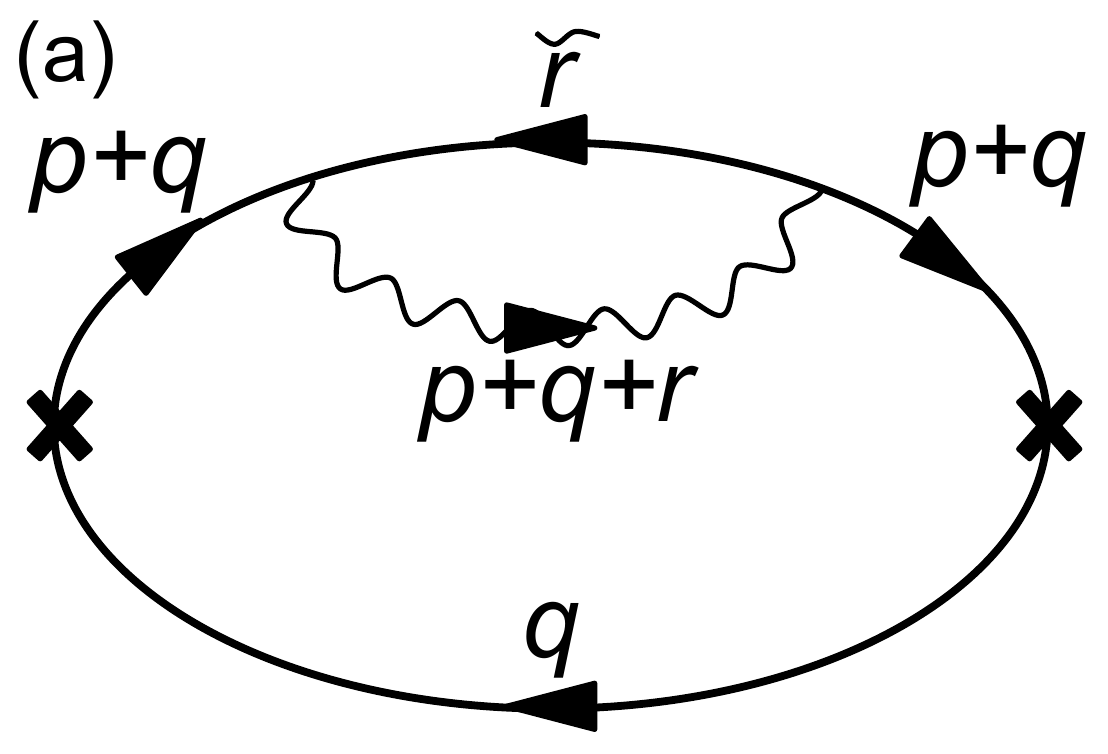}~~\includegraphics[height=90pt]{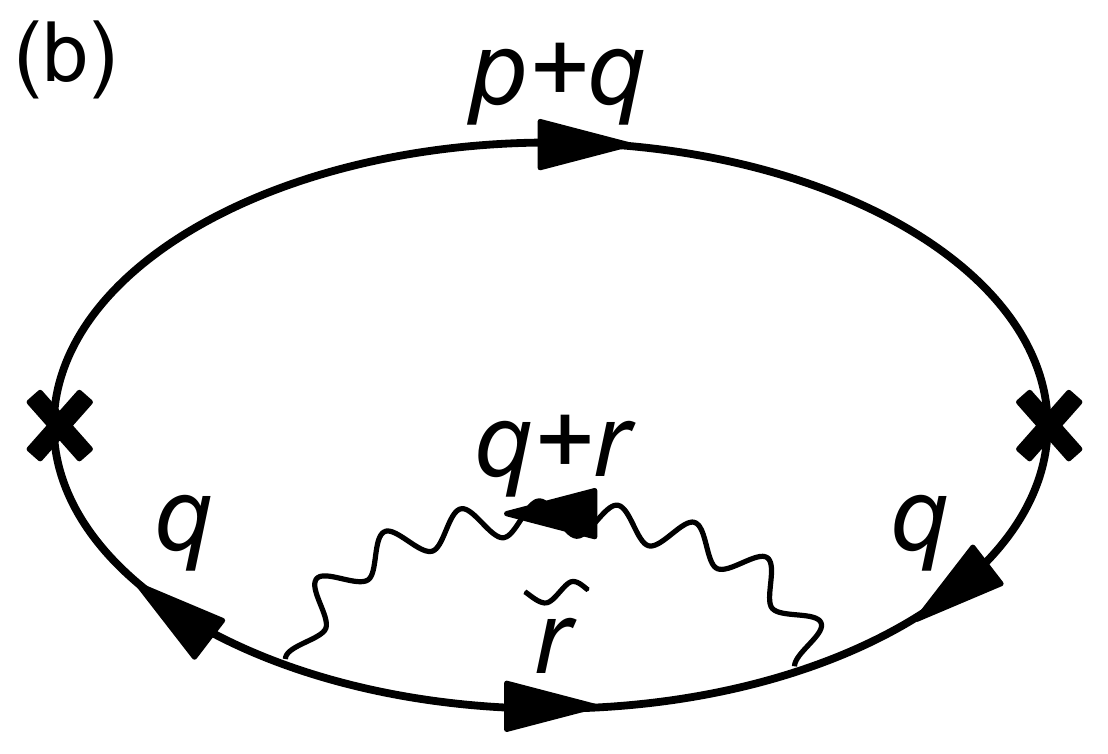}~~\includegraphics[height=90pt]{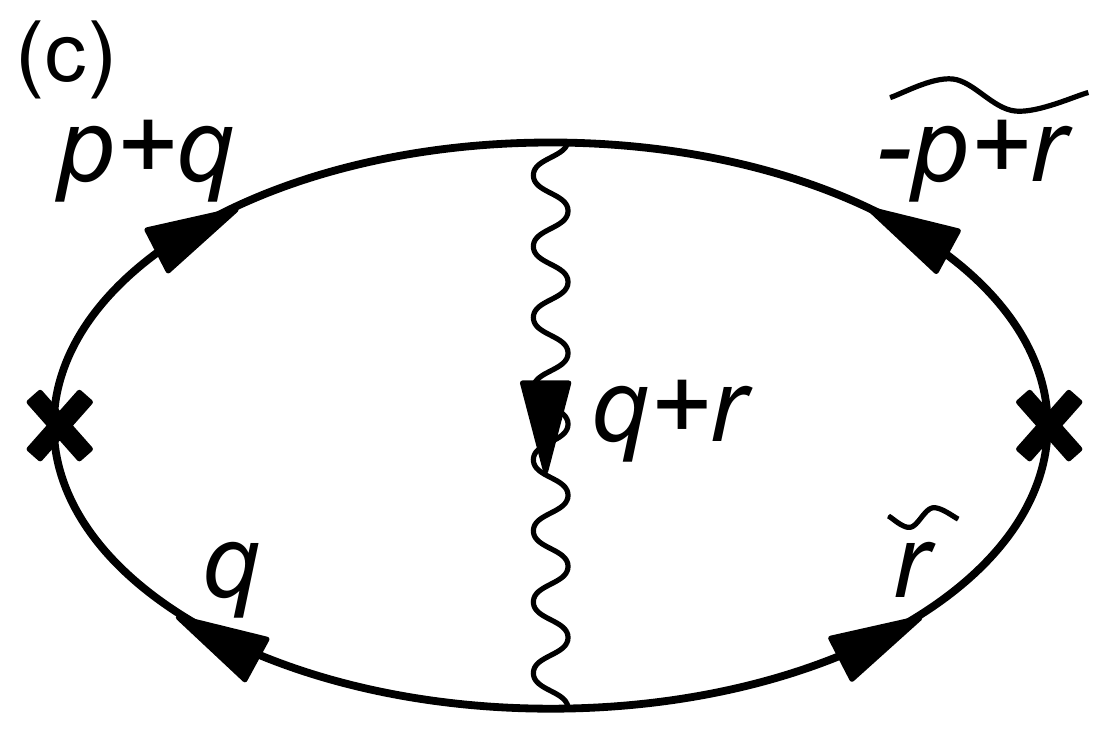}
\caption{The three leading order corrections in $\lambda_2$ to the conductivity due to cold fermions. A few fermionic arrows are reversed compared to figure \ref{fig:twoloopnematic}.\label{fig:twoloopsc}}
\end{center}
\end{figure}

Finally we turn to the case of the interaction with the Cooper pair operator. As already noted, in contrast to the CDW case, such scattering is efficient everywhere on the Fermi surface. The diagrams giving the leading order correction to the conductivity are shown in figure \ref{fig:twoloopsc}.
The fermions $\psi$ and $\widetilde \psi$ now live in antipodal patches and therefore again have opposite Fermi velocities.  Thus, using the two patch current (\ref{eq:twopatchcurrent}) we can evaluate these graphs to obtain
\be
\dd^{(a+b)} \Pi_{ij}^\text{cold}(\w) = -2 \lambda_2^2 \int \frac{d^3q}{(2\pi)^3}  \frac{d^3r}{(2\pi)^3}  v_i^\star v_j^\star G(q) \widetilde G(r) G(p+q) \Big(G(p+q) C(p+q+r) + G(q) C(q+r) \Big) \,.
\ee
and
\be
\dd^{(c)} \Pi_{ij}^\text{cold}(\w) = 2 \lambda_2^2 \int \frac{d^3q}{(2\pi)^3}  \frac{d^3r}{(2\pi)^3}  v_i^\star v_j^\star G(q) \widetilde G(r) G(p+q) \widetilde G(-p+r)  C(q+r)  \,.
\ee
These expressions have various minus signs in different places compared to the previous corresponding formulae for the CDW case. These are seen to lead to a cancellation upon using the identity (\ref{eq:identity}) and thus
\be
\dd \Pi_{ij}^\text{cold}(\w) = \dd^{(a+b)} \Pi_{ij}^\text{cold}(\w) + \dd^{(c)} \Pi_{ij}^\text{cold}(\w) = 0 \,,
\ee
similar to the first case we considered. Analogous cancellations in the fluctuation conductivity of clean superconductors are noted in Refs. \onlinecite{LSV,scbook}.
Each term takes the form
\bea\label{eq:scabc}\lefteqn{\dd^{(a+
b)} \Pi_{ij}^\text{cold}(\omega) = -\dd^{(c)} \Pi_{ij}^\text{cold}(\w) = \frac{2 \lambda_2^2}{\w^2} \int \frac{d^3q}{(2\pi)^3}  \frac{d^3r}{(2\pi)^3}  v_i^\star v_j^\star 
G(q) \widetilde G(r) } \nonumber \\
& & \qquad \qquad \qquad \qquad \qquad \times \Big(C(p+q+r)+C(-p+q+r) - 2 C(q+r) \Big) \,.
\eea
The leading self energy correction due to Cooper pair scattering does not therefore directly push forward to the conductivity. Fermi surface curvature terms are needed to obtain a nonzero contribution to the conductivity, which will consequently be suppressed by powers of momentum.

\subsection{Fermi surface curvature and leading corrections to the conductivity}
\label{sec:curvature}

We found in Section~\ref{sec:cancel} that the leading order conductivity due to scattering off a neutral, zero momentum scalar operator $\ocal$ vanished due to a cancellation. To obtain the leading nonzero contribution we must keep the curvature terms in the cold fermion current operator (\ref{eq:coldcurrents}). The relevant Feynman diagrams remain those of figure \ref{fig:twoloop}, except that now there will be extra powers of momentum at the current insertions. One can check that the contribution to the conductivity due to the cross term $\langle J_\perp^\text{cold} J_\parallel^\text{cold} \rangle$ again vanishes upon adding the three diagrams. In verifying the cancellation one must use the symmetries $C(\w,\vec p\,) = C(\w, - \vec p\,)$ and $C(\w,\vec p\,) = C(-\w, \vec p\,)$. These follow for neutral operators from parity symmetry. The contribution $\langle J_\parallel^\text{cold} J_\parallel^\text{cold} \rangle$ does not vanish however. Summing the three diagrams gives the following result
\bea\label{eq:leading}
\lefteqn{\dd \Pi_{ij}^\text{cold}(\w)  =
- \frac{\lambda_1^2}{m^{\star 2} \w^2} \int \frac{d^3q}{(2\pi)^3}  \frac{d^3r}{(2\pi)^3}  (q_\parallel-r_\parallel)_i (q_\parallel-r_\parallel)_j 
G(q) G(r) }  \\
& & \qquad \qquad \qquad \times \Big(C(p+q-r) + C(-p+q-r) - 2 C(q-r) \Big) \,. \nonumber
\eea
Note how in this expression we are taking $q_\parallel$ to be a vector whose direction in general changes with the patch on the Fermi surface. Here, and in (\ref{eq:nonzerok}) above, we should strictly also bear in mind that the couplings $\lambda$ and Fermi velocities and curvatures also depend on the Fermi surface patch.

An analogous expression to (\ref{eq:leading}) exists for scattering off charged operators. The expression follows immediately from (\ref{eq:scabc}). Because of the symmetry in the argument of the $C$ correlators in (\ref{eq:scabc}) it is not necessary to use any symmetry property of $C$ in this case. In section \ref{sec:cond} we will evaluate the integrals in (\ref{eq:leading}) to obtain the corresponding scaling dependence of the conductivity. Before this, we turn to a computation of the self energy due to the scattering processes we are considering. The main results of this section have been the general formulae (\ref{eq:nonzerok}) and (\ref{eq:leading}) for the conductivity resulting from scattering off $2k_F$ and zero momentum operators respectively.

\section{Self energy from composite operators}
\label{sec:selfen}

The cold fermion self energy acquired by scattering off one of the bosonic operators in the previous section, with any of the three couplings $\lambda_i$, is
\be\label{eq:sigma}
\Sigma(q) \equiv  \lambda_i^2 \int \frac{d^3r}{(2\pi)^3} G(r) C(q-r) \,.
\ee
The self energy gives a measure of the effect of scattering off hot modes prior to any cancellations that occur in computing the conductivity.
In this section we compute the self energy due to various operators at the hot spots and then use these results in the following section to compute the corresponding conductivity.
The self energy is of course also of interest in itself, especially insofar as one obtains deviations from Fermi liquid theory.

\subsection{Self energy for charge and momentum preserving scattering}
\label{sec:self}

Consider first the case in which the operator $\ocal$ is a zero momentum neutral operator of the critical theory. Under scale transformations
\be
\ocal(\tt,\vec x) \to s^\Delta \ocal(s^z \tau, s \vec x) \,.
\ee
Here $z$ is the dynamical critical exponent, which we keep arbitrary for the moment, and $\Delta$ is the scaling dimension of the operator $\ocal$.
The most important example, on which we concentrate below, is the operator $\phi^2$ of the critical theory (\ref{eq:theaction}). This operator perturbs the theory away from the critical point and determines the correlation length exponent $\nu$ via
\be\label{eq:nu}
\frac{1}{\nu} = z + 2 - \Delta \,.
\ee
At a mean field `Hertz-Moriya-Millis' \cite{hertz,moriya,millis,vrmp} level, $z=2$, $\nu = 1/2$ and $\Delta=2$. An important feature of our treatment here is that we can also think of $z$ and $\Delta$ as free variables and thereby consider the dependence of physical quantities on $z$ and $\Delta$. Within the specific model of (\ref{eq:theaction}), the values of $z$ and $\Delta$ are in principle determined from a difficult computation in the strongly interacting critical theory. The mean field values provide an estimate at best.
In general, the scaling dimension $\Delta$ determines the scaling form of the correlator 
(\ref{eq:cwp}) to be
\be\label{eq:scaling}
C(\w,\vec p) = \frac{1}{\w^{(z+2-2 \Delta)/z}} \widetilde C\left(\frac{\vec p}{\w^{1/z}}\right) \,.
\ee
Here we have kept only the singular contribution to $C$. The singular term is greater than the analytic terms at low energy if $z+2 \geq 2 \Delta$. The tilde over $\widetilde C$ is not related to opposite patches but indicates that $\widetilde C$ is dimensionless.

The self energy (\ref{eq:sigma}) can be written explicitly, using (\ref{eq:cold}) and shifting integration variables, as
\be\label{eq:selfone}
\Sigma(\w,\vec p) = \lambda_1^2 \int \frac{dq_\tau d^2q}{(2\pi)^3}  \frac{C(q_\tau, \vec q{\,})}{i (\w-q_\tau) - v^\star (p_\perp-q_\perp) - (p_\parallel-q_\parallel)^2/2 m^\star } \,.
\ee

The main contribution to the self energy (\ref{eq:sigma}) comes from the regime $q_\tau \sim |\vec q|^z$ in which the critical fluctuations are on shell. Conservation of energy and momentum implies that the fermion can absorb the momentum of the critical fluctuations and stay on the Fermi surface only if $q_\tau \sim v^\star q_\perp + q_\parallel^2/2 m^\star$. If $z>1$ consistency with the critical regime then requires $q_\perp \sim \text{max}(q_\parallel^2,q_\parallel^z) \ll q_\parallel$ and therefore the momentum of the critical fluctuations is nearly tangent to the Fermi surface. This condition is familiar from the study of quantum critical points or phases involving the interaction of the Fermi surface with a bosonic field carrying zero wavevector \cite{joe, AIM}. The difference in the present case is that the operator $\ocal$ is a composite of the `elementary' critical excitations. We note that a condition $z<3$ is necessary to ignore terms such as $q_\perp q_\parallel^2$ and $q_\parallel^3$ in the fermion dispersion. Thus the results that follow are valid for $1 < z < 3$.

The upshot of the previous paragraph is that in the regime when the boson is `on shell' we can ignore the dependence of $C$ on $q_\perp$. It is then easy to perform the integral over $q_\perp$ in (\ref{eq:selfone}) to give
\bea
\Sigma(\w,\vec p) & = & \frac{i \pi \lambda_1^2}{v^\star} \int \frac{dq_\tau dq_\parallel}{(2\pi)^3} \text{sgn}(q_\tau-\w) C(q_\tau,q_\parallel) \\
 & = & - \frac{i \lambda_1^2 \text{sgn}(\w)}{(2\pi)^2 v^\star} \int_0^{|\w|} dq_\tau \int_{-\infty}^\infty dq_\parallel C(q_\tau,q_\parallel) \,. \label{eq:stepback}
\eea
To obtain the second line we used the fact that $C$ is an even function of $q_\tau$. The momentum dependence of the self energy has disappeared. Now using the scaling form (\ref{eq:scaling}) and changing variables to $q_\tau = |\w| x$ and $q_\parallel = |\w|^{1/z} y$ gives
\be\label{eq:dimensionless}
\Sigma(\w,\vec p) = - \frac{i \lambda_1^2 \text{sgn}(\w)}{2\pi^2 v^\star} |\w|^{(2 \Delta-1)/z} \int_0^1 \frac{dx}{x^{(z+2-2\Delta)/z}} \int_0^{\Lambda/|\w|^{1/z}} dy \widetilde C\left(\frac{y}{x^{1/z}} \right) \,.
\ee
As in section \ref{sec:hotspot} above, we introduced a cutoff on the momentum of the critical fluctuations along the Fermi surface. It is clear from the scaled integral (\ref{eq:dimensionless}) that the non-cutoff-sensitive contribution to the self energy scales like
\be\label{eq:sigmascaling}
\Sigma(\w) = - i \frac{\lambda_1^2 c}{v^\star} \text{sgn}(\w) |\w|^\kk \,,
\ee
with exponent
\be\label{eq:kap}
\kk = \frac{2 \Delta -1}{z} \,.
\ee
In (\ref{eq:sigmascaling}), $c$ is a real number. As a quick check of this expression we can note that substituting $z=3$ and $\Delta = 3/2$, as appropriate for the Hertz theory of a nematic transition with $\ocal(x)$ being the nematic order parameter, we recover the expected $\kk = 2/3$. In general, if the exponent $\kk > 1$, the fermionic excitations in the cold regions of the Fermi surface remain well-defined quasiparticles. However, the lifetime of these quasiparticles is generically different from the Fermi liquid $\G \sim \w^2$ form.

Returning to the example in our theory of $\ocal = \phi^2$, we can write in terms of the critical exponent (\ref{eq:nu})
\be
\kk = 2 + \frac{3}{z} - \frac{2}{\nu z}. \label{eq:kappanuz}
\ee
In Hertz-Moriya-Millis theory this implies $\kk = \frac{3}{2}$.
Therefore at this level the cold fermions acquire a non-Fermi liquid dissipation, albeit surviving as well-defined excitations. The fact that the low energy behavior of the propagator is not disrupted indicates that the effective $\lambda_1$ coupling is irrelevant in this case. We will give a more sophisticated characterisation of the relevancy or irrelevancy of our $\lambda_i$ couplings in section \ref{sec:condfluc}.

\begin{figure}[h]
\begin{center}
\includegraphics[height=90pt]{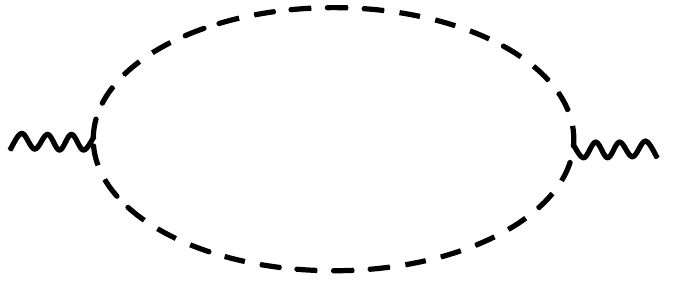}
\caption{The one loop contribution to the correlation function of the $\ocal = \phi^2$ operator.}\label{fig:phi21loop}
\end{center}
\end{figure}

We can confirm the scaling (\ref{eq:sigmascaling}) with $\kk= \frac{3}{2}$ via an explicit calculation. Indeed, at one loop level, the correlation function of the $\phi^2$ operator is given by Fig.~\ref{fig:phi21loop}.
Using the RPA propagator (\ref{eq:bosonD}), we obtain (see Appendix \ref{sec:phi4}), 
\beq C_{\phi^2}(\omega, \vec{q}) = \frac{3}{2\pi^2 N^2 \gamma} \left(\log \frac{\Lambda^2}{\gamma |\omega| + \vec{q}^2} + X(\w,\vec q) \right) \,,\label{Cphi20}\eeq
where
\be
X(\w,\vec q) = \frac{\gamma |\omega|}{\vec{q}^2}\log \left(\frac{\gamma |\omega|}{\gamma |\omega| + \vec{q}^2}\right) - \frac{1}{2} \text{Li}_2\left(\frac{\vec{q}^2}{2\gamma |\omega| + \vec{q}^2}\right)+ \frac{1}{2} \text{Li}_2\left(\frac{-\vec{q}^2}{2\gamma |\omega| + \vec{q}^2}\right) \,.
\ee
Upon substituting Eq.~(\ref{Cphi20}) into Eq.~(\ref{eq:stepback}), we recover the expected scaling (\ref{eq:sigmascaling}) with $\kk= \frac{3}{2}$.

We have also performed a complete analysis of the correlation function $C_{\phi^2}$ of the $\phi^2$ operator within the Hertz-Moriya-Millis theory.  In Hertz theory, $C_{\phi^2}$ does not strictly satisfy the scaling form (\ref{eq:scaling}) due to the presence of a marginally irrelevant perturbation and the fact that the scaling dimension of the correlator $C_{\phi^2}$ is $2\Delta - z - 2 = 0$ in this case. A detailed calculation (see Appendix \ref{sec:phi4}) gives,
\be C_{\phi^2}(\omega, \vec{q}) \propto \left(\log \frac{\Lambda}{|\vec{q}|}\right)^{1/11} \bigg[1 + \frac{1}{22 \log \Lambda/|\vec{q}|}  \Big(- \log\left(1+\frac{\gamma|\omega|}{\vec{q}^2}\right)
+ X(\w,\vec q) + \text{const}\Big)\bigg] + \text{const} \,,
\ee
and from (\ref{eq:stepback}) we obtain,
\beq \Sigma(\w) \sim  i \text{sgn}(\w) |\w|^{3/2} \left(\log\frac{\Lambda}{\omega}\right)^{-10/11} \,.
\eeq

We remind the reader that the spin density wave transition in two dimensions is not described by Hertz-Moriya-Millis theory. As shown in Ref. \onlinecite{Metlitski:2010vm}, the critical theory describing this transition is, in fact, strongly coupled and the exponents $z$ and $\nu$ receive corrections from mean field values. Thus, the discussion of the correlator in Hertz theory is given above for illustrative purposes only, and in a full theory we expect to obtain the scaling form (\ref{eq:sigmascaling}) for the self energy with the value of $\kappa$ (\ref{eq:kappanuz}) renormalized compared to the Hertz result of $\kk = \frac{3}{2}$.

We must also worry about the cutoff-dependent contribution to the integral (\ref{eq:dimensionless}). This regime, $q_\parallel \sim \Lambda \gg q_\tau^{1/z}$, is most easily accessed by taking a step back to equation (\ref{eq:stepback}). At leading order we can ignore the frequency dependence of $C(q_\tau,q_\parallel)$. The contribution thus comes from exchanging static critical fluctuations with a large momentum and gives
\be\label{eq:renor}
\Sigma(\w) = - \frac{i \lambda_1^2 \, \w}{2 \pi^2 v^\star} \int_0^\Lambda dq_\parallel C(0,q_\parallel) \,.
\ee
This contribution simply renormalises the quasiparticle residue and Fermi velocity. We may also ask about the higher order frequency dependence of the self energy coming from the `UV' part of the integral (\ref{eq:stepback}). In principle, the function $C(q_\tau,q_\parallel)$ need not have an analytic expansion in powers of $q_\tau$ for $q_\tau \ll |\vec q|^z$. Hence, one may obtain non trivial power law contributions to the quasiparticle width from this regime. For the Hertz-Moriya-Millis theory, the first subleading correction, as found in Appendix \ref{sec:phi4}, behaves as $q_\tau^2 \log |q_\tau|$, leading to $\Sigma(\w) \sim \w^3 \log \w$, which is smaller than the contribution (\ref{eq:sigmascaling}) from the $q_\tau \sim |\vec q|^z$ regime.


The main result of this subsection is the scaling form (\ref{eq:sigmascaling}) for the self energy of cold electrons scattering off neutral and zero momentum quantum critical modes. In general the exponent $\kk$ is given by (\ref{eq:kap}).

\subsection{Self energy due to $2 k_F$ scattering}
\label{sec:2kfself}

In Section~\ref{sec:noncancellation} we found that neutral hot operators carrying a net $2 k_F$ momentum could scatter cold fermions without the self energy correction to the conductivity being cancelled by a vertex correction. In this subsection, we consider the self energy due to such a scattering process. Despite the absence of a cancellation, there are various obstacles potentially preventing interesting consequences of this scattering.

Firstly, as already discussed in Section~\ref{sec:noncancellation}, this process cannot efficiently scatter most of the cold fermions due to the Fermi surface being non-nested. Only lukewarm fermions in the energy range $k_\parallel^{z+b} \ll \omega \ll k_\parallel^{z}$ are strongly affected by the scattering, with the exponent $b$ determined by the RG flow (\ref{eq:kappaflow}) of the curvature perturbation (\ref{eq:curvexp}). At tree level, $b = 1$. 

Secondly, as we mentioned above, the simplest critical operator to which these considerations apply is $\ocal = \psi^{1\dagger}_2 \psi^3_2$. At tree level, this operator has dimension $\Delta = 3$. Thus, with $z=2$, 
the `singular' self energy scales with a power $\Sigma \sim \w^{5/2}$ in (\ref{eq:sigmascaling}) which is even weaker than that due to the usual Fermi liquid interaction. Therefore, unless there are important renormalizations of the operator dimension and dynamical critical exponent, the hot $2k_F$ operator does not seem to significantly influence either the self energy or conductivity.

Nevertheless, it instructive to consider scattering in the $2k_F$ channel in the theory (\ref{eq:theaction}). As per the discussion above we will find that this process is dominated by scattering off lukewarm rather than hot fermions. Thus, the formalism in Sec.~\ref{sec:gen} will not be directly appicable in this case. Even so, the interplay of the singular SDW interaction and the Fermi surface curvature will give rise to a number of new physical effects. We proceed to discuss two such effects, the importance of effectively one-dimensional scattering and modification of the leading low energy scaling of certain observables due to curvature.

\subsubsection{One-dimensional scattering}

\begin{figure}[h]
\begin{center}
\includegraphics[height=260pt]{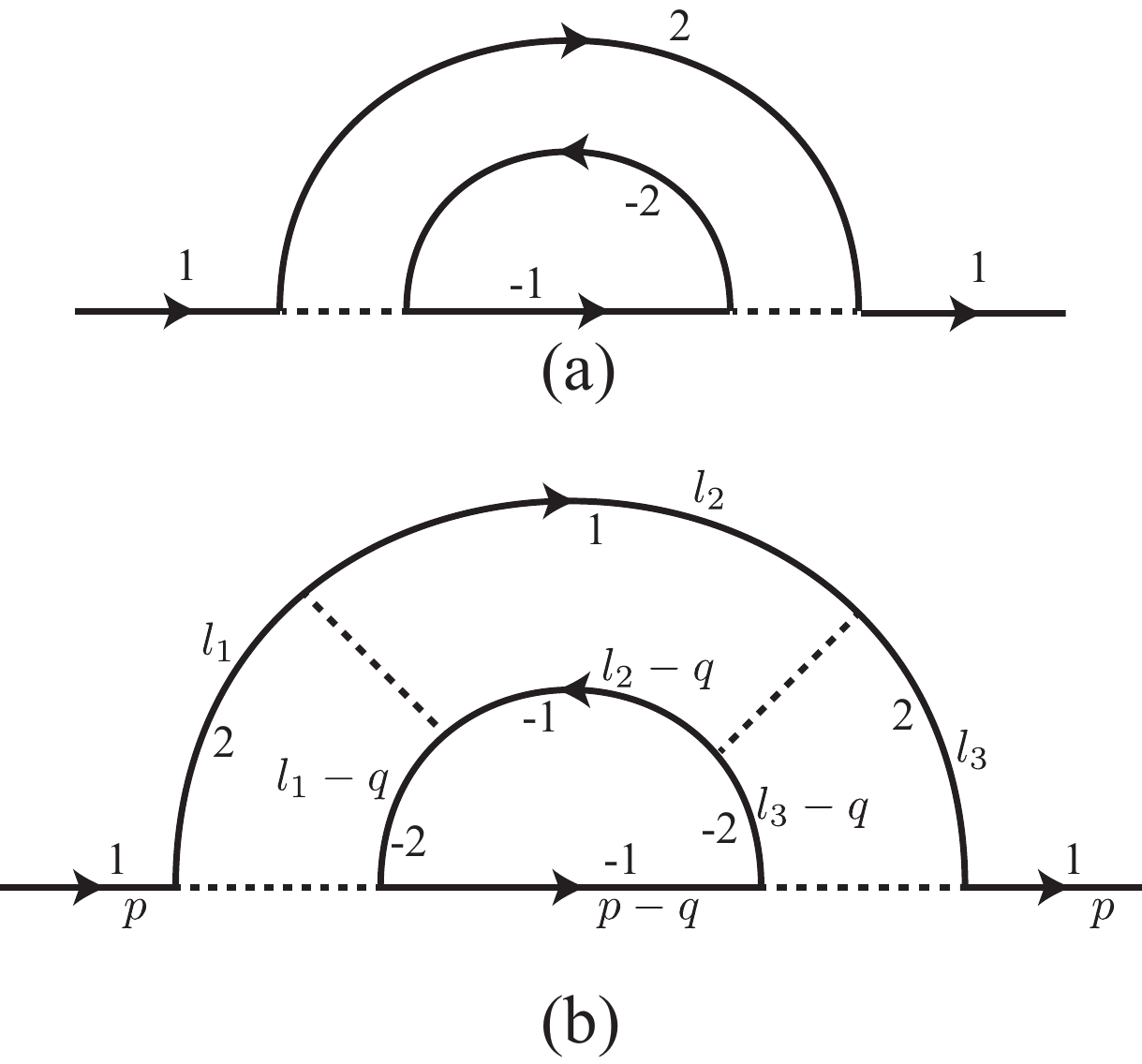}
\caption{Fermion self energy in the $2k_F$ channel. The scattering processes are those shown in Fig.~\ref{fig:umklapp};
The $1$ fermion refers to $\psi_1^1$, the $2$ to $\psi_2^1$, the $-1$ to $\psi_1^3$ and the $-2$ to $\psi_2^3$.  These fermions
are also referred to as $\psi_1$, $\psi_{\bar{1}}$, $\psi_{-1}$, and $\psi_{-\bar{1}}$ respectively.  The momentum labels in (b) correspond
to the expression in Eq.~(\ref{eq:2kFint}).
}\label{fig:2kFself}
\end{center}
\end{figure}
The two lowest order self energy diagrams in the $2k_F$ channel are shown in Fig.~\ref{fig:2kFself}. The graph in Fig.~\ref{fig:2kFself}a) is already taken into account in the rainbow approximation of Section~\ref{sec:hotspot} and gives rise to an $\omega^2$ Fermi liquid like self energy for lukewarm fermions. We now turn our attention to the graph in Fig.~\ref{fig:2kFself}b). We will consider the contribution to this graph from the region where all external and internal fermions are lukewarm. Thus, as a starting point we use the rainbow fermion propagators in the lukewarm region,
\beq G_a(\omega, \vec{p}) = \frac{Z(\hat{v}_{\bar{a}} \cdot \vec{p})}{i \omega - v^*(\hat{v}_{\bar{a}}\cdot \vec{p}) \hat{v}_a \cdot \vec{p}} \label{eq:Gcold2kF}\eeq
where $Z(l) = l/\Lambda$, $v^*(l) = |\vec{v}|l/\Lambda$ and $\Lambda$ is given by Eq.~(\ref{Lambdadef}). As shown in appendix \ref{sec:oned}, the resulting fermion self energy is
\beq \Sigma(\omega, \vec{k}) \sim -\frac{i}{N} \frac{\omega}{k_\parallel} \log\left(\frac{k^2_\parallel}{\gamma |\omega|}\right)\label{eq:Sigma1D}\eeq
This self energy is of non-Fermi liquid form and, in fact, implies that the fermionic quasiparticles are not well-defined even in the lukewarm region $\omega \ll k^z_\parallel$.  It is interesting that Eq.~(\ref{eq:Sigma1D}) has the `marginal Fermi liquid' form,\cite{varma} although it must be kept
in mind that we have not determined whether this logarithm survives the resummation of higher order terms.

\begin{figure}[h]
\begin{center}
\includegraphics[height=140pt]{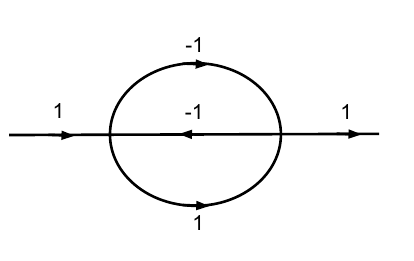}
\caption{Fermion self energy in the $2k_F$ channel. The $2$ and $-2$ fermions in Fig.~\ref{fig:2kFself}b) are assumed to be off shell and are integrated out. The resulting process is effectively one-dimensional. The interaction between the $1$ and $-1$ fermions
has a singular ($\sim 1/k_\parallel^2$) dependence upon their distance ($k_\parallel$) from the hot spot.}\label{fig:2kF1D}
\end{center}
\end{figure}

The origin of the infra-red divergence in Eq.~(\ref{eq:Sigma1D}) lies in the fact that at the level of the propagator (\ref{eq:Gcold2kF}) the Fermi surface is flat. Indeed, the self energy (\ref{eq:Sigma1D}) comes from the region where the $2$ and $-2$ fermions are off shell and can be integrated out to give an effective interaction between the $1$ and $-1$ fermions, which has a singular dependence on the momenta along the Fermi surface of the latter. The resulting process, Fig.~\ref{fig:2kF1D}, is effectively one-dimensional - hence the logarithmic divergence familiar from one-dimensional physics. The prefactor of $1/k_\parallel$ in Eq.~(\ref{eq:Sigma1D}) comes from the $k_\parallel$ dependence of the effective interaction
between the one-dimensional excitations.

Clearly, the divergence in Eq.~(\ref{eq:Sigma1D}) will be cut off by the Fermi surface curvature. One source of curvature is the explicit perturbation $\kappa_0$ in the Lagrangian, Eq.~(\ref{eq:curvexp}). This perturbation is expected to give rise to a physical Fermi surface curvature $\kappa(k_\parallel) \sim \kappa_0 k^{b-1}_\parallel$ (here we are using the fact that the `natural' scale of curvature $\kappa \sim k^{-1}_\parallel$). Still assuming the quasiparticle form in Eq.~(\ref{eq:lukewarm}), following the argument in section \ref{sec:noncancellation}, the divergence (\ref{eq:Sigma1D}) will be cut off below $\omega_\kappa(k_\parallel) \sim \kappa_0 k^{z+b}_\parallel$. However, even if $\kappa_0 = 0$ the Fermi surface of the theory (\ref{eq:theaction}) will dynamically develop a curvature. Indeed, we expect the Fermi surface shape to be controlled by the RG flow of the variable $\alpha$, related to the angle between the Fermi surfaces of fermions $1$ and $2$. If this flow has a stable fixed point at $\alpha = \alpha_*$,
\beq \frac{d \alpha}{d \ell} = -b' (\alpha - \alpha^*) \label{eq:alphaflow}\eeq
then the physical Fermi surface curvature $\kappa(k_\parallel) \sim (\alpha - \alpha_*) k^{b'-1}_\parallel$ and we expect the one-dimensional divergence in Eq.~(\ref{eq:Sigma1D}) to be cut off at $\omega_\kappa(k_\parallel) \sim (\alpha - \alpha_*) k^{z + b'}_\parallel$. Hence, the curvature physics will be dominated by the smaller of the two correction-to-scaling exponents $b$ and $b'$. Note that if $b_\kappa = \min (b,b') < 1$ the physical Fermi surface curvature diverges at the hot spot.

In Refs.~\onlinecite{advances,Metlitski:2010vm} the one-loop RG flow of $\alpha$ was determined to be,
\beq \frac{d \alpha}{d \ell} =  - \frac{3}{\pi N} \frac{\alpha^2}{1+\alpha^2}\label{eq:alphaoneloop}\eeq
which has a fixed point at $\alpha^* = 0$. The flow towards the fixed-point is logarithmic, resulting in the Fermi surface shape,
\beq k_y = \frac{3}{\pi N} k_x \log(\Lambda/|k_x|)\eeq
whose curvature $\kappa(k_y) \sim |k_y|^{-1} \log^{-2}(\Lambda/|k_y|)$. Hence, in this case we may write $b' \to 0^+$. However, it was found in Ref.~\onlinecite{Metlitski:2010vm} that all anomalous dimensions diverge in the limit $\alpha \to 0$ so we are not in a position to perform a careful analysis of the physics at the $\alpha^* = 0$ fixed point. Since the $1/N$ expansion in Refs.~\onlinecite{advances,Metlitski:2010vm} is uncontrolled, it is quite possible that higher order calculations give a flow of the more conventional form (\ref{eq:alphaflow}) with $b' \neq 0$ and $\alpha^* \neq 0$.

We conclude that for lukewarm fermions there exists a window $k^{b_\kappa + z}_\parallel \ll \omega \ll k^z_\parallel$ in which the Fermi surfaces can be regarded as straight and effectively one-dimensional divergences of the form (\ref{eq:Sigma1D}) appear.\footnote{If the one-loop flow of $\alpha$, Eq.~(\ref{eq:alphaoneloop}), is assumed, this window is only logarithmically wide.} In a truly one-dimensional system such logarithmic divergences sum up to a power-law giving rise to Luttinger-liquid physics. A more detailed analysis is needed to conclude whether an analogous phenomenon takes place in our system or if the divergences of the type (\ref{eq:Sigma1D}) lead to an instability. In either case, the Green's function will not have the quasiparticle form (\ref{eq:lukewarm}) in the lukewarm region $\omega \ll k^{z}_\parallel$. Well-defined quasiparticles may reemerge at the lowest energies where the Fermi surface curvature plays a role, however, the behavior of the quasiparticle residue and Fermi velocity will differ from Eq.~(\ref{eq:vZ}). Also, our estimate of the lower boundary of the intermediate energy window $\omega_\kappa \sim k^{b_\kappa + z}_\parallel$ was based on the assumption that fermionic quasiparticles are well defined in this window. As argued above this assumption is likely incorrect and a detailed understanding of the physics in this window is needed in order to estimate the energy scale at which curvature effects become important.

In Section~\ref{sec:cond2kf} below, we will discuss the contribution of these one-dimensional scatterings to the optical conductivity. Before doing this, we note that in addition to the `one-dimensional' contribution to the self energy in Fig.~\ref{fig:2kFself} b), there is also a `two-dimensional' contribution from the region where both the intermediate $1$, $-1$ and $2$,$-2$ are on shell. This contribution is not sensitive to whether the Fermi surface is curved, but is sensitive to whether it is nested. We will discuss this contribution in the remainder of the present section.


\subsubsection{Low energy effects of Fermi surface curvature}

As noted in section \ref{sec:noncancellation}, the explicit Fermi surface curvature perturbation, (\ref{eq:curvexp}), generates a new energy scale $\omega \sim \kappa_0 k^{z+b}_\parallel$. At this scale the Fermi surface realizes that it is not nested with respect to the $2 k_F$ wavevector and the BCS-like divergences in the particle-hole channel are cut off. Note that if the exponents $b$, $b'$ in Eqs.~(\ref{eq:kappaflow}), (\ref{eq:alphaflow}) satisfy $b < b'$, then this is also the energy scale at which the Fermi surface can no longer be treated as flat. Furthermore, it was enhancements at this scale that lead to the $2k_F$ instability in Refs.~\onlinecite{Metlitski:2010vm, Metlitski:2010zh}. In the remainder of this section we will look at the physics of $2 k_F$ scattering in the regime where this curvature scale is being saturated. We will find that, while formally irrelevant as per the discussions of section \ref{sec:noncancellation}, the inclusion of curvature perturbation (\ref{eq:curvexp}) leads to low energy scalings that are stronger than those arising from na\"ive dimensional analysis in the hot spot theory.

\begin{figure}[h]
\begin{center}
\includegraphics[height=67pt]{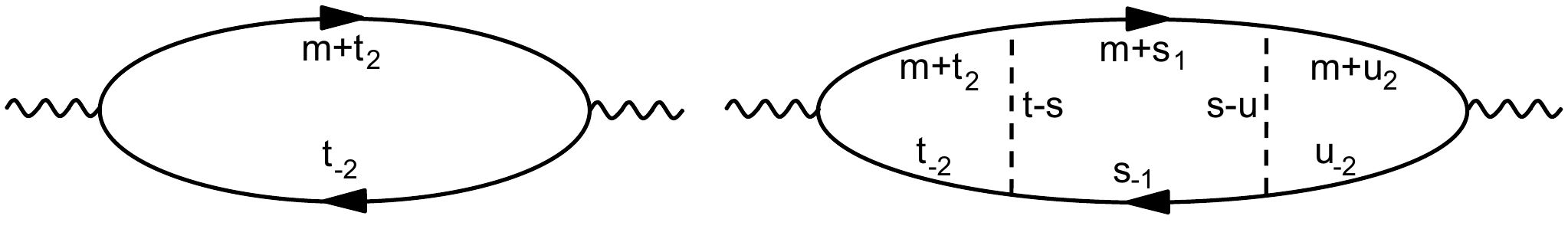}
\caption{Tree level (left) and leading correction (right) to the propagator for the operator $\psi^{1\dagger}_2 \psi^3_2$. The subscripts show the hot spot involved. $1$ and $-1$ correspond to the $a=1$ hot spots of the $\ell=1$ and $\ell = 3$ pair of hot spots, respectively. Similarly, 
$2$ and $-2$ correspond to the $a=2$ hot spots of the $\ell=1$ and $\ell = 3$ pair. \label{fig:prop}}
\end{center}
\end{figure}

The basic object illustrating how curvature generates new energy scalings in the $2k_F$ channel is the propagator for the CDW operator $\psi^{1\dagger}_2 \psi^3_2$. The tree level and leading corrected propagator for this density channel mode are shown in figure \ref{fig:prop}.
Two vertex corrections are necessary because there is mixing between $\psi^{1\dagger}_2 \psi^3_2$ and $\psi^{1\dagger}_1 \psi^3_1$. These graphs are entirely analogous to those of figure \ref{fig:2kFself} above. Here we are considering the density channel propagator as a quantity in its own right. 

As in the previous subsection we can use the lukewarm inverse propagators following from (\ref{eq:fermionG}) and (\ref{eq:lukewarm}). As we are now interested in saturating the scale at which curvature terms violate Fermi surface nesting, we will include such a curvature explicitly in the fermion propagators
\be\label{eq:lukekappa0}
G^{-1}_{\pm 2}(q) = \mp v q_\perp + \frac{3 \g v}{16 N} \frac{i q_\tau}{|q_\parallel|} - \kk q_\parallel^2 \,.
\ee
In this equation we are defining parallel and perpendicular with respect to the $\ell=1,a=2$ Fermi surface.
We are being somewhat simpleminded here in just adding the irrelevant curvature coupling without systematically checking how this term backreacts on our previous results. The results of this subsection should be taken as a preliminary investigation into new scaling effects arising from the interplay of hot spot criticality with Fermi surface curvature.

Much of the interesting physics is already visible in the leading tree level term, left in Fig.~\ref{fig:prop}. This diagram is evaluated in Appendix \ref{sec:2kfprop}. The difference relative to previous computations in this paper is that the curvature in (\ref{eq:lukekappa0}) leads to a new low energy contribution due to fermions saturating the curvature scale. We focus on this contribution, which in fact gives the most singular terms at low energy and momenta
\be\label{eq:czero0}
C^{(0)}(m) = - \frac{c_1}{2} \frac{(\g |m_\tau|)^{2/3}}{(c_2 \kk/v)^{2/3}} 
\sum_{i} \frac{3+2 r^2 x_i + 4 r x_i^2}{r^2 + 8 r x_i + 12 x_i^2} \log \frac{- x_i \kk^2 (\g |m_\tau|)^{2/3}}{(c_2 \kk /v)^{2/3}} \,.
\ee
Here $r$ is the ratio of momentum and energy
\be\label{eq:mratio0}
r = \frac{m_{\perp,2}}{(\g |m_\tau|)^{2/3}} \frac{c_2^{2/3}}{(\kk / v)^{1/3}} \,,
\ee
and the $x_i$ are the three roots of the polynomial
\be
4 x^3 + 4 r x^2 + r^2 x + 1 = 0 \,.
\ee
The notation $m_{\perp,2}$ refers to the component of $m$ perpendicular to the $\ell=1,a=2$ Fermi surface.
The constants
\be
c_1 = \frac{4 N}{3 \pi^2 \g v^2} \,, \qquad c_2 = \frac{16 N}{3} \,.
\ee

The physical point here is that the scaling $m_\tau^{2/3}$ of the result (\ref{eq:czero0}) is both stronger than the dimensional analysis scaling for this dimension $\Delta = 3$ operator, which from (\ref{eq:scaling}) would be $C(m) \sim m_\tau$, and also weaker than the full BCS Fermi surface singularity $C(m) \sim \Lambda_\parallel^2 \log m_\tau$. This intermediate result is due to the interplay of Fermi surface curvature in the particle channel with the criticality of the lukewarm fermions. In particular the overall scaling power $m_\tau^{2/3}$ arises from the dimensionality of the operator together with the fact that, from (\ref{eq:lukekappa0}), the curvature kicks in at the scale
\be\label{eq:saturate}
\frac{v \g}{N} m_\tau \; \sim \; v m_\perp \, m_\parallel \; \sim \; \kk \, m_\parallel^3 \,.
\ee

This tree level term does not capture, however, the physics of the enhancement of the particle-hole channel that leads to a pairing instability\cite{Metlitski:2010vm, Metlitski:2010zh}. These effects are contained in the vertex corrections of the right hand diagram in figure \ref{fig:prop}. This diagram is also studied in Appendix \ref{sec:2kfprop}. The most singular low energy terms, at leading order in singular logarithms, are found to be
\bea\label{eq:cmfinaltext}
C^{(1)}(m, \Lambda_\parallel) & = & \frac{c_1^3}{32} \frac{(\g |m_\tau|)^{2/3}}{(c_2 \kk/v)^{2/3}} \log^2 \tilde \Lambda_\parallel^2  \log^2 \tilde \kk^{8/3} \tilde \Lambda_\parallel^2 \\
&& \times  \sum_i  \left(\log (-x_i) x_i - \log(-y_i) \frac{3 + 2 \bar r^2 y_i + 4 \bar r y_i^2}{\bar r^2 + 8 \bar r y_i + 12 y_i^2}  \right)
 \,. \nonumber
\eea
Here we introduced the analogous ratio to (\ref{eq:mratio0}) for the $\ell=1,a=1$ Fermi surface
\be\label{eq:rratiotext}
\bar r = \frac{m_{\perp,1}}{(\g |m_\tau|)^{2/3}} \frac{c_2^{2/3}}{(\kk/v)^{1/3}} \,,
\ee
and the $y_i$ are now the three roots of the polynomial
\be\label{eq:xpoltext}
4 y^3 + 4 \bar r y^2 + \bar r^2 y + 1 = 0 \,.
\ee
Finally, we set
\be
\tilde \Lambda_\parallel^2 =  \frac{(c_2 \kk/v)^{2/3}}{(\g |m_\tau|)^{2/3}} \Lambda_\parallel^2 \,, \qquad
\tilde \kk^2 = \g |m_\tau| \left(\frac{c_2 \kk}{v}\right)^2 \,.
\ee
We must keep track of the dependence on the momentum cutoff $\Lambda_\parallel$. The above result describes the leading order singular behaviour in $m_\tau$ with the ratios $r, \bar r$ held fixed. As with the $2k_F$ vertex\cite{Metlitski:2010vm}, the bosonic vertex corrections lead to singular logarithmic enhancements upon saturating the curvature scale (\ref{eq:saturate}).
Here there are surviving overall powers of momentum that will then be evaluated at this scale, leading to the overall power of $m_\tau^{2/3}$. Note that (\ref{eq:cmfinaltext}) depends upon both momenta $m_{\perp,1}$ and $m_{\perp,2}$.

To summarize: the $2k_F$ propagator (\ref{eq:cmfinaltext}) contains two physical effects. The first is the $m_\tau^{2/3}$ (rather than $m_\tau$) scaling, due the combination of hot spot scaling and saturation of the curvature scale, and the second is the presence of logarithmic enhancements due to both cut off BCS divergences and also vertex corrections.

We can now turn to the self energy (\ref{eq:selfone}) due to scattering off the enhanced $2 k_F$ mode. As discussed above, curvature will restrict efficient scattering off this mode to be near the hot spot, so we take the intermediate fermion to be lukewarm rather than cold. Thus 
\be\label{eq:selfq}
\Sigma(q) = - \frac{\lambda^2_2}{\kk^{2/3}} \int \frac{d^3m}{(2\pi)^3} \frac{|m_\tau|^{2/3} \widetilde C\Big(\frac{m_{\perp,1}}{\kk^{1/3} m_\tau^{2/3}}, \frac{m_{\parallel,1}}{\kk^{1/3} m_\tau^{2/3}}\Big)}
{\frac{\g v}{c_2} \frac{i (q_\tau - m_\tau)}{|q_\parallel - m_{\parallel,1}|} + v(q_\perp - m_{\perp,1}) - \kk (q_\parallel - m_{\parallel,1})^2} \,.
\ee
We must be careful here to keep track of the different components of $m$ relative to the $\pm 1$ and $\pm 2$ Fermi surfaces. The scaling of $\widetilde C$ suggested in (\ref{eq:selfq}) is violated by logarithmic terms in (\ref{eq:czero0}) and (\ref{eq:cmfinaltext}). The above expression for the self energy is rather similar to the one we encountered previously in (\ref{eq:scaling}) and (\ref{eq:selfone}) with an effective $z=3/2$ induced by Fermi surface curvature effects. The crucial difference however is that the fermion propagator in (\ref{eq:selfq}) is on the $-1$ Fermi surface, while it is contributing to the self energy of a fermion on the $+1$ Fermi surface. In particular the two fermions have opposite Fermi velocities. If we were to put both of the fermions precisely on their Fermi surfaces (i.e. $q_\perp = \pm q_\parallel^2$, respectively) the missing momentum gets sent through the critical mode and moves its momentum away from $2 k_F$, thereby spoiling the enhancement we are looking for. Instead we can expect that the most efficient scattering will again occur upon saturating the curvature, i.e. when the external fermion has
\be\label{eq:qscaling}
\frac{q_\tau}{q_\parallel} \sim q_\perp \sim \kk q_\parallel^2 \,. 
\ee

Inspecting (\ref{eq:selfq}) we can now ask what scaling regime of the $m$ energy and momenta will dominate the integral. The immediate choices are either to respect the scaling of the critical mode $\widetilde C$ for both components of the momentum, or to make the denominator have an overall scaling. The latter possibility in fact takes us back to the one-dimensionality of the previous subsection, so here we consider the former case. Thus we take
\be\label{eq:mmscaling}
m_\tau = q_\tau x \,, \qquad m_{\perp,1} = q_\tau^{2/3} \kk^{1/3} y \qquad m_{\parallel,1} = q_\tau^{2/3} \kk^{1/3} z \,.
\ee
This implies that in the regime (\ref{eq:qscaling}) $m_{\parallel,1} \ll q_\parallel$.
We thereby obtain (taking $q_\tau>0$ for simplicity)
\be\label{eq:sqfinal}
\Sigma(q) = - \frac{\lambda^2_2}{\kk^{1/3}} q_\tau^{2+1/2} \int \frac{dxdydz}{(2\pi)^3} \frac{|x|^{2/3} \widetilde C\Big(\frac{y}{x^{2/3}}, \frac{z}{x^{2/3}} \Big)}
{\frac{\g v}{c_2} i (1 - x)  \frac{q_\tau^{1/3}}{\kk^{1/3} |q_\parallel|} + v\left(\frac{q_\perp}{\kk^{1/3} q_\tau^{2/3}} - y\right) - \frac{\kk^{2/3} q_\parallel^2}{q_\tau^{2/3}}} \,.
\ee
The denominator here is order one within the scaling regime and for external momenta satisfying (\ref{eq:qscaling}).

To make sense of (\ref{eq:sqfinal}) we should recall that in our lukewarm regime the coupling $\lambda_2 \sim 1/q_\parallel^2$. This is because we collapsed a boson propagator (\ref{eq:bosonD}) in generating the interaction (\ref{eq:uneff}). This collapsing involved taking $q_\parallel$ to be larger than the momenta running in the various loops that we computed to obtain (\ref{eq:cmfinaltext}). Thus the cutoff in (\ref{eq:cmfinaltext}) is $\Lambda_\parallel \sim q_\parallel$. Combined with the scaling regime we are considering, given by (\ref{eq:qscaling}) and (\ref{eq:mmscaling}), we can see that this cutoff dependence on $q_\parallel$ will remove two of the four powers of $\log q_{\tau}$ that are violating the scaling of $\widetilde C$. Thus in fact the logarithmic enhancements in the propagator due to vertex corrections (as opposed to BCS logs) do not contribute to this observable.
Putting these facts together we can write schematically
\bea
\Sigma(\w, \vec q) & \sim & - i \text{sgn}(\w) \frac{1}{\kk^{1/3}} \frac{|\w|^{2+1/3}}{q_\parallel^4} \log^2 |\w| \, F\left(\frac{q_\parallel}{(\omega/\kk)^{1/3}}, \frac{q_\perp}{(\kk\, \omega^2)^{1/3}}\right) \\
& \sim & - i \text{sgn}(\w) \frac{|\w|^{2}}{q_\parallel^3} \log^2 |\w| \, \widetilde F\left(\frac{q_\parallel}{(\omega/\kk)^{1/3}}, \frac{q_\perp}{(\kk\, \omega^2)^{1/3}}\right) \,. \label{eq:log4}
\eea
In the second line we have rescaled away the explicitly singular dependence on the curvature $\kk$ using $q_\parallel \sim (\omega/\kk)^{1/3}$. This modifies the overall function: $F \to \widetilde F$. We are restricting ourselves to the regime (\ref{eq:qscaling}) where the overall functions $F$ and $\widetilde F$ are of order $1$.

The self energy (\ref{eq:log4}) is logarithmically stronger than the result of the rainbow approximation Eq.~(\ref{eq:lukewarmdamp}). It is also stronger than the na\"ive hot spot scaling $\Sigma \sim \w^{5/2}/q^4_\parallel$ for scattering off the hot operator $\psi_2^{1\dagger} \psi_2^3$. Because of the instability in the $2 k_F$ channel\cite{Metlitski:2010vm, Metlitski:2010zh}, resumming these logarithms presumably leads to a singularity below a critical frequency.



\section{Conductivity from composite operators}
\label{sec:cond}

\subsection{Conductivity due to neutral zero momentum operators}

We are now in a position to compute the contribution to the conductivity of the various processes we have discussed.
Consider first the conductivity due to scattering off neutral and zero momentum operators. We have derived the formula (\ref{eq:leading}) as giving the leading low energy contribution to the current-current correlator in this case. As a warmup and for later use we can first compute the (ultimately cancelled) contribution due to self energy corrections (\ref{eq:equalopposite}). Using (\ref{eq:sigma}), this can be written in terms of the self energy as
\be\label{eq:condselfA}
\dd^{(a+b)} \Pi_{ij}^\text{cold}(\w) =  \frac{2}{\w^2} \int \frac{d^3q}{(2\pi)^3} v_i^\star v_j^\star 
G(q) \Big(\Sigma(p+q)+ \Sigma(-p+q) - 2 \Sigma(q) \Big) \,. 
\ee
Using the propagator for the cold fermions (\ref{eq:cold}) and the singular contribution to the self energy (\ref{eq:sigmascaling}), 
we first perform the integral over $q_\perp$, and then pick out the non-analytic term in the integral over $q_\tau$ to obtain
\be\label{eq:abG}
\dd^{(a+b)} \Pi_{ij}^\text{cold}(\w) = - \frac{4 \, c \, |\w|^{\kk-1}}{\kk + 1} \int \frac{d q_\parallel}{(2\pi)^2} \frac{\lambda_1^2\, v^\star_i v^\star_j}{v^{\star\,2}} \,.
\ee
Here $q_\parallel$ should be thought of as the integral along the different patches on the Fermi surface. Continuing to real time, it is then clear from the definition of the electrical conductivity in (\ref{eq:hotc}) that this contribution to the conductivity would have scaled as
\be\label{eq:wouldbe}
\dd^{(a+b)} \sigma(\Omega) \sim \lambda_1^2 \Lambda_\parallel \Omega^{\kk-2} \,.
\ee
For general $\kk$ this contribution contains both real and imaginary parts. Here $\Lambda_\parallel$ is a UV cutoff on the $q_\parallel$ integral in (\ref{eq:abG}). We can estimate its value as the Fermi momentum $\Lambda_\parallel \sim m^\star v^\star$.

As we saw above, the actual leading contribution to the conductivity is (\ref{eq:leading}) while the contribution we have just discussed gets cancelled. It is easy to see what the effect of the extra two insertions of momenta along the Fermi surface in (\ref{eq:leading}) will be without doing any additional integrals. The momenta appear as $q_\parallel^2$ in (\ref{eq:stepback}). Upon moving to dimensionless variables as in equation (\ref{eq:dimensionless}) and below, the effect of these two extra powers of momentum will be to shift the exponent $\kk \to \kk + 2/z$ in what was previously the self energy. Then, following the same manipulations as we have above, we can conclude that the leading nonzero conductivity due to cold fermions scattering off neutral zero momentum operators is
\be\label{eq:is}
\delta \sigma(\Omega) \sim \frac{\lambda_1^2}{v^\star  m^\star} \Omega^{\kk+2/z-2} \,.
\ee
Here we used the estimate $\Lambda_\parallel \sim m^\star v^\star$.
Returning once again to the simplest example $\ocal = \phi^2$, at the level of Hertz-Moriya-Millis theory, our previous observation that $\kk = \frac{3}{2}$ implies that
$\dd \sigma(\Omega) \sim \Omega^{1/2}$. This is a weak low frequency dependence, comparable to the hot contribution to the conductivity (\ref{sigmaf}). Presumably operators which are higher order polynomials in fields and derivatives have increasingly higher values of $\kappa$ and are thereby increasingly irrelevant. Scattering off the $\ocal = \phi^2$ operator becomes more significant if $z$ and $\Delta$ are renormalised such that $z$ becomes larger and $\Delta$ smaller than the Hertz-Moriya-Millis values. At present a controlled computation of $z$ and $\Delta$ in the theory is not available. Given $z$ and $\Delta$ our formulae for the conductivities are presumably reliable so long as our effective interactions do not destroy the cold fermion quasiparticles. This is potentially compatible with non-Fermi liquid behavior of optical conductivity.

A check of the formulae (\ref{eq:wouldbe}) and (\ref{eq:is}) comes from comparing with Ref.~\onlinecite{MIT}. The theory studied in that paper had $\kk = 2/3$ and $z=3$. It was found that the individual self energy and vertex corrections to the conductivity scaled as $\Omega^{-4/3}$, while the total conductivity scaled as $\Omega^{-2/3}$. These results are in agreement with our expressions.

Also following Ref.~\onlinecite{MIT} it is tempting to note that the leading scaling behavior (\ref{eq:is}) can be obtained somewhat na\"ively from a modified Drude formula. Introduce the scattering rate
\be
\frac{1}{\tau} \sim \text{Im} \Sigma(\Omega) \sim \Omega^{\kk} \,.
\ee
The transport scattering rate involves the usual extra factor of $1 - \cos\theta$, with $\theta$ the angle between the initial and final fermion wavevectors. At low momenta and with the boson on shell as above $1 - \cos\theta \sim p^2/k_F^2 \sim \Omega^{2/z}$. Thus the transport scattering rate is
\be\label{eq:transport}
\frac{1}{\tau_\text{tr.}} \sim \Omega^{\kk+2/z} \,.
\ee
Now write down a generalized Drude formula
\be\label{eq:Drude}
\text{Re} \, \sigma(\Omega) \sim \frac{\tt_\text{tr.}^{-1}}{\Omega^2 + \tt_\text{tr.}^{-2}} \,.
\ee
As long as $\kappa + 2/z > 1$, at low frequency we may drop $\tt_\text{tr.}^{-2}$ in the denominator of Eq.~(\ref{eq:Drude}). Using (\ref{eq:transport}) the leading answer is then
\be
\text{Re} \, \dd \sigma(\Omega) \sim \Omega^{\kk + 2/z - 2} \,,
\ee
in agreement with our previous (\ref{eq:is}).

It is further tempting, in the formulation of the previous paragraph, to take the $\Omega \to 0$ limit of the Drude formula (\ref{eq:Drude}) and further exchange the frequency dependence of the transport scattering rate (\ref{eq:transport}) for temperature dependence. Doing this one would obtain a resistivity $\rho \sim T^{\kk + 2/z}$. For the case $\ocal = \phi^2$ at a mean field level this translates into the weak resistivity $\rho \sim T^{5/2}$. 
This reasoning is however unlikely to be correct: it does not account for the 
key physics of how the delta function in the conductivity at $\Omega=0$ is resolved through the
umklapp scattering which is implicitly included in our theory (\ref{eq:theaction}), because pseudospin symmetry has played no role.

Before moving on, it is instructive to rewrite our result (\ref{eq:is}) as
\be\label{eq:sigmaabc}
\delta \sigma(\Omega) \sim \Omega^{(2 \Delta + 1)/z-2} \,.
\ee
This expression shows how a moderately large renormalised value of $z$ enhances the effect of scattering off composite hot bosonic modes. In particular at large values of $z$ the cancellation between self energy and vertex corrections becomes increasingly insignificant. This observation does not immediately appear to be related to the $z=\infty$ non Fermi liquids recently discussed holographically and otherwise \cite{MITagain, Sachdev:2010um}. We can note that $\Delta \sim z/2$ at large $z$ will give a linear in temperature resistivity according to the simple discussion of the previous paragraph. While the usual unitarity bound on operator dimensions at a fixed point requires a full conformal symmetry, at least in some holographic settings the dimensions of scalar operators (in two spatial dimensions) are bounded by  $\Delta \geq z/2$. This follows from adapting the finite action argument of Ref.~\onlinecite{Klebanov:1999tb} to the case\cite{Kachru:2008yh} of a general $z$. These holographic setups do not include the spatial anisotropy for the critical theory we are studying in this paper.

\subsection{Conductivity due to $2 k_F$ modes}
\label{sec:cond2kf}

\subsubsection{One-dimensional contribution}

In section \ref{sec:noncancellation} we observed that there was no cancellation between vertex and self energy corrections for scattering off hot $2 k_F$ CDW fluctuations. However, as discussed in section \ref{sec:2kfself} in our theory the self energy in the $2 k_F$ channel, Fig.~\ref{fig:2kFself}b), is actually not dominated by the hot contribution, so the formalism in section \ref{sec:noncancellation} is not directly applicable. Nevertheless, as a first estimate we will ignore vertex corrections in the computation of the conductivity here. Then, similarly to the previous subsection,
\beq \delta \Pi_{ij}(\omega) = \frac{1}{\omega^2} N v_i v_j \int \frac{d^3 k}{(2\pi)^3} Z^2(k_\parallel) G(k) (\Sigma(k+\omega) + \Sigma(k-\omega) - 2 \Sigma(k)) \,. \eeq
Using the self energy, Eq.~(\ref{eq:Sigma1D}), we compute the integral above as in Eq.~(\ref{eq:condselfA}) to 
estimate,\footnote{Strictly speaking Eq.~(\ref{eq:Sigma1D}) represents only the self energy on the Fermi surface. However, inclusion of the dependence on momentum perpendicular to the Fermi surface does not qualitatively modify the result (\ref{eq:Pi2kf}).}
\beq \delta \Pi(\omega) \sim \frac{1}{\gamma} \int {d k_\parallel} k_\parallel \log\frac{k^2_\parallel}{\gamma |\omega|} \,. \label{eq:Pi2kf}\eeq  
The integral over the momentum $k_\parallel$ along the Fermi surface in Eq.~(\ref{eq:Pi2kf}) will be cut-off by the Fermi surface curvature. As estimated in section \ref{sec:2kfself}, curvature effects set in once $\omega \lesssim k^{z + b_\kappa}_\parallel$. Hence, setting $z = 2$ and ignoring logarithmic corrections,
\beq \delta \sigma(\Omega) \sim (-i \Omega)^{{-b_\kappa}/({2 + b_\kappa})} \,. \label{eq:cond2kf}\eeq
Note that Eq.~(\ref{eq:cond2kf}) holds as $\Omega \rightarrow 0$ (at $T=0$); the energy scale set by the curvature, $\omega_\kappa (k_\parallel)
\sim k_\parallel^{b_\kappa + z} \rightarrow 0$ as $k_\parallel \rightarrow 0$, and so the one-dimensional contribution appears at the lowest
energy scales.

Clearly Eq.~(\ref{eq:cond2kf}) is sensitive to the exponent $b_\kappa$, which controls the physical Fermi surface curvature. As discussed in section \ref{sec:2kfself} our understanding of this exponent is rather limited. If one assumes the one-loop flow of $\alpha$, Eq.~(\ref{eq:alphaoneloop}), then $b_\kappa \to 0^+$ and the correction in Eq.~(\ref{eq:cond2kf}) is expected to be only logarithmic.

One should ask if vertex corrections strongly modify the result in Eq.~(\ref{eq:cond2kf}). This question is particularly pressing due to the one-dimensional nature of the self energy divergence (\ref{eq:Sigma1D}), which translates into the conductivity (\ref{eq:cond2kf}). In a purely one-dimensional system away from half-filling, at leading order in energy the electrical current would be proportional to the conserved electron momentum and so the finite frequency conductivity would vanish. Indeed, the simplified process in Fig.~\ref{fig:2kF1D} naively looks like a forward-scattering process, which cannot contribute to the conductivity. However, unlike in a purely one-dimensional system, the intermediate $1$ and $-1$ fermions in Fig.~\ref{fig:2kF1D} possess a coordinate along the Fermi surface. The physical `on shell' current vertex $Z(k_\parallel) \Gamma(\omega, k_\parallel)$ depends on this coordinate and hence we don't expect the cancellation between the self energy and vertex corrections to occur. Note that in the calculation leading to (\ref{eq:cond2kf}) we've used the bare current vertex $\Gamma_i(\omega, k_\parallel) = v_i$. In principle, a systematic analysis of the current vertex which takes into account both the one-dimensional divergences in Section~\ref{sec:2kfself} and the rainbow graphs in Section~\ref{sec:hotspot} is needed. Such an analysis is, however, outside the scope of this paper.

\subsubsection{`Enhanced critical umklapp'}

With similar caveats to the discussion immediately above, we can estimate the contribution to the conductivity from the $2k_F$ channel when fermion loops are saturating the Fermi surface curvature scale. Using our result (\ref{eq:log4}) of $\Sigma(\w) \sim \w^2/q_\parallel^3 \log^2 |\w|$ and
repeating the arguments of the previous subsection, or directly using the scaling (\ref{eq:qscaling}), leads to
\be\label{eq:omega2kf}
\dd \sigma(\Omega) \sim v^2 \log^2 \Omega \int^{\Omega^{1/3}} \frac{Z(q_\parallel)^2}{q_\parallel^3} d q_\parallel \sim v^2 \log^3 \Omega \,.
\ee
This is a stronger than Fermi liquid conductivity.
To accurately describe this contribution we would have to consider the effect of resumming logarithms,
and also the effects of Fermi surface curvature on the vertex correction computation in Section~\ref{sec:hotspot}.

\subsection{Conductivity due to fluctuating critical modes}
\label{sec:condfluc}

As well as the diagrams we have been considering, studies of quantum critical response (e.g. Ref.~\onlinecite{MIT}) or fluctuating superconductivity response (e.g. Ref.~\onlinecite{scbook}) also consider Aslamazov-Larkin type diagrams in which two separate fermion loops are connected via a pair of bosons. This graph is higher order in the couplings $\lambda_i$. We have seen above that in the mean field Hertz-Moriya-Millis level these couplings appear to in fact be irrelevant. Therefore we expect such higher order graphs to be suppressed in our low frequency $T=0$ computations. However, because the parameters $z$ and $\Delta$ will get renormalised, it is of interest to see what scaling is obtained from these graphs in general and whether there are cases in which this contribution can dominate. For instance in Ref.~\onlinecite{MIT} such fluctuating criticality contributions were found to scale equally to the vertex and self energy corrections we have considered so far. At the end of this subsection we will give a scaling argument showing that the condition for the $\lambda_i$ couplings to be relevant is that $\Delta < \frac{3}{2}$, independently of $z$. Let us first see this emerge from an explicit computation.

\begin{figure}[h]
\begin{center}
\includegraphics[height=100pt]{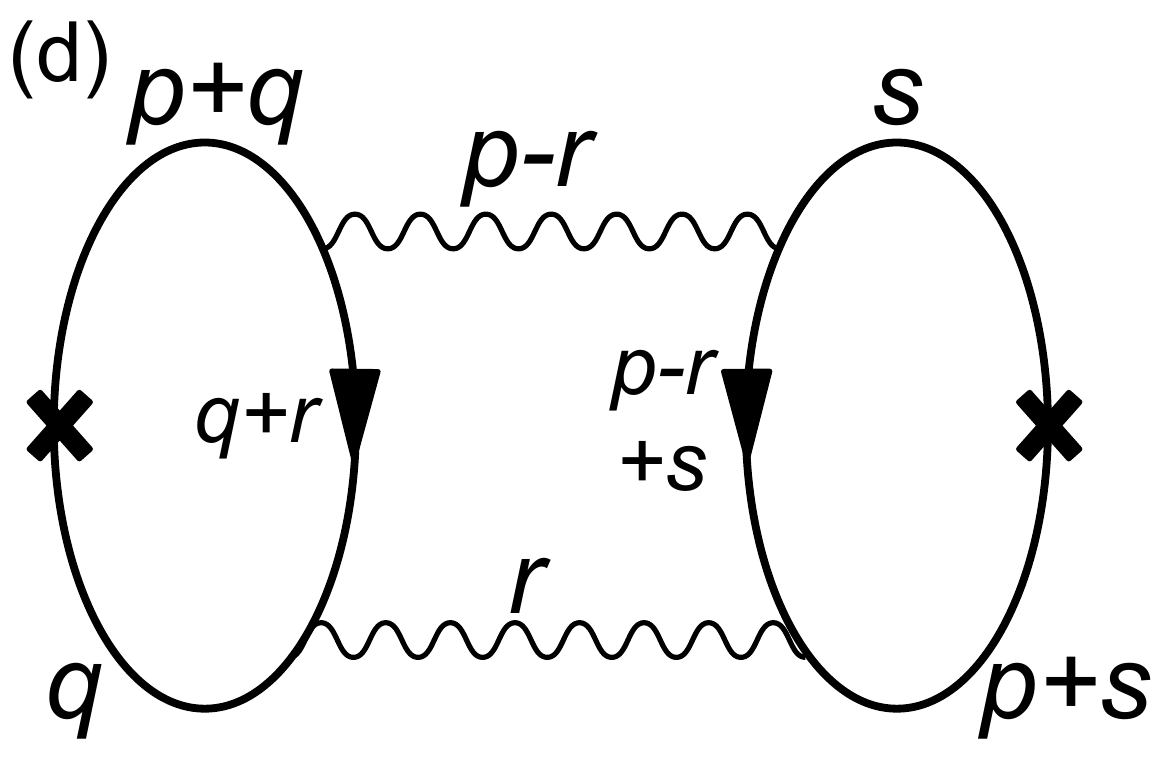}\hspace{1cm}\includegraphics[height=100pt]{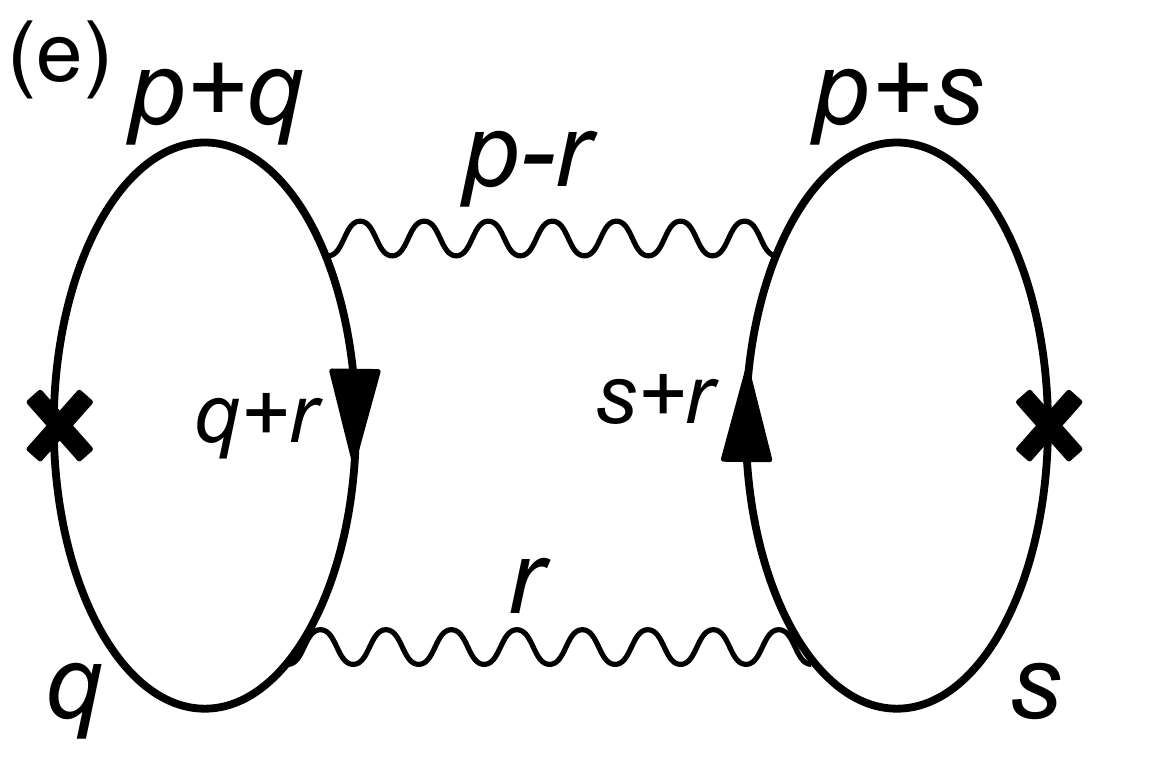}
\caption{The two diagrams describing the contribution to the conductivity of the cold fermions due to critical fluctuations. The wavy lines denote propagators of the neutral bosonic quantum critical operator $\ocal$. \label{fig:threeloop}}
\end{center}
\end{figure}

We can consider first the case of neutral and zero momentum critical operators. The two Feynman diagrams we are interested in are shown in figure \ref{fig:threeloop}.
These two contributions can be written, using the leading order momentum-independent term in the currents (\ref{eq:coldcurrents}),
\bea
\lefteqn{\dd^{(d+e)} \Pi_{ij}^\text{cold}(\w) \propto \lambda_1^4 \int \frac{d^3q}{(2\pi)^3} \frac{d^3r}{(2\pi)^3} \frac{d^3s}{(2\pi)^3} v^\star_i v^\star_j G(p+q)G(q) G(p+s) G(s) G(q+r)} \nonumber \\
&& \qquad \qquad \qquad \qquad \qquad \qquad \quad \times C(r) C(p-r) \Big(G(p-r+s) + G(r+s) \Big) \,.
\eea
We will not keep track of numerical prefactors in this subsection.
As above, we must recall that the spatial components of the external momentum $p$ vanish, so that $p=(\w,0,0)$. This allows us to use the identity (\ref{eq:identity}) twice on the first four Green's functions in the above equation. Relabeling momentum integrals then shows that the two diagrams precisely cancel, so that
\be
\dd^{(d+e)} \Pi_{ij}^\text{cold}(\w) = 0 \,.
\ee
An analogous cancellation was observed in Ref.~\onlinecite{MIT}. In order to obtain a nonzero contribution we will need to include the momentum-dependent term in the current, the second term in (\ref{eq:coldcurrents}), as we did in section \ref{sec:curvature} for the vertex and self energy corrections which also cancelled. As in section \ref{sec:curvature} one finds that the cross term $\langle J_\perp^\text{cold} J_\parallel^\text{cold} \rangle$ also gives a vanishing contribution (upon using evenness of $C(r)$) while the term $\langle J_\parallel^\text{cold} J_\parallel^\text{cold} \rangle$ gives the nonvanishing answer
\bea\label{eq:efneutral}
\lefteqn{\dd^{(d+e)} \Pi_{ij}^\text{cold}(\w) \propto - \frac{1}{\w^2} \frac{\lambda_1^4}{m^{\star 2}} \int \frac{d^3q}{(2\pi)^3} \frac{d^3r}{(2\pi)^3} \frac{d^3s}{(2\pi)^3}  (r_\parallel)_i (r_\parallel)_j C(r) C(p-r)} \nonumber \\
&& \qquad \qquad \qquad \qquad \quad \times G(s) G(q) G(q+r) \Big( G(s+r) - G(s+r-p) \Big) \,.
\eea
Before evaluating this expression we turn to the case of finite momentum critical modes, which once again is seen to exhibit no cancellation.

\begin{figure}[h]
\begin{center}
\includegraphics[height=100pt]{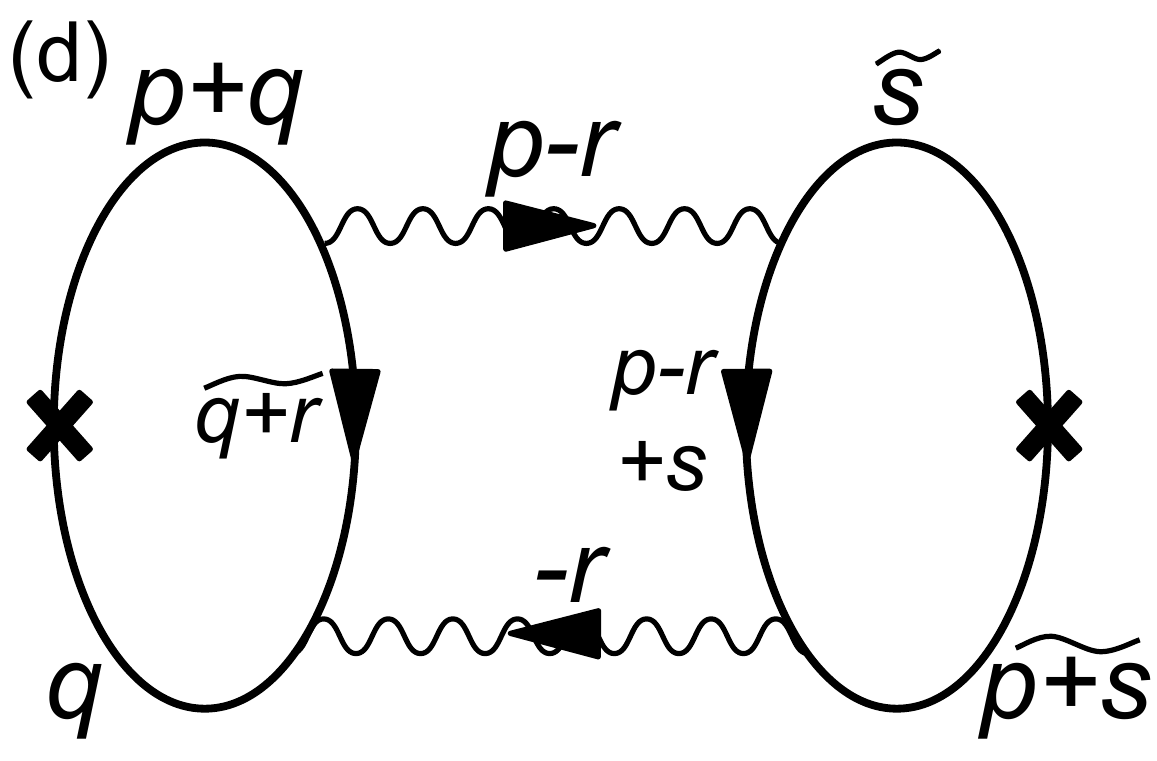}\hspace{1cm}\includegraphics[height=100pt]{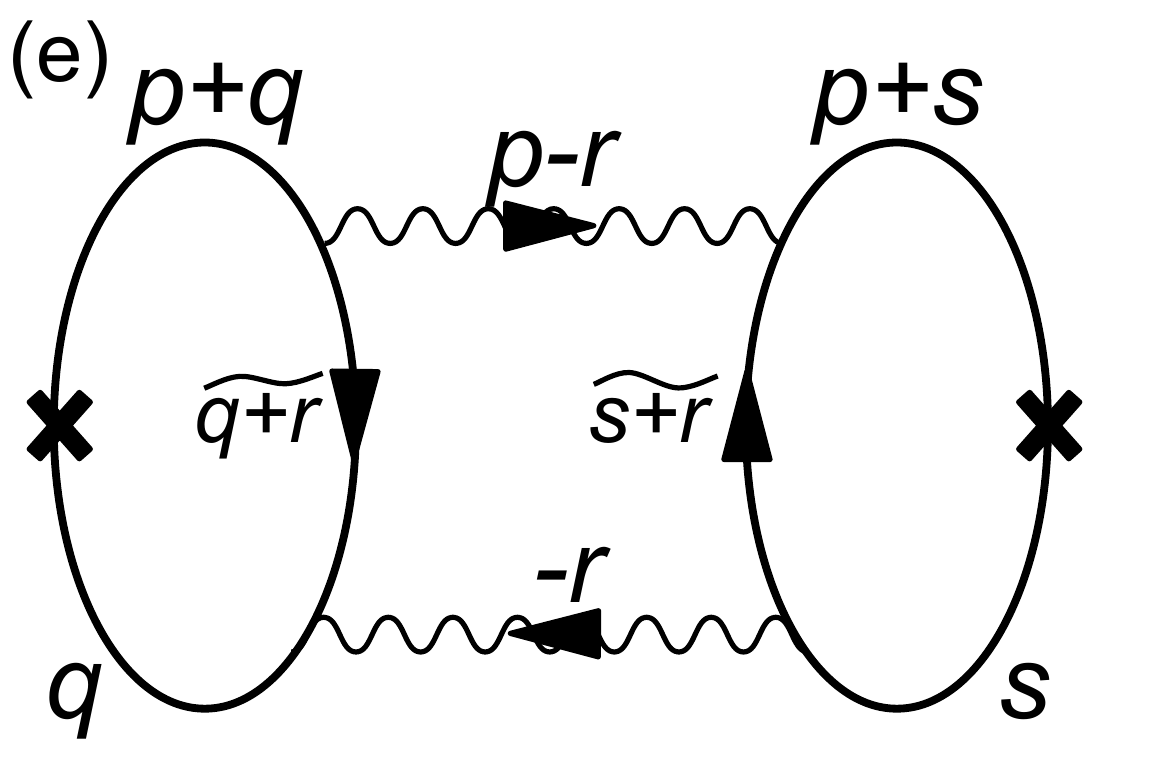}
\caption{The two diagrams describing the contribution to the conductivity of the cold fermions due to neutral critical fluctuations carrying momentum $2 k_F$. Tildes over momenta indicate that the fermion propagator in question corresponds to the opposite patch. The hot operator $\ocal$ is now complex and so its propagators carry an arrow. \label{fig:threeloopnematic}}
\end{center}
\end{figure}

The two diagrams describing critical fluctuations of modes carrying momentum $2 k_F$ are shown in figure \ref{fig:threeloopnematic}.
To properly account for the two opposite patches of the Fermi surface involved, we must again use the expression (\ref{eq:twopatchcurrent}) for the current. As before, the relative minus sign in the contribution to the current from the two patches is crucial. The diagrams in the figure evaluate to
\bea
\lefteqn{\dd^{(d+e)} \Pi_{ij}^\text{cold}(\w) \propto \lambda_2^4 \int \frac{d^3q}{(2\pi)^3} \frac{d^3r}{(2\pi)^3} \frac{d^3s}{(2\pi)^3} v^\star_i v^\star_j G(p+q)G(q) \widetilde G(q+r)} \nonumber \\
&& \qquad \times C(-r) C(p-r) \Big(\widetilde G(p+s) \widetilde G(s) G(p-r+s) - G(p+s) G(s) \widetilde G(r+s) \Big) \,.
\eea
Performing the same operations as outlined above for zero momentum critical operators, we find that the terms now add rather than cancel. The result can be expressed as
\bea
\lefteqn{ \dd^{(d+e)} \Pi_{ij}^\text{cold}(\w) \propto \frac{\lambda_2^4}{\w^2} \int \frac{d^3q}{(2\pi)^3} \frac{d^3r}{(2\pi)^3} \frac{d^3s}{(2\pi)^3} v^\star_i v^\star_j C(-r) C(p-r) G(q) G(s)} \nonumber \\
&& \qquad \qquad \qquad \times \left(\widetilde G(s+r) -\widetilde G(s+r-p) \right)
 \left(\widetilde G(q+r) -\widetilde G(q+r-p) \right) \,. \qquad \qquad \qquad
\eea

\begin{figure}[h]
\begin{center}
\includegraphics[height=100pt]{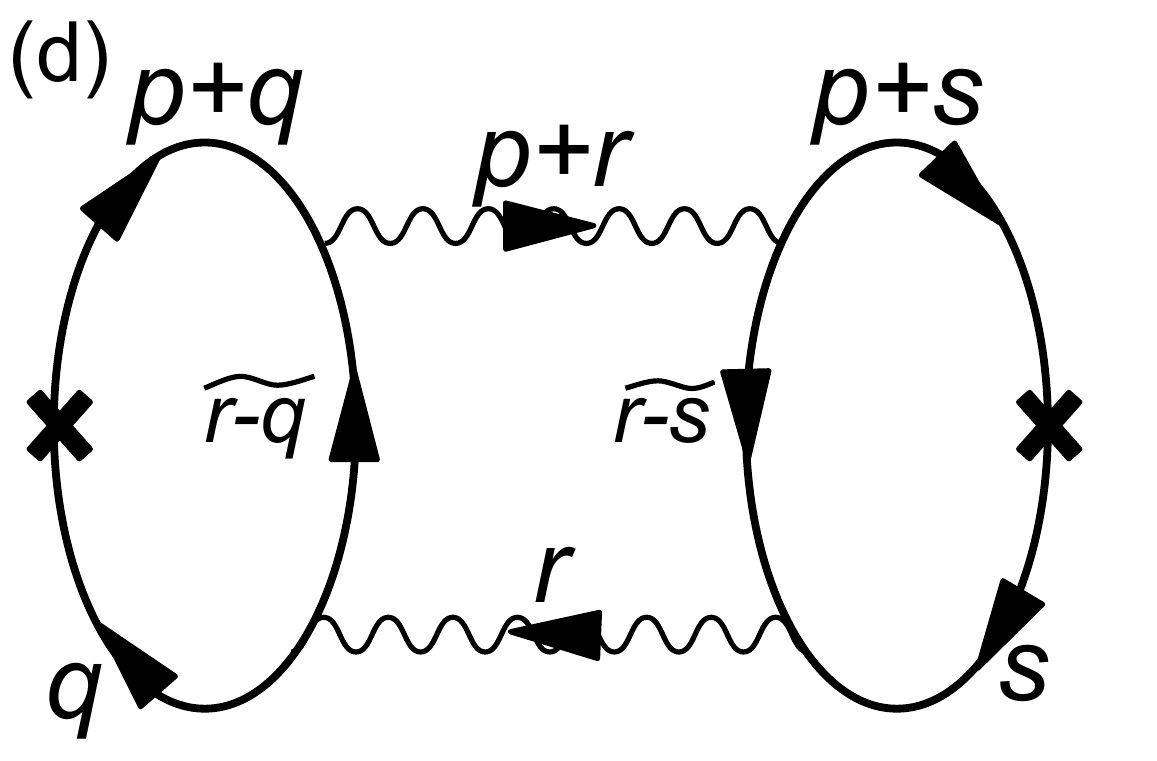}\hspace{1cm}\includegraphics[height=100pt]{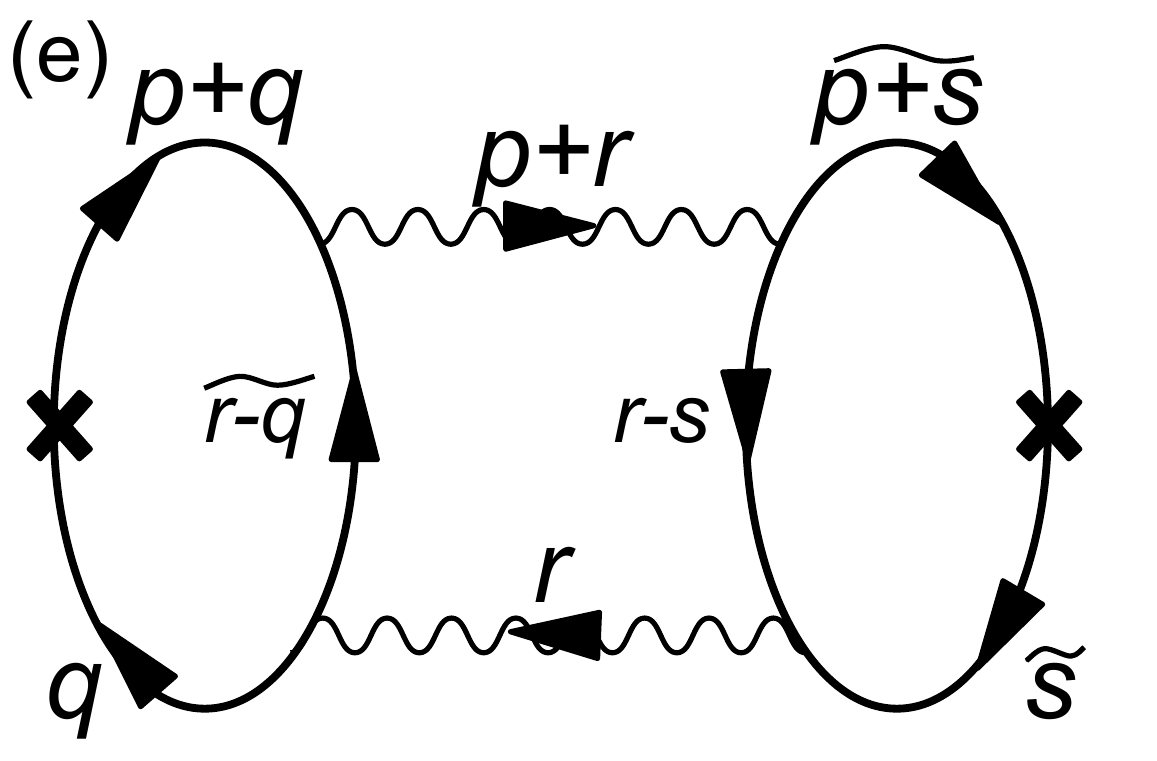}
\caption{The two diagrams describing the contribution to the conductivity of the cold fermions due to charged Cooperon-like critical fluctuations.\label{fig:threeloopsc}}
\end{center}
\end{figure}

Finally, we can consider fluctuations of zero momentum but charged critical modes. The two relevant diagrams are shown in figure \ref{fig:threeloopsc}.
While we will not write out the expressions explicitly in this case, one can easily check that there is again a cancellation between the two diagrams. Therefore curvature terms in the current operator must be included and an expression similar to (\ref{eq:efneutral}) is obtained for the current correlator.

We can now evaluate the momentum integrals to obtain the frequency scaling of these contributions to the conductivity. We will only evaluate the neutral and zero momentum case. Using the cold fermion Green's function (\ref{eq:cold}), the scaling form (\ref{eq:scaling}) for the critical modes, and taking the critical modes to have momentum tangent to the Fermi surface as per the discussion in the paragraph below equation (\ref{eq:selfone}) above, it is easy to perform the $\{q_\perp,q_\tau, s_\perp,s_\tau, r_\perp \}$ integrals in (\ref{eq:efneutral}) to obtain
\bea
 \dd^{(d+e)} \Pi_{ij}^\text{cold}(\w) \propto \frac{i}{\w^2} \frac{\lambda_1^4}{v^{\star 4}  m^{\star \, 2}} \int \frac{dr_\parallel ds_\parallel dq_\parallel}{(2\pi)^3} \frac{(r_\parallel)_i (r_\parallel)_j }{r_\parallel (s_\parallel - q_\parallel)/m^\star v^\star + i \w/v^\star} \times \nonumber \\
 \int_0^{\w} \frac{dr_{\tau}}{r_{\tau}^{(2-2\Delta)/z} (\w-r_{\tau})^{(2-2\Delta)/z} }
\widetilde C \left( \frac{r_\parallel}{r_{\tau}{}^{1/z}} \right) \widetilde C \left( \frac{r_\parallel}{(\w-r_{\tau})^{1/z}} \right) \,.
\eea
This expression is not yet ready to be put in a scaling form because of the curvature terms that have survived in the denominator of the last term in the first line. However, we can now symmetrise in $q_\parallel$ and $s_\parallel$ and then perform the $s_\parallel$ integral which is convergent.
Then setting $r_\tau = \w x$ and $r_\parallel = \w^{1/z} y$ we obtain
\bea
 \dd^{(d+e)} \Pi_{ij}^\text{cold}(\w) \propto - \frac{\w^{(4 \Delta -2)/z}}{\w} \frac{\lambda_1^4}{v^{\star 3}  m^{\star}} \int \frac{dy dq_\parallel}{(2\pi)^2} \frac{y_i y_j}{y} \times \nonumber \\ 
 \qquad \qquad \int_0^{1} \frac{dx}{[x (1-x)]^{(2-2\Delta)/z} }
\widetilde C \left( \frac{y}{x^{1/z}} \right) \widetilde C \left( \frac{y}{(1-x)^{1/z}} \right) \,.
\eea
Taking the frequency scaling from the above formulae and using the expression for the conductivity (\ref{eq:hotc}) we obtain the singular (i.e. scaling) conductivity due to fluctuating critical modes
\be
 \dd^{(d+e)} \sigma(\Omega) \sim \Omega^{(4 \Delta-2)/z - 2} \,.
\ee
Comparing with our previous result (\ref{eq:sigmaabc}) for the self energy and vertex contribution, we see that the fluctuating criticality contribution dominates if $\Delta \leq \frac{3}{2}$, independently of $z$. In particular when $\Delta = \frac{3}{2}$, the two contributions are equal, in agreement with the results of Ref. \onlinecite{MIT}. While, as above, for $\ocal = \phi^2$ this contribution is not strong with the mean field values of exponents, that is $z=2, \Delta = 2$, it is interesting to note in passing that if the dimension of this operator was renormalized down to the holographically motivated `bound' we mentioned above, $\Delta = z/2 = 1$, then we would obtain a conductivity $\sigma = \Omega^{-1}$.

The condition $\Delta \leq \frac{3}{2}$ for the effective $\lambda_i$ couplings to be relevant, and hence strong at low energies, can be derived from the following scaling argument. Exchange of a critical boson generates an effective four fermion interaction. The total cold fermion action is
\be
S_{\psi} \sim \int d^3x \left[ \psi^2 \left(- i \w + v^\star k_\perp + k_\parallel^2/2m^\star \right) + \lambda^2 \psi^4 \w^{-(z+2-2\Delta)/z} \widetilde C\left(k_\parallel/\w^{1/z} \right) \right] \,.
\ee
We are being schematic here, as we only wish to keep track of power counting.
The natural scaling to consider, given this action, is
\be
t \to s^z\, t \,, \quad x_\parallel \to s x_\parallel \,, \quad x_\perp \to s^2 x_\perp \,. 
\ee
From the quadratic term we see that the fermion field $\psi$ has dimension $(z+1)/2$ under this scaling action. The effective coupling after scaling is thus easily seen to be $\lambda^2 s^{3 - 2 \Delta}$. At low energies, $s \to \infty$, the interaction therefore becomes relevant if $\Delta \leq \frac{3}{2}$.

\section{Conclusions}
\label{sec:conc}

We have analyzed the fermion spectrum and the optical conductivity at the quantum critical point
describing the onset of spin density wave order in a two-dimensional metal. We focused on the universal
character of the results as described by a recently developed low energy theory \cite{Metlitski:2010vm}.

An appealing property of this low energy theory is that it has a finite d.c. conductivity at non-zero temperatures. 
This is a consequence
of an emergent pseudospin symmetry which decouples the electrical current from the conserved
momentum; umklapp processes are implicitly included in our continuum theory.
Consequently there is no delta function in the conductivity at $\omega=0$, and
the d.c. conductivity is dominated by interactions among low energy excitations, leading to hopes
of a universal non-Fermi liquid behavior.
In other continuum theories of fermions at non-zero density,
there is inevitably a strong delta function at $\omega=0$ even at non-zero temperatures, and its resolution depends
sensitively upon how terms beyond those in the leading critical theory will relax the momentum.

Going beyond our low energy critical theory, there are, of course, corrections which violate the emergent pseudospin
symmetry. Some of these terms would, by themselves, induce a weak delta function at $\omega=0$, but there are other
(also formally irrelevant) umklapp terms which would broaden the delta function at non-zero temperatures. 
We did not examine the delicate interplay
between various irrelevant terms for the zero frequency conductivity.

Here, we limited ourselves
to the simpler problem of the zero temperature optical conductivity. In the leading rainbow approximation of fermions scattering
off fluctuations of $\phi$, the spin density wave order parameter, we found that the optical conductivity
was suppressed relative to contribution expected from naive scaling arguments or previous 
computations \cite{advances}; these results appeared in Section~\ref{sec:hotspot}. 

Next, we examined the influence of composite operators \cite{vicari} on the optical conductivity: we found particularly
strong effects from an operator representing $2k_{F}$ density fluctuations with an Ising-nematic character \cite{Metlitski:2010vm,Metlitski:2010zh}.
Scattering generated from this operator, which arises at higher orders in our low energy theory, leads to the umklapp
process shown in Fig.~\ref{fig:umklapp}. While analyzing this process, we have encountered an intermediate energy window for fermions in the hot spot vicinity, where Luttinger-liquid like one-dimensional divergences appear. We estimate that electrons in this window may give rise to a large optical conductivity, as summarized in Section~\ref{sec:cond2kf}. Furthermore, we found that in the $2 k_F$ channel, BCS-like divergences cut off at the Fermi surface curvature scale lead to a more singular low energy scaling of the density wave correlator than would follow from simple power counting in the hot spot scaling theory.

We also studied the nature of the electron spectral functions around the Fermi surface. There is strong non-Fermi liquid
damping of the quasiparticles at the hot spots, as was already discussed in earlier work \cite{Metlitski:2010vm}. Here we showed that
non-Fermi liquid behavior is present at all locations on the Fermi surface, arising from scattering off composite operators. 
Such results appear in Section~\ref{sec:selfen}.

Further progress on the difficult issues left open by our analysis will likely require new approaches.  
More complete solutions of simpler problems
involving Fermi surface reconstruction would be useful. A holographic realization
of a field theory like that in Eq.~(\ref{eq:theaction}) would be valuable. Our analysis indicates that it would be best to
work in a holographic setting in which the pseudospin symmetry is explicitly realized. This would ensure there are no
zero frequency delta function contributions to the conductivity at non-zero temperatures. 
Such delta functions are a ubiquitous feature of existing
holographic studies of fermion systems at non-zero density, which are often ignored without justification.

\subsection*{Acknowledgements}

We thank A.~Chubukov, W.~Metzner, T.~M.~Rice, A.~Rosch, and D.~Scalapino for useful discussions.
This research was supported by the National Science Foundation under grant DMR-0757145 
and by a MURI grant from AFOSR (M.A.M. and S.S.) and by DOE grant DE-FG02-91ER40654, the FQXi foundation and the Center for the Fundamental Laws 
of Nature at Harvard University (S.A.H. and D.M.H).

\appendix

\section{Current vertex in the rainbow approximation}
\label{sec:appvert}
In this appendix we present the details of the analysis of the integral equations (\ref{Gminit}) for the current vertex.

Let us set $q = (\omega, 0)$ and without loss of generality assume $\omega > 0$. It is convenient to change variables to $l_a = \hat{v}_a \cdot \vec{l}$. Then  Eq.~(\ref{Gminit}) for the current vertex $\Gamma^-$ becomes,
\bea \Gamma^-_a(\omega, p) &=& (-1)^{a+1} + \frac{3 \pi v^2 \gamma}{4 N} \int \frac{d l_\tau dl_a dl_{\bar{a}}}{(2\pi)^3}
\left(i l_\tau + \frac{3 i v \sin 2\varphi}{8 N} \mathrm{sgn}(l_\tau)(\sqrt{\gamma |l_\tau| + l^2_a}-  |l_a|) - v l_{\bar{a}}\right)^{-1} \nn
\\
&\times& \left(i (l_\tau + \omega) + \frac{3 i v \sin 2\varphi}{8 N} \mathrm{sgn}(l_\tau + \omega) (\sqrt{\gamma |l_\tau + \omega| + l^2_a}-  |l_a|) - v l_{\bar{a}}\right)^{-1} \nn\\
&\times& \left(\gamma|l_\tau - p_\tau| + \csc^2 2\varphi ((l_a -p_a)^2 + (l_{\bar{a}} - p_{\bar{a}})^2+ 2 (l_a-p_a) (l_{\bar{a}}-p_{\bar{a}}) \cos 2 \varphi)\right)^{-1} \Gamma^-_{\bar{a}}(\omega, l)\nn\\\label{app:Gammalong}\eea
Let us perform the integral over $l_{\bar{a}}$ in Eq.~(\ref{app:Gammalong}). In the $N \to \infty$ limit, the main contribution comes from the poles of the fermion propagators. Since these have $l_{\bar{a}} \sim O(1/N)$, we will set $l_{\bar{a}}$ to zero in the rest of Eq.~(\ref{app:Gammalong}),
\bea \Gamma^-_a(\omega, p) &=& (-1)^{a+1} + 2 \pi \gamma \sin 2 \varphi \int \frac{d l_a}{2 \pi} \int_{-\omega}^0 \frac{d l_\tau}{2 \pi} \frac{1}{\sqrt{\gamma |l_\tau + \omega| + l^2_a}+\sqrt{\gamma |l_\tau| + l^2_a} -  2 |l_a| + \frac{\gamma \omega}{2 \Lambda}} \nn\\&\times&\frac{1}{(l_a - p_a - p_{\bar{a}} \cos 2 \varphi)^2 + \sin^2 2 \varphi (\gamma |l_\tau - p_\tau| + p^2_{\bar{a}})} \Gamma^-_{\bar{a}}(\omega,l_\tau, l_a, l_{\bar{a}} = 0)\nn\\\eea
where the UV momentum scale $\Lambda$ is given by Eq.~(\ref{Lambdadef}). Hence, to determine $\Gamma^-_a(\omega,p)$ for all $p$ it is sufficient to know its behavior for $p_a = 0$, i.e. we need to find the current vertex with external momentum on the Fermi surface. Moreover, introducing the variable $\nu = p_\tau + \omega$ we can restrict our attention to $0 < \nu < \omega$. Thus,
\beq \Gamma^-_a(\omega, \nu, p_a = 0, p_{\bar{a}} = p) = (-1)^{a+1} \Gamma^-(\omega, \nu, p) \eeq
where $\Gamma^-$ satisfies,
\bea \Gamma^-(\omega, \nu, p) &=& 1 - 2 \pi \gamma \sin 2 \varphi \int \frac{d l}{2\pi} \int_0^{\w} \frac{d \nu'}{2 \pi} 
\frac{1}{\sqrt{\gamma \nu' + l^2} + \sqrt{\gamma (\omega - \nu') + l^2} - 2 |l| +\frac{\gamma \omega}{2 \Lambda}}\nn \\
&& \frac{1}{(l - p \cos 2 \varphi)^2 + \sin^2 2 \varphi(\gamma |\nu' - \nu| + p^2)} \Gamma^-(\omega, \nu', l) \label{app:Gammaminusf}\eea
We note that repeating the same steps for $\Gamma^+_a$ gives,
\beq \Gamma^+_a(\omega, \nu, p_a = 0, p_{\bar{a}} = p) = \Gamma^+(\omega, \nu, p) \eeq
with 
\bea \Gamma^+(\omega, \nu, p) &=& 1 + 2 \pi \gamma \sin 2 \varphi \int \frac{d l}{2\pi} \int_0^{\w} \frac{d \nu'}{2 \pi} 
\frac{1}{\sqrt{\gamma \nu' + l^2} + \sqrt{\gamma (\omega - \nu') + l^2} - 2 |l| + \frac{\gamma \omega}{2 \Lambda}}\nn \\
&& \frac{1}{(l - p \cos 2 \varphi)^2 + \sin^2 2 \varphi(\gamma |\nu' - \nu| + p^2)} \Gamma^+(\omega, \nu', l) \label{app:Gammaplusf}\eea


We now proceed to study the vertex $\Gamma^-$. At low momentum and frequency, we may drop the term $\frac{\gamma \omega}{2 \Lambda}$ in the kernel of Eq.~(\ref{app:Gammaminusf}) (this term originates from the analytic part of the fermion propagators). Then,
\bea \Gamma^-(\omega, \nu, p) &=& 1 - 2 \pi \gamma \sin 2 \varphi \int \frac{d l}{2\pi} \int_0^{\w} \frac{d \nu'}{2 \pi} 
\frac{1}{\sqrt{\gamma \nu' + l^2} + \sqrt{\gamma (\omega - \nu') + l^2} - 2 |l|}\nn \\
&& \frac{1}{(l - p \cos 2 \varphi)^2 + \sin^2 2 \varphi(\gamma |\nu' - \nu| + p^2)} \Gamma^-(\omega, \nu', l)\label{app:Gm2}\eea
The integral in Eq.~(\ref{app:Gm2}) now requires regularization and should be cut-off at $l \sim \Lambda$. 

Let us begin by determining $\Gamma^-(p) = \lim_{\omega, \nu \to 0} \Gamma^-(\omega, \nu, p)$. Expanding the kernel of Eq.~(\ref{app:Gm2}) for $\omega \to 0$ and performing the integral over $\nu'$ we obtain,
\beq \Gamma^-(p) =  1 - \frac{\sin 2 \varphi}{\pi} \int dl \frac{|l|}{(l- p \cos 2 \varphi)^2 + \sin^2 2\varphi p^2} \Gamma^-(l) \label{app:Gmw0}\eeq
We expect that 
\beq \Gamma^-(p) \sim |p|^{r_0},  \quad |p| \ll \Lambda \label{app:Gmw0scal}\eeq
Note that the power $r_0$ cannot be negative. Indeed, if $r_0 < 0$ then for $p \to 0$ we may neglect the constant $1$ in Eq.~(\ref{app:Gmw0}). Assuming that $\Gamma^-(l) > 0$ for all $l$, this leads to a contradiction. Hence, $r_0 > 0$, which means,
\beq \frac{\sin 2 \varphi}{\pi} \int dl \frac{\Gamma^-(l)}{|l|} = 1 \eeq
i.e.,
\beq \Gamma^-(p) = - \frac{\sin 2 \varphi}{\pi} \int dl \left(\frac{|l|}{(l- p \cos 2 \varphi)^2 +  \sin^2 2\varphi p^2}  - \frac{1}{|l|}\right) \Gamma^-(l)\label{app:Gmw0conv}\eeq
We can now substitute the scaling form (\ref{app:Gmw0scal}) into Eq.~(\ref{app:Gmw0conv}). The integral converges in the UV for $r_0 < 2$. Evaluating,
\beq I(r) = \frac{\sin 2 \varphi}{\pi} \int dy \left(\frac{|y|}{(y-\cos 2 \varphi)^2 + \sin^2 2 \varphi} -  \frac{1}{|y|}\right) y^r = -\frac{\cos((\pi/2 - 2 \varphi)(r+1))}{\sin \pi r/2} \label{app:Irdef}\eeq
we obtain the condition,
\beq I(r_0) = -1\eeq
which after some algebra may be rewritten as,
\beq \cos(\varphi(r_0+1)) \cos((\pi/2 - \varphi)(r_0+1)) = 0 \eeq
giving
\beq r_0 = -1 + \frac{\pi}{2 \varphi} (2 n - 1), \quad \mathrm{or}\quad r_0 = -1 + \frac{\pi}{\pi - 2 \varphi} (2 n - 1), \quad n \in \mathbb{Z} \label{app:r0sol}\eeq
We expect that the IR behavior of $\Gamma^-$ will be dominated by the smallest positive exponent $r_0$; the other values of $r_0$ give corrections to scaling. Hence,
\beq r_0 = \left\{\begin{array}{c} \frac{2 \varphi}{\pi - 2 \varphi}, \quad 0 < \varphi < \pi/4\\
\frac{\pi - 2\varphi}{2 \varphi}, \quad \pi/4 < \varphi < \pi/2\end{array}\right.\label{app:r0}\eeq
We note that Eq.~(\ref{app:Gm2}) is invariant under $\varphi \to \pi/2 - \varphi$, so below we only consider $0 < \varphi < \pi/4$. Fig.~\ref{fig:r0} shows the behavior of the exponent $r_0$ as a function of $\varphi$. Note that $0 < r_0 \leq 1$. We also briefly point out that at the special value $\varphi=\pi/4$, $r_0 = 1$ and we expect a logarithmic correction $\Gamma^-(p) \sim |p| \log(\Lambda/|p|)$ to Eq.~(\ref{app:Gmw0scal}).
 
Next, we come back to discuss the frequency dependence of $\Gamma^-$. We expect $\Gamma^-$ to obey the following scaling form for $\omega \ll \Lambda_\omega$, $p \ll \Lambda$,
\beq \Gamma^-(\omega, \nu, p) = C(\omega/\Lambda_\omega) \gamma^-\left(\frac{\nu}{\omega}, \frac{p}{\sqrt{\gamma \omega}}\right)\label{app:Gmscal}\eeq
For $p \gg \sqrt{\gamma \omega}$, we expect to recover the behavior (\ref{app:Gmw0scal}), hence,
\beq \gamma^-(x,y) \sim y^{r_0}, \quad y \to \infty\eeq
and consequently $C(\omega/\Lambda_\omega) \sim \left(\frac{\omega}{\Lambda_\omega}\right)^{r_0/2}$, i.e,
\beq \Gamma^-(\omega, \nu, p) \sim \left(\frac{\omega}{\Lambda_\omega}\right)^{r_0/2} \gamma^-\left(\frac{\nu}{\omega}, \frac{p}{\sqrt{\gamma \omega}}\right)\label{app:Gmscal2}\eeq
Hence, the vertex $\Gamma^-$ aquires an anomalous dimension. Unlike, $\Gamma^+$, which is enhanced at low energy, $\Gamma^-$ is suppressed. 

To determine $\gamma^-$ let us improve the UV convergence properties of the kernel in Eq.~(\ref{app:Gm2}).
\bea &&\Gamma^-(\omega, \nu, p) = 1 - 2 \pi \gamma \sin 2 \varphi \int \frac{d l}{2 \pi} \int_0^{\omega} \frac{d\nu'}{2 \pi} \frac{2}{\gamma \omega \sqrt{\gamma \omega + l^2}} \Gamma^-(\omega, \nu', l)\nn\\
&-& 2 \pi \gamma \sin 2 \varphi \int \frac{d l}{2 \pi} \int_0^{\omega} \frac{d \nu'}{2 \pi}
\Bigg(\frac{1}{(\sqrt{\gamma \nu' + l^2} + \sqrt{\gamma (\omega - \nu') + l^2} - 2 |l|)} \nn\\
&&~~~\times
\frac{1}{(l - p \cos 2 \varphi)^2 + \sin^2 2 \varphi(\gamma |\nu' - \nu| + p^2)} -  \frac{2}{\gamma \omega \sqrt{\gamma \omega + l^2}}\Bigg)\Gamma^-(\omega, \nu', l)\eea
We can now replace $\Gamma^-$ in the second integral above by its scaling form (\ref{app:Gmscal}). The integral over $l$ will converge as $r_0 < 2$. Note that the term $\frac{2}{\gamma \omega \sqrt{\gamma \omega + l^2}}$ that we have added and subtracted from the kernel is somewhat arbitrary; we could have used any $\nu$ and $p$ independent term with the same  UV behavior. Thus, we obtain,
\bea \gamma^-(x,y) &=& 1 -  \int_0^{\infty} d y' \int_0^{1} d x' K(x, y; x',y') \gamma^-(x', y')\label{app:gammaint}\eea
with the kernel,
\bea && K(x,y; x', y') =  \frac{\sin 2 \varphi}{2 \pi} \Bigg(\frac{1}{\sqrt{x' + y'^2} + \sqrt{(1 - x') + y'^2} - 2 y'}  \nn\\
&&~\times \left(\frac{1}{(y' - y \cos 2 \varphi)^2 + \sin^2 2 \varphi(|x' - x| + y^2)} + \frac{1}{(y' + y \cos 2 \varphi)^2 + \sin^2 2 \varphi(|x' - x| + y^2)}\right) \nn \\ 
&& \quad \quad -  \frac{4}{ \sqrt{1 + y'^2}}\Bigg)\label{app:Kernel}\eea
and
\beq C = 1 - \frac{4 \pi \sin 2 \varphi}{\omega} \int \frac{d l}{2 \pi} \int_0^{\omega}\frac{d\nu'}{2 \pi} \frac{1}{\sqrt{\gamma \omega + l^2}} \Gamma^-(\omega, \nu', l) \eeq

Although we don't have an explicit analytic solution of Eq.~(\ref{app:gammaint}), by studying the behavior of $\gamma^-(x,y)$ for $y \gg 1$ we will obtain the `sum rule' (\ref{sumrule}) that will allow us to compute the optical conductivity. We expect that in the above limit $\gamma^-(x,y)$ can be expanded as a power series in $y$. We, therefore, write
\beq \gamma^-(x,y) = g_0(x) y^{r_0} + g_1(x) y^{r_1} + g_2(x) y^{r_2} + \ldots, \quad y \gg 1 \label{app:gmlargey}\eeq
We have already determined the exponent $r_0$ (Eq.~(\ref{app:r0})) of the leading term in the above expansion; we now turn to the analysis of the subleading terms. Let us introduce a scale $\mu$ such that $1 \ll \mu \ll y$. We will divide the integration range in Eq.~(\ref{app:gammaint}) into two intervals $y' < \mu$ and $y' > \mu$. For $y' < \mu \ll y$ we expand the kernel (\ref{app:Kernel}) in powers of $y^{-2}$, 
\beq K = K^0_< + K^2_< + \cdots \eeq
\bea K^0_<(x,y;x',y') &=& -\frac{2 \sin 2 \varphi}{\pi} \frac{1}{\sqrt{1+y'^2}}\\
K^2_<(x,y;x',y') &=& \frac{\sin 2 \varphi}{\pi} \frac{1}{\sqrt{x' + y'^2} + \sqrt{(1 - x') + y'^2} - 2 y'} y^{-2}\eea
In the opposite range, $y' > \mu \gg 1$, we expand the kernel (\ref{app:Kernel}) in powers of $y^{-2}$ and $(y')^{-2}$,
\beq K = K^0_> + K^2_> + \cdots \eeq
\bea && K^0_>(x,y;x',y') =
  \frac{\sin 2 \varphi}{\pi}\left(\frac{y'}{(y' - y \cos 2 \varphi)^2 + y^2 \sin^2 2 \varphi } + \frac{y'}{(y' + y \cos 2 \varphi)^2 + y^2 \sin^2 2 \varphi } - \frac{2}{y'}\right)\nn\\
&& K^2_>(x,y;x',y') = \nn\\
&&~  \frac{ (x'^2 + (1-x')^2) \sin 2\varphi}{4 \pi}  \left(\frac{1}{(y' - y \cos 2 \varphi)^2 + y^2 \sin^2 2 \varphi } + \frac{1}{(y' + y \cos 2 \varphi)^2 + y^2 \sin^2 2 \varphi}\right)\frac{1}{y'} \nn\\
&&~- \frac{|x-x'| \sin^3 2 \varphi }{\pi} \left(\frac{y'}{((y' - y \cos 2 \varphi)^2 + y^2 \sin^2 2 \varphi)^2} + \frac{y'}{((y' + y \cos 2 \varphi)^2 + y^2 \sin^2 2 \varphi)^2}\right)  \nn\\
&&~+ \frac{\sin 2 \varphi}{\pi} \frac{1}{y'^3}\eea
Note that in the range $y' > \mu \gg 1$ we can use the expansion (\ref{app:gmlargey}). Hence, to the present order,
\bea  && g_0(x) y^{r_0} + g_1(x) y^{r_1} + g_2(x) y^{r_2} = 1 - \int_0^{\mu} d y' \int_0^{1} d x' K^0_<(x, y; x',y') \gamma^-(x', y') \nn\\ &-& \int_0^{\mu} d y' \int_0^{1} d x' K^2_<(x, y; x',y') \gamma^-(x', y')\nn\\
&-& \int_{\mu}^{\infty} d y' \int_0^{1} d x' K^0_>(x, y; x',y') (g_0(x') y'^{r_0} + g_1(x') y'^{r_1} + g_2(x') y'^{r_2})\nn\\ 
&-& \int_{\mu}^{\infty} d y' \int_0^{1} d x' K^2_>(x, y; x',y') g_0(x') y'^{r_0}\nn\\ \label{app:gmintexp}\eea
Now, for $-4 < r < 2$, and $y \gg \mu$, 
\beq \int_{\mu}^{\infty} dy' K^0_>(x,y;x',y') y'^{r} = I(r) y^r + \frac{2 \sin 2\varphi}{\pi} \frac{\mu^r}{r} - \frac{2 \sin 2\varphi}{\pi} \frac{\mu^{r+2}}{r+2}y^{-2} + O(\mu^{r+4} y^{-4})\eeq
with $I(r)$ given by Eq.~(\ref{app:Irdef}).
Likewise, for $-2 < r < 2$ and $y \gg \mu$,
\bea \int_{\mu}^{\infty} dy' K^2_>(x,y;x',y') y'^r &=& \frac{1}{4}(x'^2 + (1-x')^2) \left(I(r-2) y^{r-2} - \frac{2 \sin 2 \varphi}{\pi} \frac{\mu^r}{r} y^{-2}\right) \nn\\  &-& |x-x'| J(r) y^{r-2} - \frac{\sin 2 \varphi}{\pi} \frac{\mu^{r-2}}{r-2} + O(\mu^{r+2} y^{-4})\eea
with
\beq J(r) =  \frac{1}{2 \sin \pi r/2}(r \sin 2 \varphi \cos((\pi/2 -2 \varphi)r) + \cos 2\varphi \sin ((\pi/2 - 2 \varphi)r)\eeq
We see that the integration over the range $0 < y' < \mu$ on the right-hand-side of Eq.~(\ref{app:gammaint}) gives a series of terms $y^{-2 m}$ with integer $m \geq 0$. The integration over the range $y' > \mu$ also gives such a contribution, but in addition produces a series of terms $y^{r-2m}$ with integer $m \geq 0$ and $r$ - one of the exponents in the expansion (\ref{app:gmlargey}). Hence, we conclude that all exponents appearing in the expansion (\ref{app:gmlargey}) are either `primary' exponents satisfying
\beq I(r) = -1 \label{app:Ir}\eeq
or `descendant' exponents of the form $r - 2m$ with $r$ - a primary exponent and an integer $m \geq 1$. Note that terms of the form $y^{-2m}$, with integer $m \ge 0$ do not appear in the expansion (\ref{app:gmlargey}), as these produce terms of the form $y^{-2m} \log y$ upon convolution with $K^0_>$. 

As already discussed, the solutions to Eq.~(\ref{app:Ir}) take the form (\ref{app:r0sol}). We have already determined the largest primary exponent $r_0$, Eq.~(\ref{app:r0}), hence,  we only admit solutions to Eq.~(\ref{app:r0sol}) which are smaller than $r_0$ as subleading primary exponents. So the next largest primary exponent is 
\beq r_2 = -1 - \frac{\pi}{\pi - 2 \varphi} = - 2 -r_0 \label{app:r2app}\eeq
At this point the relation (\ref{app:r2app}) between $r_2$ and $r_0$ appears accidental. We will see later that this relation is necessary for a consistent definition of the anomalous dimension of the current operator. We also point out that the first descendant exponent 
\beq r_1 = r_0 - 2\eeq
satisfies $r_2 <  r_1 < r_0$, which explains the labeling that we have chosen. 

Coming back to Eq.~(\ref{app:gmintexp}), by matching the coefficients of the terms $y^{r_0}$, $y^{r_0 -2}$, $y^{r_2}$,  appearing on the right and left hand sides we conclude that the coefficients $g$ of the terms with the primary exponents are independent of $x$,
\beq g_0(x) = g_0, \quad g_2(x) = g_2 \eeq
while
\beq g_1(x) = \left(- \frac{(1+2 J(r_0)) I(r_0 - 2)}{6(1+I(r_0 -2))} + \frac{1}{2} (x^2 + (1-x)^2)J(r_0)\right) g_0 \label{app:g1}\eeq
Moreover, by matching the coefficients of $y^0$ and $y^{-2}$ we obtain the sum-rules,
\beq \int_0^{\mu} dy \int_0^1 dx \frac{\gamma^-(x,y)}{\sqrt{x + y^2} + \sqrt{1 -x +y^2} - 2 y}  = a_0 \mu^{r_0+2} + a_1 \mu^{r_0} + a_2 \mu^{r_2 + 2} + \cdots \label{app:sumrule}\eeq
\beq \int^{\mu}_0 dy \int_0^{1} dx \frac{\gamma^{-}(x,y)}{\sqrt{1+y^2}} = - \frac{\pi}{2 \sin 2 \varphi} + b_0 \mu^{r_0} + 
b_1 \mu^{r_0 - 2} + b_2 \mu^{r_2} + \cdots\eeq
with
\bea a_0 &=& \frac{2}{r_0 + 2} g_0 \\
a_1 &=& \frac{2 J(r_0) +1}{3 r_0 (1+I(r_0-2))}g_0\\
a_2 &=& \frac{2}{r_2 + 2} g_2\\
b_0 &=& \frac{1}{r_0} g_0 \\
b_1 &=&  \frac{1}{r_0 -2} \left(\frac{2 J(r_0) - I(r_0-2)}{6 (1 + I(r_0 - 2))} - \frac{1}{2}\right) g_0\\
b_2 &=& \frac{1}{r_2} g_2 \eea
The sum rule (\ref{app:sumrule}) is crucial for the computation of the conductivity in Eq.~(\ref{Pixxgm}). We note that we expect $g_0 \geq 0$ so that both $a_0 \geq 0$ and $a_1 \geq 0$.

\section{Constant contribution to the conductivity in the rainbow approximation}
\label{app:const}
In this appendix, we compute the constant contribution to the conductivity, Eq.~(\ref{C1}). We work in the rainbow approximation and take the limit $N \to \infty$. To determine $C_1$, we must solve the integral equation, Eq.~(\ref{GammaF}), for the current vertex $\Gamma^-$ in the regime $p \sim \Lambda$. It will be sufficient to compute $\Gamma^-$ to linear order in frequency. We expect,
\beq \Gamma^-(\omega, \nu, p) = f_0(p/\Lambda) + \frac{\gamma \omega}{\Lambda^2} f_1(\nu/\omega, p/\Lambda) \label{app:gammaf0f1}\eeq
At zeroth order in frequency we obtain from (\ref{GammaF}) an equation for the static vertex $f_0$,
\bea f_0(y) &=& 1 - \int_0^{\infty} dy' R(y,y') f_0(y')\label{app:f0}\eea
with the kernel,
\beq R(y,y') = \frac{\sin 2 \varphi}{\pi}\frac{y'}{1+y'}\left(\frac{1}{(y' - y \cos 2 \varphi)^2 + y^2 \sin^2 2 \varphi } + \frac{1}{(y' + y \cos 2 \varphi)^2 + y^2 \sin^2 2 \varphi } \right)\eeq
Next, expanding Eq.~(\ref{GammaF}) to first order in frequency, after some algebra one finds,
\beq f_1(x,y) = A(y) + (x^2 + (1-x)^2 - \frac{2}{3}) B(y)\label{app:f1}\eeq
Here, $B(y)$ is expressed in terms of the static vertex $f_0$,
\bea && B(y) = \frac{\sin^3 2 \varphi }{2 \pi} \int_0^{\infty} dy' \frac{y'}{1+y'}\Bigl(\frac{1}{((y' - y \cos 2 \varphi)^2 + y^2 \sin^2 2 \varphi)^2} \nn \\
&& \quad \quad + \frac{1}{((y' + y \cos 2 \varphi)^2 + y^2 \sin^2 2 \varphi)^2}\Bigr) f_0(y')\eea
while $A(y)$ satisfies,
\beq A(y) = C(y) - \int_0^{\infty} dy' R(y,y') A(y')\label{app:A}\eeq
and $C(y)$ is expressed in terms of $f_0$,
\beq C(y) = \frac{2}{3} B(y) - \frac{1}{6} \int_0^{\infty} dy' \frac{1}{y'^2 (1+y')}R(y,y')f_0(y')\eeq
From Eqs.~(\ref{app:Gmscal2}), (\ref{app:gmlargey}), we expect the functions $f_0$, $f_1$ to have the following asymptotic behavior in the limit $y \to 0$,
\bea f_0(y) &\to& g_0 y^{r_0}\nn\\
f_1(x,y) &\to& g_1(x) y^{r_0 -2}\label{app:fscal}\eea  
with $g_1(x)$ given by Eq.~(\ref{app:g1}). 

\begin{figure}[t]
\begin{center}
\includegraphics[width = 3in]{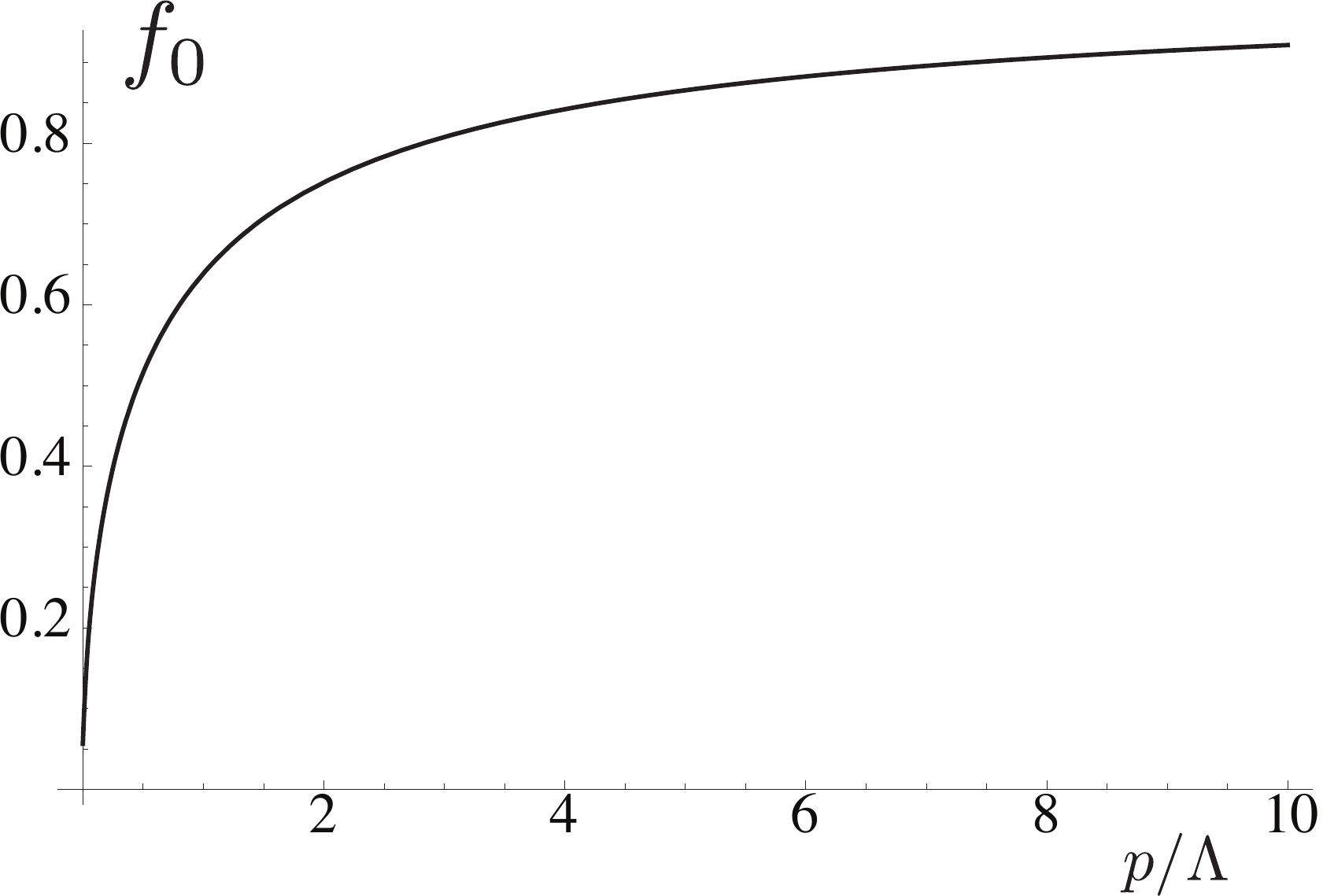}\\~\\
\includegraphics[width = 3in]{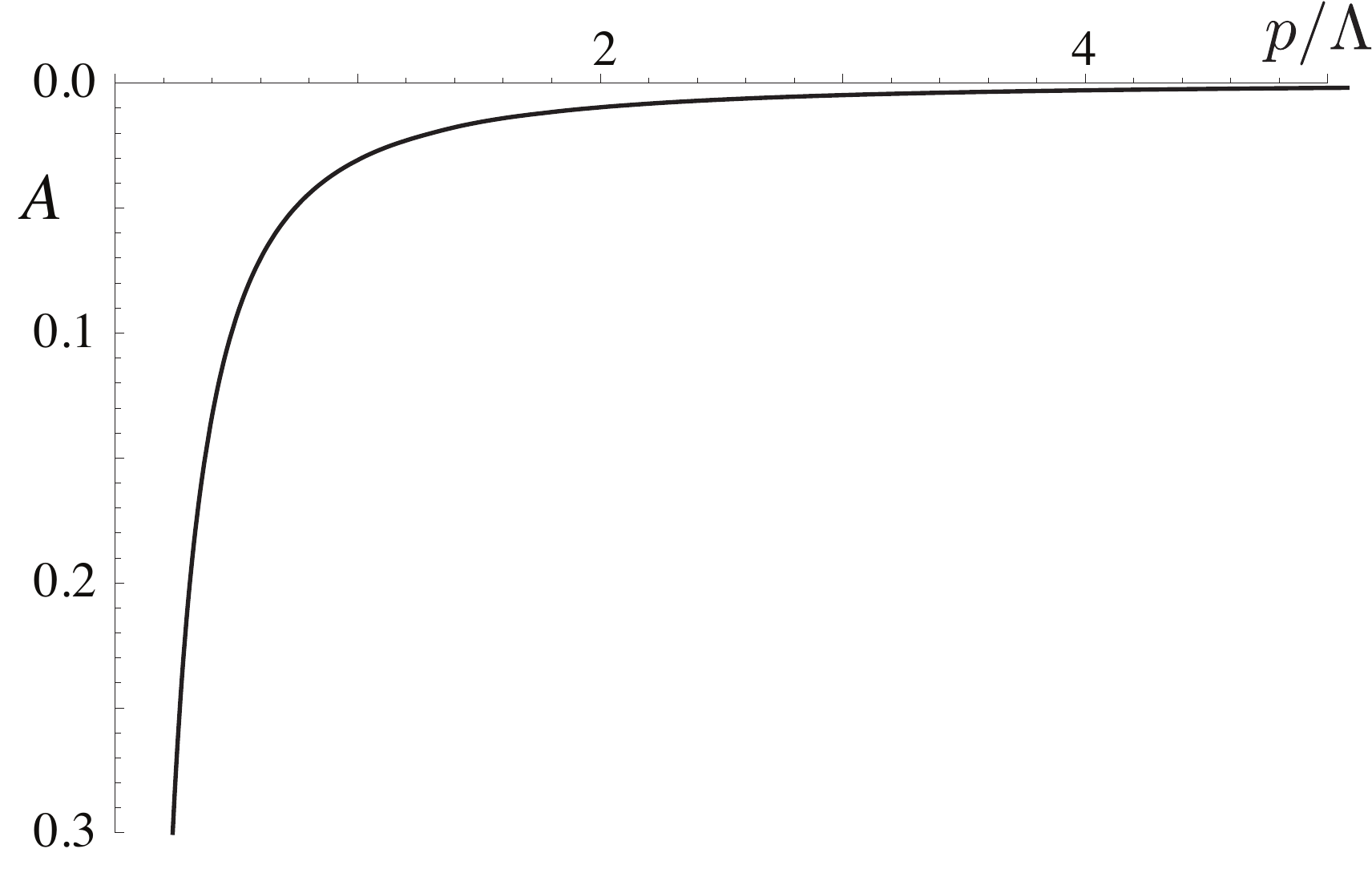}
\caption{Numerical solution for the current vertex in the rainbow approximation. Top: $f_0$ - current vertex in the static limit. 
Bottom: function $A$, Eq.~(\ref{app:f1}), which enters the first frequency dependent correction to the current vertex. }\label{fig:vertnum}
\end{center}
\end{figure}

We have numerically solved Eqs.~(\ref{app:f0}), (\ref{app:A}) for the functions $f_0(y)$, $A(y)$. Sample plots are shown in Fig.~\ref{fig:vertnum}. The numerical solution is in good agreement with the asymptotic forms (\ref{app:fscal}) in the limit $p \ll \Lambda$. 

Next, we proceed to the optical conductivity itself. Expanding the integrand in Eq.~(\ref{PixxGm}) to linear order in $\omega$ and using Eqs.~(\ref{app:gammaf0f1}),~(\ref{app:f0}),
\beq \Pi^{\ell=1}_{xx}(\omega) = \frac{1}{2} N^2 \omega C_1(\varphi) \label{app:Pixx}\eeq 
with
\beq C_1(\varphi) = \frac{16 N^2}{3 \pi^2 \sin^2 \varphi} \left(\int_0^{\infty} dy \frac{y}{1+y} A(y)+  \frac{1}{6}\int_0^{\infty} dy \frac{1}{y (1+y)^2} f_0(y)\right)\label{app:C1}\eeq
Note that we have dropped a constant contribution in Eq.~(\ref{app:Pixx}) which renormalizes the Drude weight. Finally, performing a sum over hot spot index $\ell$ and going to real frequency, we obtain from (\ref{app:Pixx}) 
\beq \mathrm{Re}\sigma_{ij}(\Omega) = N^2 C_1(\varphi)\delta_{ij}\eeq
We have performed the integral (\ref{app:C1}) using the numerical solutions for $f_0$ and $A$. The result is shown in Fig.~\ref{fig:C1}. 

\section{The correlation function $\langle \phi^2(x) \phi^2(0) \rangle$ in Hertz-Moriya-Millis theory}
\label{sec:phi4}

For completeness we calculate the two point function of the operator $O(x) = \vec{\phi}^2(x)$ in Hertz-Moriya-Millis theory. We start with the non-local action,
\beq S = \frac{1}{2} \int \frac{d^2 \vec{q} d\omega}{(2\pi)^3} (|\omega| + \frac{\vec{q}^2}{\gamma} + r) |\vec{\phi}(\vec{q},\omega)|^2 + \frac{u \gamma}{4} \int d^2\vec{x} d \tau (\vec{\phi}^2)^2\eeq
With our choice of normalizations, the coupling $u$ is dimensionless, while $\gamma$ has dimensions $\vec{q}^2/\omega$. Note that the normalization of $\phi$ differs here by a factor of $\sqrt{\gamma}$ compared to Eq.~(\ref{eq:bosonD}). We tune the coefficient $r$ to the critical value. At the critical point, the theory requires the following renormalizations,
\beq \gamma = Z_\gamma \gamma_r, \quad u = \frac{Z_u}{Z_\gamma} u_r, \quad [\vec{\phi}^2]_r = Z_2 [\vec{\phi}^2] \eeq
The field strength of $\vec{\phi}$ requires no renormalization. The renormalization constants $Z_\gamma$, $Z_u$ and $Z_2$ are functions of $\Lambda/\mu$ and $u_r$ only, with $\Lambda$ - the UV cut-off and $\mu$ - the renormalization scale. We define the $\beta$-functions and anomalous dimensions,
\beq \beta(u_r) = \mu \frac{\dd u_r}{\dd \mu}|_{u, \gamma}, \quad b_\gamma(u_r) = - \mu \frac{\dd \log Z_\gamma}{\dd \mu}|_{u, \gamma}, \quad \eta_2 = \mu \frac{\dd\log Z_2}{\dd\mu}|_{u, \gamma}  \eeq
To leading order in $u_r$ \cite{HertzMillisGamma},
\beq Z_u =1+ \frac{11}{2 \pi^2} u_r \log \Lambda/\mu, \quad Z_\gamma = 1 + \frac{5 (12 - \pi^2)}{16 \pi^4} u^2_r \log \Lambda/\mu, \quad Z_2 = 1+ \frac{5}{2 \pi^2} u_r \log \Lambda/\mu\ \eeq
\beq \beta(u_r) = c_u u^2_r, \quad b_\gamma = c_\gamma u^2_r, \quad \eta_2 = c_2 u_r \eeq
\beq c_u = \frac{11}{2 \pi^2}, \quad c_\gamma = \frac{5 (12 - \pi^2)}{16 \pi^4}, \quad c_2 = - \frac{5}{2 \pi^2}\eeq

The two-point function $C$ of the $\vec{\phi}^2$ operators requires an additional additive renormalization,
\beq C(\vec{q}, \omega, u, \gamma, \Lambda) = Z^{-2}_2 C_r(\vec{q}, \omega, u_r, \gamma_r, \Lambda) + B(\Lambda/\mu, u,\gamma)  \eeq
The renormalized two-point function $C_r$ then satisfies the RG equation,
\beq \left(\mu \frac{\dd}{\dd \mu} + \beta(u_r) \frac{\dd}{\dd u_r} + b_\gamma(u_r) \gamma_r \frac{\dd}{\dd \gamma_r} - 2 \eta_2(u_r)\right) C_r(\omega, \vec{q}, 
u_r, \gamma_r, \mu) = \gamma_r X(u_r)\eeq
where
\beq \gamma_r X(u_r) = -Z^2_2 \mu \frac{\dd}{\dd \mu} B|_{u, \gamma}\eeq
Solving the RG equation, we obtain,
\beq C_r(\vec{q}, \omega, u_r, \gamma_r, \mu) = Z_2^{-2}(\lambda) C_r(\vec{q},\omega, u_r(\lambda), Z_\gamma(\lambda) \gamma_r, \lambda \mu)-\gamma_r\int_1^{\lambda} \frac{d \lambda'}{\lambda'} Z^{-2}_2(\lambda') Z_\gamma(\lambda') X(u_r(\lambda'))\eeq
\bea 
&&\lambda \frac{d u_r}{d \lambda} = \beta(u_r(\lambda))\\
&&\lambda \frac{d \log Z_2(\lambda)}{d \lambda} = \eta_2(u_r(\lambda))\\
&&\lambda \frac{d \log Z_\gamma(\lambda)}{d \lambda} = b_\gamma(u_r(\lambda))\eea
By dimensional analysis,
\beq C_r(\vec{q}, \omega, u_r, \gamma_r, \mu) = \gamma_r L(\frac{|\vec{q}|}{\mu}, \frac{\gamma_r |\omega|}{\mu^2}, u_r)\eeq
So,
\beq L(\tilde{q}, \tilde{\omega}, u_r) = Z^{-2}_2(\tilde{q}) Z_\gamma(\tilde{q}) L(1, Z_\gamma(\tilde{q})\frac{ \tilde{\omega}}{\tilde{q}^2}, u_r(\tilde{q})) - \int_1^{\tilde{q}} \frac{d \lambda'}{\lambda'} Z^{-2}_2(\lambda') Z_\gamma(u_r(\lambda')) X(u_r(\lambda')) \label{Gsol}\eeq
Now solving the flow equations using the leading order anomalous dimensions, we obtain,
\bea &&u_r(\lambda) = \frac{u_r}{1 - c_u u_r \log \lambda}\\
&&Z_2 = (1 - c_u u_r \log \lambda)^{-c_2/c_u}\\
&&Z_\gamma = 1 + \frac{c_\gamma}{c_u} \frac{u^2_r \log \lambda}{1 - u_r \log \lambda}\eea
Thus, for $\tilde{q} \to 0$ and to leading order in $u_r$,
\beq L(\tilde{q}, \tilde{\omega},u_r) \approx (c_u u_r |\log \tilde{q}|)^{2 c_2/c_u} \left(L(1, \frac{\tilde{\omega}}{\tilde{q}^2}, 0) + \frac{X(0)}{c_u u_r}\right) + \frac{X(0)}{(2c_2/c_u + 1) c_u u_r} (c_u u_r |\log \tilde{q}|^{2c_2/c_u + 1} -1) \label{Gsol2}\eeq
We see that the second term Eq.~(\ref{Gsol2}), which originates from the additive renormalization, dominates over the first term. In fact corrections to this second term coming from considering higher order contributions in $u_r$ to $\eta_2$ and $X$ are of the same order as the first term in Eq.~(\ref{Gsol2}). Nevertheless, the first term contains the dependence on $\tilde{\omega}$ and this is the reason we have kept it. 

To complete the calculation, we need to compute $C$ in the non-interacting theory (Fig.~\ref{fig:phi21loop}), 
\beq C_0(\vec{q}, \omega) = 6 \int \frac{d^2 \vec{l} dl_\tau}{(2 \pi)^3} D(l) D(l+q) \eeq
Using the Feynman trick,
\bea \label{eq:feyntrick} C_0(\vec{q}, \omega) &=& 6 \gamma^2 \int_0^{1} du \int \frac{d^2 \vec{l} dl_\tau}{(2 \pi)^3} \frac{1}{((\vec{l} - u \vec{q})^2 + \gamma (u |l_\tau| + (1-u) |l_\tau - \omega|) + u(1-u) \vec{q}^2)^2}\nn\\
&=& \frac{3\gamma^2}{(2 \pi)^2} \int_0^1 du \int_{-\infty}^{\infty} dl_\tau \frac{1}{\gamma(u |l_\tau - \omega| + (1-u) |l_\tau|) + u(1-u) \vec{q}^2}\nn\\
&=& \frac{3 \gamma}{2\pi^2} \int_0^1 du \left[\log \frac{\Lambda^2}{\gamma u |\omega| + u(1-u) \vec{q}^2}  + \frac{1}{2(1-2u)}\log \frac{\gamma(1-u)|\omega| + u(1-u) \vec{q}^2}{\gamma u |\omega| + u(1-u) \vec{q}^2}\right]\nn\\\eea
where we have cut-off the $l_\tau$ integral at $l_\tau = \gamma \Lambda^2$. Performing the $u$ integral,
\bea C_0(\vec{q}, \omega) &=& \frac{3 \gamma}{2\pi^2} \Bigl(\log \frac{\Lambda^2}{\gamma |\omega| + \vec{q}^2} + \frac{\gamma |\omega|}{\vec{q}^2}\log \left(\frac{\gamma |\omega|}{\gamma |\omega| + \vec{q}^2}\right) \nn \\
&~& \quad - \frac{1}{2} {\rm Li}_2\left(\frac{\vec{q}^2}{2\gamma |\omega| + \vec{q}^2}\right)+ \frac{1}{2} {\rm Li}_2\left(\frac{-\vec{q}^2}{2\gamma |\omega| + \vec{q}^2}\right)\Bigr)\eea
with ${\rm Li}_2$ - the polylogarithm function. After renormalization,
\bea C_{0r}(\vec{q}, \omega) &=& \frac{3 \gamma}{2\pi^2} \Bigl(\log \frac{\mu^2}{\gamma |\omega| + \vec{q}^2} + \frac{\gamma |\omega|}{\vec{q}^2}\log \left(\frac{\gamma |\omega|}{\gamma |\omega| + \vec{q}^2}\right) \nn \\
&~& \quad - \frac{1}{2} {\rm Li}_2\left(\frac{\vec{q}^2}{2\gamma |\omega| + \vec{q}^2}\right)+ \frac{1}{2} {\rm Li}_2\left(\frac{-\vec{q}^2}{2\gamma |\omega| + \vec{q}^2}\right)\Bigr)\eea
Giving the final result,
\bea L(\tilde{q}, \tilde{\omega}, u_r) &=& \frac{33}{\pi^2} (c_u u_r)^{-10/11} |\log \tilde{q}|^{1/11} \bigg[1 + \frac{1}{22 |\log{\tilde{q}}|} \Big( \frac{|\tilde{\omega}|}{\tilde{q}^2} \log \frac{|\tilde{\omega}|}{\tilde{q}^2}-\left(1 + \frac{|\tilde{\omega}|}{\tilde{q}^2}\right) \log\left(1+\frac{|\tilde{\omega}|}{\tilde{q}^2}\right) \nn\\
&-& \frac{1}{2}{\rm Li}_2\left(\frac{\tilde{q}^2}{2 |\tilde{\omega}| + \tilde{q}^2}\right) + \frac{1}{2}{\rm Li}_2\left(\frac{-\tilde{q}^2}{2 |\tilde{\omega}| + \tilde{q}^2}\right) + \frac{2}{c_u u_r}\Big)\bigg] - \frac{33}{\pi^2 c_u u_r}\eea

\section{Self energy in the $2 k_F$ channel (one-dimensional contribution)}\label{sec:oned}
In this section we compute the self energy in Fig.~\ref{fig:2kFself}b). We take all the external and internal fermions to be lukewarm and use the fermion propagator \ref{eq:Gcold2kF}. The self energy is given by,
\bea \Sigma(p) &=& - 42 N \lambda^8 \int \frac{d^3 q}{(2\pi)^3} \frac{d^3 l_1}{(2\pi)^3} \frac{d^3 l_2}{(2\pi)^3} \frac{d^3 l_3}{(2\pi)^3} D(p-l_1) D(l_1-l_2)D(l_2-l_3) D(l_3 -p) \nn\\
&& ~~~~~~~~~~~~~~\times G^{-1}(p-q) G^2(l_1) G^{-2}(l_1 - q) G^1(l_2) G^{-1}(l_2 -q) G^2(l_3) G^{-2}(l_3 -q) \nn\\\label{eq:2kFint}\eea
We take the boson propagators to be static and ignore their dependence on fermion momenta perpendicular to the Fermi surface,
\bea D(p - l_1) &\approx& N^{-1} \sin^2 2 \varphi \left((\hat{v}_2 \cdot \vec{p})^2 + (\hat{v}_1 \cdot \vec{l}_1)^2 - 2 \cos 2\varphi (\hat{v}_2 \cdot \vec{p}) (\hat{v}_1 \cdot \vec{l}_1)\right)^{-1}\nn\\
D(l_1-l_2) &\approx& N^{-1} \sin^2 2 \varphi \left((\hat{v}_1 \cdot \vec{l}_1)^2 +(\hat{v}_2 \cdot \vec{l}_2)^2- 2 \cos 2\varphi (\hat{v}_1 \cdot \vec{l}_1) (\hat{v}_2 \cdot \vec{l}_2)\right)^{-1}\nn\\
D(l_2-l_3) &\approx& N^{-1} \sin^2 2 \varphi \left((\hat{v}_2 \cdot \vec{l}_2)^2+ (\hat{v}_1 \cdot \vec{l}_3)^2 -2 \cos 2\varphi (\hat{v}_2 \cdot \vec{l}_2) (\hat{v}_1 \cdot \vec{l}_3)\right)^{-1}\nn\\
D(p-l_3) &\approx& N^{-1} \sin^2 2 \varphi \left((\hat{v}_2 \cdot \vec{p})^2 + (\hat{v}_1 \cdot \vec{l}_3)^2 - 2 \cos 2\varphi (\hat{v}_2 \cdot \vec{p}) (\hat{v}_1 \cdot \vec{l}_3)\right)^{-1}
\eea
Then integrals over $l_{i \tau}$ and components of $\vec{l}_i$ perpendicular to the Fermi surface in Eq.~(\ref{eq:2kFint}) can be performed,
\beq \int \frac{d^3 l}{(2\pi)^3} G^a(l) G^{-a}(l-q) \approx -\frac{|\vec{v}|^2}{16 \pi^2 v_x v_y} \int d (\hat{v}_{\bar{a}} \cdot \vec{l}) \frac{Z^2(\hat{v}_{\bar{a}} \cdot \vec{l})}{v^*(\hat{v}_{\bar{a}} \cdot \vec{l})} \log \left(\frac{\Lambda^2_{FL}(\hat{v}_{\bar{a}} \cdot \vec{l})}{q^2_\tau + v^*(\hat{v}_{\bar{a}} \cdot \vec{l})^2 (\hat{v}_a \cdot \vec{q})^2}\right)\eeq
Here, $\Lambda_{FL}(l) \sim l^2/\gamma$ is the upper energy cut-off of the region where the rainbow propagator (\ref{eq:fermionG0}) has the quasiparticle form (\ref{eq:Gcold2kF}). Hence,
\bea \Sigma(\omega, \vec{p}) &=& 42 \lambda^8 N (8 \pi^2 \sin 2 \varphi v \Lambda)^{-3} \int \frac{d^3 q\, d (\hat{v}_1 \cdot \vec{l}_1)d(\hat{v}_2 \cdot \vec{l}_2)d(\hat{v}_1 \cdot \vec{l}_3)}{(2\pi)^3} \nn\\
&& D(p-l_1)D(l_1-l_2)D(l_2-l_3)D(l_3 - p) |\hat{v}_1 \cdot \vec{l}_1||\hat{v}_2 \cdot \vec{l}_2||\hat{v}_1 \cdot \vec{l}_3| \nn\\
&&\frac{Z(\hat{v}_2 \cdot \vec{p})}{i (\omega - q_\tau) + v^*(\hat{v}_2 \cdot \vec{p}) \hat{v}_1 \cdot (\vec{p}-\vec{q})} \log\left(\frac{\Lambda^2_{FL}(\hat{v}_2 \cdot \vec{l}_2)}{q^2_\tau + v^*(\hat{v}_2 \cdot \vec{l}_2)^2 (\hat{v}_1 \cdot \vec{q})^2}\right)\nn\\
&&\log\left(\frac{\Lambda^2_{FL}(\hat{v}_1 \cdot \vec{l}_1)}{q^2_\tau + v^*(\hat{v}_1 \cdot \vec{l}_1)^2 (\hat{v}_2 \cdot \vec{q})^2}\right)\log\left(\frac{\Lambda^2_{FL}(\hat{v}_1 \cdot \vec{l}_3)}{q^2_\tau + v^*(\hat{v}_1 \cdot \vec{l}_3)^2 (\hat{v}_2 \cdot \vec{q})^2}\right)\label{eq:2kFfact}\eea
The crucial observation is that the integrals over $\hat{v}_1 \cdot \vec{q}$ and $\hat{v}_2 \cdot \vec{q}$ in Eq.~(\ref{eq:2kFfact}) factorize. This is due to our treatment of the Fermi surfaces as flat. The leading contribution to the integral over $\hat{v}_2 \cdot \vec{q}$ comes from $2$ and $-2$ fermions away from the Fermi surface and we may approximate,
\bea && \int d (\hat{v}_2 \cdot \vec{q})\, \log\left(\frac{\Lambda^2_{FL}(\hat{v}_1 \cdot \vec{l}_1)}{q^2_\tau + v^*(\hat{v}_1 \cdot \vec{l}_1)^2 (\hat{v}_2 \cdot \vec{q})^2}\right)\log\left(\frac{\Lambda^2_{FL}(\hat{v}_1 \cdot \vec{l}_3)}{q^2_\tau + v^*(\hat{v}_1 \cdot \vec{l}_3)^2 (\hat{v}_2 \cdot \vec{q})^2}\right) \nn\\
&\sim& \min\left(\frac{\Lambda_{FL}(\hat{v}_1 \cdot \vec{l}_1)}{v^*(\hat{v}_1 \cdot \vec{l}_1)},\frac{\Lambda_{FL}(\hat{v}_1 \cdot \vec{l}_3)}{v^*(\hat{v}_1 \cdot \vec{l}_3)}\right) \sim \frac{1}{N} \min(|\hat{v}_1 \cdot \vec{l}_1|,|\hat{v}_1 \cdot \vec{l}_3|), \label{eq:dq2}\eea
because the domain of integration is limited to the regime where the arguments of the logarithms are large.
On the other hand, integration over $\hat{v}_1 \cdot \vec{q}$ and $q_\tau$ to logarithmic accuracy gives,
\bea && \int dq_\tau d(\hat{v}_1 \cdot \vec{q}) \frac{1}{i (\omega - q_\tau) - v^*(\hat{v}_2 \cdot \vec{p}) \hat{v}_1 \cdot \vec{q}} \log\left(\frac{\Lambda^2_{FL}(\hat{v}_2 \cdot \vec{l}_2)}{q^2_\tau + v^*(\hat{v}_2 \cdot \vec{l}_2)^2 (\hat{v}_1 \cdot \vec{q})^2}\right) \nn\\&=& -\frac{2 \pi i \omega}{v^*(\hat{v}_2 \cdot \vec{l}_2) + v^*(\hat{v}_2 \cdot \vec{p})} \log\left(\frac{\Lambda_{FL}(\hat{v}_2 \cdot \vec{p})}{|\omega|}\right)\label{eq:dq1}\eea
Here for simplicity we have set the external momentum $\vec{p}$ to lie on the Fermi surface. We have also set the argument of $\Lambda_{FL}$ inside the logarithm on the second line of Eq.~(\ref{eq:dq1}) to $\hat{v}_2 \cdot \vec{p}$.  We will see shortly that all integrals over momenta along the Fermi surface are saturated at $l_\parallel \sim p_\parallel$, so to logarithmic accuracy the precise value of the argument is not important. Combining Eqs.~(\ref{eq:2kFfact}),(\ref{eq:dq2}), (\ref{eq:dq1}),
\bea \Sigma(\omega, \vec{p}) &\sim& -i N^3 \gamma v \omega \int d (\hat{v}_1 \cdot \vec{l}_1)d(\hat{v}_2 \cdot \vec{l}_2)d(\hat{v}_1 \cdot \vec{l}_3) D(p-l_1) D(l_1-l_2)D(l_2-l_3) D(l_3 -p) \nn\\
&\times&|\hat{v}_1 \cdot \vec{l}_1||\hat{v}_2 \cdot \vec{l}_2||\hat{v}_1 \cdot \vec{l}_3| \min(|\hat{v}_1 \cdot \vec{l}_1|,|\hat{v}_1 \cdot \vec{l}_3|) \frac{|\hat{v}_2 \cdot \vec{p}|}{|\hat{v}_2 \cdot \vec{l}_2| + |\hat{v}_2 \cdot \vec{p}|} \log\left(\frac{\Lambda_{FL}(\hat{v}_2 \cdot \vec{p})}{|\omega|}\right)\nn\\\label{eq:Sigma2kFpar}\eea
In the integrand, we can now count 7 powers of $\vec{l}$ in the numerator, and 4 powers in the denominator from 2 of the $D$ functions;
so the integral in Eq.~(\ref{eq:Sigma2kFpar}) is not singular for $\vec{l} \to 0$, which confirms our claim that the self energy in the $2k_F$ channel is not dominated by scattering off hot modes. The integral is saturated at  $\hat{v}_1 \cdot \vec{l}_1, \hat{v}_2 \cdot \vec{l}_2, \hat{v}_1 \cdot \vec{l}_3 \sim \hat{v}_2 \cdot \vec{p} \sim p_\parallel$, and all the 4 $D$ functions yield a contribution of $1/p_\parallel^2$ each;
this leads to our estimate in Eq.~(\ref{eq:Sigma1D}). 

\section{The  $2 k_F$ correlation function (curvature saturated contribution)}
\label{sec:2kfprop}

We can first compute the `tree level' lowest order contribution in figure \ref{fig:prop}. This is simply
\be\label{eq:Cdensity0}
C^{(0)}(m) = \int \frac{d^3t}{(2\pi)^3} G_2(m+t) G_{-2}(t) \,.
\ee
As indicated in the main text, we work with the lukewarm propagators following
from (\ref{eq:fermionG}) and (\ref{eq:lukewarm}), together with an additional curvature term
\be\label{eq:lukekappa}
G^{-1}_{\pm 2}(q) = \mp v q_\perp + \frac{3 \g v}{16 N} \frac{i q_{\tau}}{|q_\parallel|} - \kk q_\parallel^2 \,.
\ee
Recall that the lukewarm regime is
\be\label{eq:inequalities}
t_\parallel \; \gg \;  \sqrt{\g t_{\tau}}, N t_\perp  \,.
\ee
We can restrict ourselves to this regime because we are looking to pick out a BCS-like logarithmic divergence that is cut off in the particle channel by curvature effects. Here we are defining parallel and perpendicular with respect to the $\ell=1,a=2$ hot spot Fermi surface.

Using (\ref{eq:lukekappa}) it is possible to explicitly perform the integrals in (\ref{eq:Cdensity0}). One computes first the $t_\perp$ integral, followed by the $t_{\tau}$ integral. The $t_{\tau}$ integral is logarithmically UV sensitive, and the divergence is cut of by $t_\parallel^2/\g$.
We noted in the main text that the self energy scaling determined by hot spot scaling is weak, $\Sigma \sim \w^{5/2}$. Here we are interested in determining the effect of BCS-like contributions that will violate this scaling. We will be interested in the upper limit of the remaining $t_\parallel$ integral and thus we are free to focus on
\be\label{eq:lower}
t_\parallel \gg \sqrt{\g m_{\tau}}, |\vec m| \,.
\ee
In this regime we have
\be
\frac{1}{(2\pi)^3} \int dt_{\tau} dt_\perp G_2(m+t) G_{-2}(t) = \frac{c_1}{2} X(t_\parallel,m_{\tau},m_{\perp,2}) \,,
 \ee
where we introduced, also for future use,
\be\label{eq:X}
X(t_\parallel,m_{\tau},m_{\perp,2}) \equiv  t_\parallel \log \frac{\g^2 m_{\tau}^2 + c_2^2 t_\parallel^2 \left(m_{\perp,2} +2 \kk/v t_\parallel^2 \right)^2}{4 t_\parallel^4} \,.
\ee
The notation $m_{\perp,2}$ refers to the component of $m$ perpendicular to the $\pm 2$ Fermi surfaces.
The constants
\be
c_1 = \frac{4 N}{3 \pi^2 \g v^2} \,, \qquad c_2 = \frac{16 N}{3} \,.
\ee

The remaining integral of (\ref{eq:X}) over $t_\parallel$ can also be performed exactly. The most IR singular terms are found to be (we mean terms that are not saturating the UV cutoff scale and also that are not the IR terms with $C \sim m_{\tau}$ arising in the absence of Fermi surface curvature)
\be
C^{(0)}(m) = - \frac{c_1}{2} \frac{(\g |m_{\tau}|)^{2/3}}{(c_2 \kk/v)^{2/3}} 
\sum_{i} \frac{3+2 r^2 x_i + 4 r x_i^2}{r^2 + 8 r x_i + 12 x_i^2} \log \frac{- x_i \kk^2 (\g |m_{\tau}|)^{2/3}}{(c_2 \kk/v)^{2/3}} \,.
\ee
Here r is the ratio
\be\label{eq:rratio}
r = \frac{m_{\perp,2}}{(\g |m_{\tau}|)^{2/3}} \frac{c_2^{2/3}}{(\kk/v)^{1/3}} \,,
\ee
and the $x_i$ are the three roots of the polynomial
\be\label{eq:xpol}
4 x^3 + 4 r x^2 + r^2 x + 1 = 0 \,.
\ee
In the main text, in section \ref{sec:2kfself}, we discuss the physical interpretation of this formula.

Considering the right hand graph in figure \ref{fig:prop}, we must incorporate the vertex corrections
\be\label{eq:gamma}
\dd \Gamma(m,s) =  \int \frac{d^3 t}{(2\pi)^3} G_2(m+t) G_{-2}(t) D(t-s) \,,
\ee
so that the diagram can be written
\be\label{eq:Cdensity}
C(m) = \int \frac{d^3s}{(2\pi)^3} \dd \Gamma(m,s)^2 G_1(m+s) G_{-1}(s) \,.
\ee
We used the fact that the boson propagator $D$ is symmetric.

Nine integrals are required in principle to evaluate (\ref{eq:Cdensity}). We are interested in isolating as strong a contribution as possible in order to lead to the strongest self energy and hence dissipative conductivity. We shall take cues from
the computation of the density wave vertex correction in Ref.~\onlinecite{Metlitski:2010vm}. The complication in the case at hand is that we have an additional external energy-momentum $s$ as well as the spatial components $\vec m$.

The first lesson and simplification from Ref.~\onlinecite{Metlitski:2010vm} is that enhancement physics is dominated by lukewarm fermions.
A second lesson that appeared already in the tree level computation we have just performed is that it is important to account for the effects of Fermi surface curvature. While the curvature is an irrelevant operator of the scaling theory, it will cut off the BCS-like logarithmic divergence and therefore sets an energy scale that will be saturated in the processes we are interested in. 

Thus both the $t$ and $s$ integrals can be performed using the propagators (\ref{eq:lukekappa}) in the lukewarm regime and additionally satisfying (\ref{eq:lower}). The momentum $s$ of the integrals in (\ref{eq:Cdensity}) however lives on the $+1$ rather than the $+2$ Fermi surface and we will therefore use $s_\parallel$ and $s_\perp$ to denote the components of $s$ with respect to that $+1$ Fermi surface. In the lukewarm regime, the boson propagator (\ref{eq:bosonD}) appearing in the vertex correction (\ref{eq:gamma}) can be written
\be\label{eq:Dluke}
D(t-s) = D(t_\parallel,s_\parallel) = \frac{1}{t_\parallel^2 + s_\parallel^2 - 2 t_\parallel s_\parallel \text{cos}\, 2\varphi} \,.
\ee
We can then perform the $s_\perp$ and $s_{\tau}$ integrals in (\ref{eq:Cdensity}) in the same way as we did the $t_\perp$ and $t_{\tau}$ integrals above. We obtain
\be\label{eq:cm}
C(m) = c_1^3 \int_0^{\Lambda_\parallel} ds_\parallel X(s_\parallel,m_{\tau},m_{\perp,1}) \left( \int_0^{\Lambda_\parallel} dt_\parallel D(t_\parallel,s_\parallel)  X(t_\parallel,m_{\tau},m_{\perp,2}) \right)^2  \,.
\ee
The lower limits of the integrals are determined by (\ref{eq:lower}) but will not be important for isolating the most singular contribution. In particular, the lower limit of integration can be taken to zero for our purposes.

The dependence on the angle $2 \varphi$ between the Fermi surfaces does not in fact appear in the final leading logarithmic result for these integrals. In order to keep the size of intermediate equations down, we will quote intermediate results with $\cos 2 \varphi = 0$. The $t_\parallel$ integral can be evaluated explicitly in terms of logarithms and dilogarithms
\bea\label{eq:tdone}
\lefteqn{\int^{\Lambda_\parallel}_0 dt_\parallel D(t_\parallel,s_\parallel)  X(t_\parallel,m_{\tau},m_{\perp,2}) = 
\frac{1}{2} \log^2 \tilde s_\parallel^2 + \frac{1}{4} \log^2 \tilde \Lambda_\parallel^2 - \frac{1}{3} \log \g m_{\tau} (c_2 \kk/v)^2 \log \frac{\tilde s_\parallel^2}{\tilde \Lambda_\parallel^2}}
\nonumber \\
& & \qquad \qquad \qquad \qquad - \frac{1}{4} \sum_i \left(\log \frac{1}{\tilde s_\parallel^2+x_i}\log \frac{(-x_i)^2}{\tilde s_\parallel^2+x_i} + 2 \text{Li}_2 \frac{x_i}{\tilde s_\parallel^2+x_i} \right) \,. \qquad \qquad
\eea
Here we set
\be
\tilde s_\parallel^2 = \frac{(c_2 \kk/v)^{2/3}}{(\g m_{\tau})^{2/3}} s_\parallel^2 \,, \qquad 
\tilde \Lambda_\parallel^2 = \frac{(c_2 \kk/v)^{2/3}}{(\g m_{\tau})^{2/3}} \Lambda_\parallel^2 \,, 
\ee
and the $x_i$ are again the roots of the cubic polynomial (\ref{eq:xpol}) above. We have only quoted the result to leading order in logarithms, i.e. order logarithm squared. There are additional constant terms and terms with single logarithms.

The remaining $s_\parallel$ integral in (\ref{eq:cm}) appears more challenging to compute exactly. It is however possible to extract the leading logarithmic behavior. Recall we are interested in the most singular behaviour as $m_{\tau} \to 0$. We are keeping the ratio (\ref{eq:rratio}) fixed in taking this limit. In our result (\ref{eq:tdone}) $m_{\tau}$ appears both explicitly and through the rescaled $s_\parallel$ and $\Lambda_\parallel$. We will rescale
the $s_\parallel$ variable of integration and integrate instead over $\tilde s_\parallel$. After squaring our result (\ref{eq:tdone}) there will be a term $\sim \frac{1}{81} \log^4 m_{\tau}$, from the $m_{\tau}$ in the $\tilde \Lambda_\parallel$ terms. This will contribute at leading order in logarithms of $m_{\tau}$ and therefore we need to integrate this term over the whole range of integration. The remaining terms can only compete with this term near the upper limit of the integral, where $\tilde s_\parallel \to \tilde \Lambda_\parallel$. Here extra powers of $m_{\tau}$ can appear. Both of these two contributions can be evaluated, in the latter case by expanding the integrand at large $\tilde s_\parallel$. The final result for the most singular contribution is
\bea\label{eq:cmfinal}
\lefteqn{C(m) = c_1^3 \frac{(\g m_{\tau})^{2/3}}{(c_2 \kk/v)^{2/3}} \log^2 \tilde \Lambda_\parallel^2 } \\
&& \times \Bigg( \frac{1}{32} \log^2 \left[ (\g m_{\tau})^{4/3} (c_2 \kk/v)^{8/3} \tilde \Lambda_\parallel^2\right]  \sum_i  \left(\log (-x_i) x_i - \log(-y_i) \frac{3 + 2 \bar r^2 y_i + 4 \bar r y_i^2}{\bar r^2 + 8 \bar r y_i + 12 y_i^2}  \right) \nonumber \\
&& + \;  \frac{r+\bar r}{2160} \log \tilde \Lambda_\parallel^2 \left(40 \log^2 \g m_{\tau} (c_2 \kk/v)^2 + 75 \log \g m_{\tau} (c_2 \kk/v)^2 \log \tilde \Lambda^2_\parallel + 36 \log^2 \tilde \Lambda^2_\parallel  \right) \Bigg) \,. \nonumber
\eea
Here we introduced the analogous ratio to (\ref{eq:rratio}) for the $+1$ Fermi surface
\be\label{eq:rratiox}
\bar r = \frac{m_{\perp,1}}{(\g |m_{\tau}|)^{2/3}} \frac{c_2^{2/3}}{(\kk/v)^{1/3}} \,,
\ee
and the $y_i$ are now the three roots of the polynomial
\be\label{eq:ypol}
4 y^3 + 4 \bar r y^2 + \bar r^2 y + 1 = 0 \,.
\ee
Equation (\ref{eq:cmfinal}) is the result to leading order in logarithms, i.e. keeping $\log^5$ terms, and with an overall scaling $m_{\tau}^{2/3}$. In one of these limits a power of logarithms is lost, but the result is still accurately captured by (\ref{eq:cmfinal}) in this case. In the above expression we can note that
\be
\sum_i \frac{3 + 2 \bar r^2 y_i + 4 \bar r y_i^2}{\bar r^2 + 8 \bar r y_i + 12 y_i^2} = \bar r \,, \qquad \sum_i x_i = - r \,.
\ee
In the main text, equation (\ref{eq:cmfinaltext}), we dropped the last line of (\ref{eq:cmfinal}) as it is linear in $r, \bar r$, which replace the overall frequency dependence $m_{\tau}^{2/3}$ with momenta $m_{\perp,2}$ and $m_{\perp,1}$.

\end{document}